\documentclass[english,12pt, journal, onecolumn]{IEEEtran}
\usepackage[T1]{fontenc}
\usepackage[latin9]{inputenc}
\usepackage{amsthm}
\usepackage{amsmath}
\usepackage{amssymb}
\usepackage{graphicx}
\usepackage{setspace}
\makeatletter

\providecommand{\tabularnewline}{\\}

\theoremstyle{plain}
\newtheorem{thm}{\protect\theoremname}
\theoremstyle{definition}
\newtheorem{defn}[thm]{\protect\definitionname}
\theoremstyle{plain}
\newtheorem{cor}[thm]{\protect\corollaryname}
\theoremstyle{plain}
\newtheorem{lem}[thm]{\protect\lemmaname}
\theoremstyle{definition}
\newtheorem{example}[thm]{\protect\examplename}
\theoremstyle{plain}
\newtheorem{prop}[thm]{\protect\propositionname}

\usepackage{setspace}
\usepackage{geometry}
\makeatother

\usepackage{babel}
\providecommand{\corollaryname}{Corollary}
\providecommand{\definitionname}{Definition}
\providecommand{\examplename}{Example}
\providecommand{\lemmaname}{Lemma}
\providecommand{\propositionname}{Proposition}
\providecommand{\theoremname}{Theorem}

\begin{document}

\title{\huge Strategic Formation of Heterogeneous Networks}

\maketitle
\vspace{-2cm}

\begin{center}
Eli A. Meirom, Shie Mannor, Ariel Orda
\par\end{center}

\begin{center}
Dept. of Electrical Engineering
\par\end{center}

\begin{center}
Technion - Israel Institute of Technology
\par\end{center}

\vspace{0.5cm}
\begin{abstract}
We establish a network formation game for the Internet's Autonomous
System (AS) interconnection topology. The game includes different
types of players, accounting for the heterogeneity of ASs in the Internet.
In this network formation game, the utility of a player depends on
the network structure, e.g., the distances between nodes and the cost
of links. We also consider the case where utility (or monetary) transfers
are allowed between the players. We incorporate reliability considerations
in the player's utility function, and analyze static properties of
the game as well as its dynamic evolution. We provide dynamic analysis
of topological quantities, and explain the prevalence of some ``network
motifs'' in the Internet graph. We assess our predictions with real-world
data.
\end{abstract}

\section{Introduction}

The structure of communication networks, in particular that of the
Internet, has carried much interest. The Internet is a living example
of a large, self organized, many-body complex system. Understanding
the processes that shape its topology would provide tools for engineering
its future.

The Internet is assembled out of multiple Autonomous Systems (ASs),
which are contracted by different economic agreements. These, in turn,
compose the routing pathways among the ASs. With some simplifications,
we can represent the resulting network as a graph, where two nodes
(ASs) are connected by a link if traffic is allowed to traverse through
them. The statistical properties of this ``Internet graph'', such
as the clustering properties, degree distribution etc., have been
thoroughly investigated (\cite{Gregori2013,Vazquez2002,Siganos2003})
However, without proper understanding of the mechanism that led to
this structure, the statistical analysis alone lacks the ability to
either predict the future evolution of the Internet nor to shape its
evolution.

A large class of models, a primary example is the ``preferential
attachment'' model \cite{Barabasi1999}, use probabilistic rules
in order to simulate the network evolution and recover some of its
statistical properties, e.g., its degree distribution. Yet, these
models fail to account for many other features of the network \cite{1019306}.
Possibly, one of the main reasons for that is they treat the ASs as
passive elements rather than economic, profit-maximizing entities.
An agent-based approach is a promising alternative. 

Game theory describes the behavior of interacting rational agents
and the resulting equilibria, and it is one of the main tools of the
trade in estimating the performance of distributed algorithms \cite{Borkar2007}.
Game theory has been applied extensively to fundamental control tasks
in communication networks, such as flow control \cite{Altman1994},
network security \cite{Roy2010}, routing \cite{Orda1993}, and wireless
network design \cite{Charilas2010}.

Recently, there has been increased activity in the field of \emph{network
formation games}. These studies aim to understand the network structure
that results from interactions between rational agents (\cite{Johari2006,Jackson1996}).
Different authors emphasized different contexts, such as wireless
networks \cite{5062080} or the inter-AS topology (\cite{Anshelevich2011,Alvarez2012}).
The theme of such research is to investigate the equilibria's properties,
e.g., establishing their existence and obtaining bounds on the ``price
of anarchy'' and ''price of stability''. These metrics measure from
above and below, correspondingly, the social cost deterioration at
an equilibrium compared with a (socially) optimal solution. Alternatively, agent-based simulations are used in order to obtain
statistical characteristics of the resulting topology \cite{Lodhi2012a}.

Nonetheless, the vast majority of these studies assume that the players
are identical, whereas the Internet is a heterogeneous mixture of
various entities, such as CDNs, minor ISPs, tier-1 ASs etc. There
are only a few studies that have explicitly considered the effects
of heterogeneity on the network structure. Some examples include \cite{Alvarez2012},
which extends a previous model of formation games for directed networks
\cite{Johari2006}, and in the context of social networks, \cite{Vandenbossche2012}. 

Most of the studies on the application of game theory to networks,
with very few exceptions, e.g., \cite{Arcaute2013}, focused on static
properties of the game. This is particularly true for network formation
games. However, it is not clear that the Internet has reached an equilibrium.
Indeed, ASs continuously draw new contracts, some merge with others
while other quit business. In fact, a dynamic inspection of the inter-AS
network suggests that the system may be far from equilibrium. Therefore,
a \emph{dynamic} study of an inter-AS network formation game is needed.

In addition, previous work ignores an important requirement that Autonomous
Systems has - reliability. Indeed, failures occur, and an AS must
face such events. While some game theoretic works addressed reliability
in other contexts (\cite{Bala,Haller2005}), to the best of our knowledge,
there are no works that considered the \emph{topological properties
}that emerge in a \emph{heterogeneous, dynamic,} network formation
game with \emph{reliability constraints.} 

We establish an analytically-tractable model, which explicitly accounts
for the heterogeneity of players as well as reliability requirements.
We base our model on the heralded Fabrikant model (\cite{Fabrikant2003,Corbo2005}).
We model the inter-AS connectivity as a network formation game with
\emph{heterogeneous} players that may share costs by monetary transfers.
We account for the inherent bilateral nature of the agreements between
players, by noting that the establishment of a link requires the agreement
of both nodes at its ends, while removing a link can be done unilaterally.
As reliability comes into play, agents may require to be connected
to other agents, or to all the other agents in the network, by at
least two disjoint paths. We investigate both the static properties
of the resulting game as well as its \emph{dynamic evolution}. 

Game theoretic analysis is dominantly employed as a ``toy model''
for contemplating about real-world phenomena. It is rarely confronted
with real-world data. In this study we go a step further from traditional
formal analysis, and we do consider real inter-AS topology data analysis
to support our theoretical findings. 

The main contributions of this paper, which summarizes the findings
in (\cite{Meirom2014,Meirom2015a}), are as follows:
\begin{itemize}
\item We evaluate static properties of the considered game, such as the
prices of anarchy and stability and characterize additional properties
of the equilibrium topologies. We introduce the concept of \textquotedbl{}\emph{price
of reliability}\textquotedbl{}, which is defined as the ratio of the
social cost with reliability constraints to the social cost with no
such additional constraints. Surprisingly, we show that this price
can be smaller than one, namely, that the additional reliability requirements
may increase the social utility.
\item We discuss the dynamic evolution of the inter-AS network, and calculate
convergence rates and basins of attractions for the different final
states. Our findings provide useful insight towards incentive design
schemes for achieving optimal configurations. Our model predicts the possible evolvement of some early emerging
minor players (say, regional ISPs) to a major player, e.g., a tier-1
or tier-2 ASs, and the  existence of a \emph{settlement-free} clique, and that most of the
other contracts between players include monetary transfers.
\item We provide dynamical analysis of topological quantities, and explain
the prevalence of some ``\emph{network motifs}'', i.e., sub-graphs
that appear frequently in the network. Through real-world data, we
provide encouraging support to our predictions.
\end{itemize}
In the next section, we describe our model. We discuss multiple variants,
corresponding to whether utility transfers (e.g., monetary transfers)
are allowed or not, and whether reliability constraints are in effect.
In Section 3, we provide static analysis. Dynamic analysis is presented
in Section 4. In section 5 we compare our theoretical predictions
with real-world data on inter-AS topologies. Finally, conclusions
are presented in Section 6.

\section{Model}

We assume that each AS is a player. While there are many types of
players, we aggregate them into two types: \emph{major league} (or
t\emph{ype-A}) players, such as major ISPs, central search engines
and the likes, and \emph{minor league} (or \emph{type-B}) players,
such as local ISP or small enterprises. Each player, regardless of its type,
may form contracts with other players, and should they reach a mutual
understanding, a link between them is formed. A player's strategy
is set by specifying which links it is interested in establishing,
and, if permissible, the price it will be willing to pay for each. 

We denote the set of type-A (type B) player by $T_{A}$ ($T_{B}$).
A link connecting node $i$ to node $j$ is denoted as either $(i,j)$
or $ij$. The total number of players is $N=|T_{A}|+|T_{B}|$, and
we assume $N\geq3$. The \emph{shortest distance} between nodes $i$
and $j$ is the minimal number of hops along a path connecting them
and is denoted by $d(i,j)$. Finally, the degree of node $i$ is denoted
by $deg(i)$ and the connected component of $i$, given the set of links $E,$
is $\Gamma(i).$ .

\subsection{Basic model}

Our cost function is based on the cost structure in \cite{Fabrikant2003}
and \cite{Corbo2005}. Players are penalized for their distance from
other players. First and foremost, players require a good, fast connection
to the major players, while they may relax their connection requirements
to minor players. Bandwidth usage and delay depends heavily on the
hop distance, and connection quality is represented by this metric.
We weight the relative importance of a major player by a factor $A>1$
in the cost function in the corresponding distance term. The link
prices represent factors such as the link\textquoteright s maintenance
costs, bandwidth allocation costs etc. Different player types may
incur different link costs, $c_{A},c_{B}$, due to varying financial
resources or infrastructure. Formally, the (dis-)utility of players
is represented as follows.
\begin{defn}
\emph{(basic mode)} The cost function,$\,\, C_{\beta}^{(bare)}(i)$,
of node $i$ of type $\beta\in\{A,B\}$, is defined as:\label{cost-definition}
\begin{eqnarray*}
C_{\beta}^{(bare)}(i) & \triangleq & deg(i)\cdot c_{\beta}+A\sum_{j\in T_{A}}d(i,j)+\sum_{j\in T_{B}}d(i,j)
\end{eqnarray*}

where $A>1$ represents the relative importance of class A nodes over
class B nodes. Then, the \emph{social cost} is defined as $\mathcal{S}=\sum_{i}C(i)$.

Set $c\triangleq\left(c_{A}+c_{B}\right)/2$. We assume $c_{A}\leq c_{B}.$
The optimal (minimal) social cost is denoted as $\mathcal{S}_{optimal}^{(bare)}$
(for brevity, $\mathcal{S}_{optimal}$).
\end{defn}
We denote the the social cost at the optimal stable solution as $\tilde{S}{}_{optimal}$.
The \emph{price of stability} is the ratio between the social cost
at the best stable solution and its value at the optimal solution,
namely $PoS=\tilde{S}{}_{optimal}/\mathcal{S}_{optimal}$. Similarly,
denote by $\tilde{S}{}_{pessimal}$ the highest social cost in an
equilibrium. Then, the \emph{price of anarchy} is the ratio between
the social cost at the worst stable solution and its value at the
optimal solution, namely $PoA=\tilde{S}{}_{pessimal}^ {}/\mathcal{S}_{optimal}^ {}$. 

We denote the change in cost of player $i$ as after the addition
(removal) of a link $(j,k)$ by $\Delta C(i,E+jk)\triangleq C\left(i,E\cup(j,k)\right)-C\left(i,E\right)$.
The establishment of a link requires the bilateral agreement of the
two parties at its ends, while removing a link can be done unilaterally.
This is known as a \emph{pairwise-stable }equilibrium \cite{Jackson1996,Arcaute2013}. 
\begin{defn}
The players' strategies are \emph{pairwise-stable }if for all $i,j\in T_{A}\cup T_{B}$,
the following hold:

a) if $ij\in E$, then $\Delta C(i,E-ij)>0$;

b) if $ij\notin E$, then either $\Delta C(i,E+ij)>0$ or $\Delta C(j,E+ij)>0$. 

The resulting graph is referred to as a \emph{stabilizable} graph.
\end{defn}
In many cases, ASs must maintain access to the Internet in case of
a single link failure. This is tantamount to the requirement that
all the players must have at least two \emph{link disjoint} paths
to each other node. If survivability constraints are in effect, we
generalize the basic model by modifying the distance term in the cost
function, and adding a term representing the distance along the backup
path. Nevertheless, if either link prices are high, crash frequencies
are low or the content of a minor AS is of little value, players may
relax their reliability requirements and demand the establishment
of disjoint paths only to the major players. This is represented by
a binary parameter $\tau$, which is set to one if two disjoint paths
are required to all nodes, and zero if the reliability requirement
holds for major players only. 

The relative weight of the primary path and the backup path is set
by the parameter $\delta.$ If failures are often, then the regular
and backup paths (in the corresponding pair of link-disjoint paths)
are used almost as frequently. As such, they must be weighted the
same in the cost function. In this case, the likelihood of using either
route is the same, and $\delta=1$. Conversely, if failures are rare,
traffic will be mostly carried across the shorter path. Therefore,
its length should carry more weight in the cost than the length of
the backup route, hence $\delta\ll1$. 
\begin{defn}
\emph{(cost function under survivability constraints)} The cost function,
$C_{\beta}^{(reliable)}(i)$, of node $i$ 
\begin{eqnarray*}
C_{\beta}^{(reliable)}(i) & \triangleq & deg(i)\cdot c_{\beta}+\frac{A}{1+\delta}\sum_{j\in T_{A}}\left(d(i,j)+\delta d'(i,j)\right)\\
 &  & +\tau\cdot\frac{1}{1+\delta}\sum_{j\in T_{B}}\left(d(i,j)+\delta d'(i,j)\right)+\left(1-\tau\right)\sum_{j\in T_{B}}d(i,j)
\end{eqnarray*}

where $d(i,j)$ and $d'(i,j)$ are the lengths of a pair of paths
between $i,j$ that minimizes the cost function. $d(i,j)$ denotes
the length of the shorter path. Formally, denote a pair of disjoint
paths connecting player $i$ and player $j$ as $\left(R_{(i,j)},R'_{(i,j)}\right)_{\alpha}$,
where $d_{\alpha}(i,j)$ $\left(d'_{\alpha}(i,j)\right)$ is the length
of shorter (correspondingly, longer) path. Set 
\[
\left(\hat{R}_{(i,j)},\hat{R'}_{(i,j)}\right)=\arg\min_{\left(R_{(i,j)},R'_{(i,j)}\right)_{\alpha}}C_{\beta}(i)
\]
 then $d(i,j)=\left\Vert \hat{R}_{(i,j)}\right\Vert $ and $d'(i,j)=\left\Vert \hat{R'}_{(i,j)}\right\Vert $.

We denote the change in cost of player $i$ as after the addition
(removal) of a link $(j,k)$ by $\Delta C(i,E+jk)\triangleq C\left(i,E\cup(j,k)\right)-C\left(i,E\right)$
(correspondingly, $\Delta C(i,E-jk)\triangleq C\left(i,E\setminus(j,k)\right)-C\left(i,E\right)$).
The abbreviation $\Delta C(i,jk)$ is often used. 
\end{defn}
If $\delta=1$, then the two routing pathways are used the same. In
this case, the shortest cycle length $d(i,j)+d'(i,j)$ is the relevant
quantity that appears in the cost function. This can be found in polynomial
time by Suurballe's algorithm (\cite{Suurballe1974,Bhandari1999}).
However, if $\delta\ll1$, routing will occur along two disjoint paths,
such that the length of the shortest between the two is shortest (among
all pairs of disjoint paths). Although the complexity of finding this
pair is NP-Hard, first finding the shortest path and then finding
the next shortest path is a heuristic that works remarkably well,
both in the real-world data analysis and on the networks obtained
in the theoretical discussion. The reason behind this is that, when
failures are rare, information is predominantly routed along the shortest
path. When players are required to establish a fall-back route, they
will establish a path that is disjoint from the \emph{current }routing
path, namely the shortest one.

The additional reliability requirements result in additional link
expenses, as for example, the degree of every node needs to be at
least two. The \emph{price of reliability} is the ratio between the
optimal social cost under the additional survivability constraint
to the optimal social cost when the additional constraints are removed.
\begin{defn}
The \emph{price of reliability }(\emph{PoR)} is the ratio between
the optimal value of the social costs among the set of corresponding
stable equilibria, $PoR=\tilde{S}_{optimal}/\tilde{S}_{optimal}^{(bare)}$.
\end{defn}
Surprisingly, we shall show that there exist scenarios in which reliability
requirements \emph{increase }the social utility, so that the price
of reliability can be smaller than one.

\subsection{Utility transfer}

In the above formulation, we have implicitly assumed that players
may not transfer utilities. However, often players are able to do
so, in particular via monetary transfers. We incorporate such possibility
by introducing an extended model, allowing for a monetary transaction
in which player $i$ pays player $j$ some amount $P_{ij}$ iff the
link $(i.j)$ is established. Player $j$ sets some minimal price
$w_{ij}$ and if $P_{ij}\geq w_{ij}$ the link is formed. The corresponding
change to the cost function is as follows.
\begin{defn}
\label{monetary cost definition}The cost function of player $i$
when monetary transfers are allowed is $\tilde{C}(i)\triangleq C(i)+\sum_{j,ij\in E}\left(P_{ij}-P_{ji}\right)$.
\end{defn}
In this definition the cost $C(i)$ may be either $C^{(reliable)}(i)$
or $C^{(bare)}(i)$. 

Note that the social cost remains the same as in Def. \ref{monetary cost definition}
as monetary transfers are canceled by summation.

Monetary transfers allow the sharing of costs. Without transfers,
a link will be established only if \emph{both} parties, $i$ and $j$,
reduce their costs, $C(i,E+ij)<0$ and $C(j,E+ij)<0$. Consider, for
example, a configuration where $\Delta C(i,E+ij)<0$ and $\Delta C(j,E+ij)>0$.
It may be beneficial for player $i$ to offer a lump sum $P_{ij}$
to player $j$ if the latter agrees to establish $(i,j)$. This will
be feasible only if the cost function of both players is reduced.
It immediately follows that if $\Delta C(i,E+ij)+\Delta C(i,E+ij)<0$
then there is a value $P_{ij}$ such that this condition is met. Hence,
it is beneficial for both players to establish a link between them.
In a game theoretic formalism, if the \emph{core} of the two players
game is non-empty, then they may pick a value out of this set as the
transfer amount. Likewise, if the core is empty, or $\Delta C(i,E+ij)+\Delta C(j,E+ij)>0$,
then the best response of at least one of the players is to remove
the link, and the other player has no incentive to offer a payment
high enough to change the its decision. Formally:
\begin{cor}
\textup{\emph{\label{lem:edges with monetary transfers.}When monetary
transfers are allowed, the link $(i,j)$ is established iff $\Delta C(i,E+ij)+\Delta C(j,E+ij)<0$.
The link is removed iff $\Delta C(i,E-ij)+\Delta C(j,E-ij)>0$.}}
\end{cor}
In the remainder of the paper, whenever monetary transfers are feasible,
we will state it explicitly, otherwise the basic model (without transfers
or survivability constraints) is assumed.

\section{\label{sec:Static-analysis}Static analysis}

In this section we discuss the properties of stable equilibria. Specifically,
we first establish that, under certain conditions, the major players
group together in a clique (section \ref{sub:The-type-A-clique}).
We shall further discuss topological properties that emerge from our
analysis (section \ref{sub:Pair-wise-equilibria}). 

As a metric for the quality of the solution we apply the commonly
used measure of the social cost, which is the sum of individual costs.
We evaluate the \emph{price of anarchy}, which is the ratio between
the social cost at the worst stable solution and its value at the
optimal solution, and the \emph{price of stability}, which is the
ratio between the social cost at the best stable solution and its
value at the optimal solution (section \ref{sub:PoA and PoS}). We
introduce utility transfers in section \ref{sec:Monetary-transfers},
and consider the impact of survivability requirements on the prices
of stability and anarchy in section \ref{sub:static-survivability}.

\subsection{\label{sub:The-type-A-clique}The type-A clique}

Our goal is understanding the resulting topology when we assume strategic
players and myopic dynamics. Obviously, if the link's cost is extremely
low, every player would establish links with all other players \cite{Fabrikant2003}.
The resulting graph will be a clique. This is not a realistic setting.

If two nodes are at a distance $L+1$ of each other, then there is
a path with $L$ nodes connecting them. By establishing a link with
cost $c$, we are shortening the distance between the end node to
$\sim L/2$ nodes that lay on the other side of the line. The average
reduction in distance is also $\approx L/2$, so by comparing $L^{2}\approx4c$
we obtain a bound on $L$, as follows:
\begin{lem}
\label{lem:The-longest-distance}The longest distance between any
node $i$ and node $j\in T_{B}$ is bounded by $2\sqrt{c_{B}}$. The
longest distance between nodes $i,j\in T_{A}$ is bounded by $\sqrt{(1-2A)^{2}+4c_{A}}-2\left(A-1\right)$.
In addition, if $c_{A}<A$ then there is a link between every two
type-A nodes.
\end{lem}
\begin{IEEEproof}
Assume $d(i,j)=k-1\geq\left\lfloor 2\sqrt{c_{B}}\right\rfloor >1$
and $i\in T_{B}$. Then there exist nodes $(x_{1}=i,x_{2},..x_{k}=j)$
such that $d(i,x_{\alpha+1})=\alpha$. By adding a link $(i,j)$ the
change in cost of node $i$, $\Delta C(i,E+ij)$ is, according to
lemma \ref{lem:shortcut benefit} in the appendix, 
\begin{align*}
 & \Delta C(i,E+ij)\\
= & c_{B}-\sum_{\alpha=1}^{k-1}d(i,x_{k})(1+\delta_{x_{\alpha},A}(A-1))\\
 & +\sum_{\alpha=1}^{k-1}d'(i,x_{\alpha})(1+\delta_{x_{\alpha},A}(A-1))\\
= & c_{B}-\sum_{\alpha=1}^{k-1}\left(d(i,x_{\alpha})-d'(i,x_{\alpha})\right)(1+\delta_{x_{\alpha},A}(A-1))\\
< & c_{B}-\sum_{\alpha=1}^{k-1}d(i,x_{\alpha})+\sum_{\alpha=1}^{k-1}d'(i,x_{\alpha})\\
< & c_{B}-\frac{k\left(k-2\right)+mod(k,2)}{4}\\
< & c_{B}-\frac{k\left(k-2\right)}{4}<0
\end{align*}
 where $d'(i,x_{\alpha})<d(i,x_{\alpha})$ is the distance after the
addition of the link $(i,j)$ and $\delta_{x_{\alpha},A}=1$ iff $x_{\alpha}\in T_{A}$.
Therefore, it is of the interest of player $i$ to add the link. 

Consider the case that $j\in T_{A}$ and 
\begin{equation}
d(i,j)=k-1\geq\sqrt{(1-2A)^{2}+4c_{B}}-2\left(A-1\right)>1\label{eq: distance}
\end{equation}
 The change in cost after the addition of the link $(i,j)$ is 
\begin{align*}
 & c_{B}-\sum_{\alpha=1}^{k-1}d(i,x_{\alpha})(1+\delta_{x_{\alpha},A}(A-1))\\
 & +\sum_{\alpha=1}^{k-1}d'(i,x_{\alpha})(1+\delta_{x_{\alpha},A}(A-1))\\
= & c_{B}-\sum_{\alpha=1}^{k-1}\left(d(i,x_{\alpha})-d'(i,x_{\alpha})\right)(1+\delta_{x_{\alpha},A}(A-1))\\
< & c_{B}-\sum_{\alpha=1}^{k-2}d(i,x_{\alpha})+\sum_{\alpha=1}^{k-2}d'(i,x_{\alpha})+A\left(d(i,x_{k})-d'(i,x_{k})\right)\\
< & c_{B}-\frac{k\left(k-2\right)+mod(k-1,2)}{4}-\left(k-1\right)A\\
 & +\left(k-1\right)+A-1\\
< & c_{B}-\frac{k\left(k-2\right)}{4}-\left(A-1\right)\left(k-2\right)\\
= & c_{B}-k^{2}/4-k(A-1)\\
< & 0
\end{align*}

Therefore it is beneficial for player $i$ to establish the link.
Similarly, if $i\in T_{A}$ then eq. \ref{eq: distance} is replaced
by $k-1\geq\sqrt{(1-2A)^{2}+4c_{A}}-2\left(A-1\right)>1$.

In particular, if we do not omit the $mod(k-1,2)$ term and set $k=3$
we get that if $2\sqrt{(-1+A)A+c_{A}}-2(A-1)<2$ the distance between
two type-A nodes is smaller than 2, in other words, they connected
by a link. The latter expression can be recast to the simple form
$c<A$.

Recall that if the cost of both parties is reduced (the change of
cost of node $j$ is obtained by the change of summation to $0..k-1$)
a link connecting them will be formed. Therefore, if $i,j\in T_{B}$
then maximal distance between then is $d(i,j)\leq max\{2\sqrt{c_{B}},1\}$
as otherwise it would be beneficial for both $i,j$ to establish a
link that will reduce their mutual distance to $1.$ Likewise, if
$i,j\in T_{A}$ then 
\[
d(i,j)\leq\sqrt{(1-2A)^{2}+4c_{A}}-2\left(A-1\right)
\]
using an analogous reasoning. If $i\in T_{A}$ and $j\in T_{B}$ then
it'll be worthy for player $i$ to establish the link only if $d(i,j)\geq\left\lfloor 2\sqrt{c_{B}}\right\rfloor $.
In this case it'll be also worthy for player $j$ to establish the
link since 
\[
d(i,j)\geq\left\lfloor 2\sqrt{c_{B}}\right\rfloor \geq\sqrt{(1-2A)^{2}+4c_{B}}-2\left(A-1\right)
\]
and the link will be established. Notice however that if 
\[
\left\lfloor 2\sqrt{c_{B}}\right\rfloor \geq d(i,j)\geq\sqrt{(1-2A)^{2}+4c_{A}}-2\left(A-1\right)
\]
 then although it is worthy for player $j$ to establish the link,
it is isn't worthy for player $i$ to do so and the link won't be
established. This concludes our proof.\end{IEEEproof}

Lemma \ref{lem:The-longest-distance} indicates that if $1<c_{A}<A$
then the type $A$ nodes will form a clique (the ``nucleolus'' of
the network). The type $B$ nodes form structures that are connected
to the type $A$ clique (the network nucleolus). These structures
are not necessarily trees and will not necessarily connect to a single
point of the type-A clique only. This is indeed a very realistic scenario,
found in many configurations. We shall compare this result to actual
data on the inter-AS interconnection topology in section \ref{sec:Data-Analysis}. 

If $c>A$ then the type-A clique is no longer stable. This setting
does not correspond to the observed nature of the AS topology and
we'll focus in all the following sections on the case $1<c_{A}<A$.

\subsection{\label{sub:Pair-wise-equilibria}Equilibria's properties}

Here we describe common properties of all pair-wise equilibria. We
start by noting that, unlike the findings of several other studies
\cite{Arcaute2013,5173479,NisanN.RoughgardenT.TardosE.2007}, in our
model, at equilibrium, the type-B nodes are not necessarily organized
in trees. This is shown in the next example.

\begin{figure}
\centering{}\includegraphics[width=0.75\columnwidth]{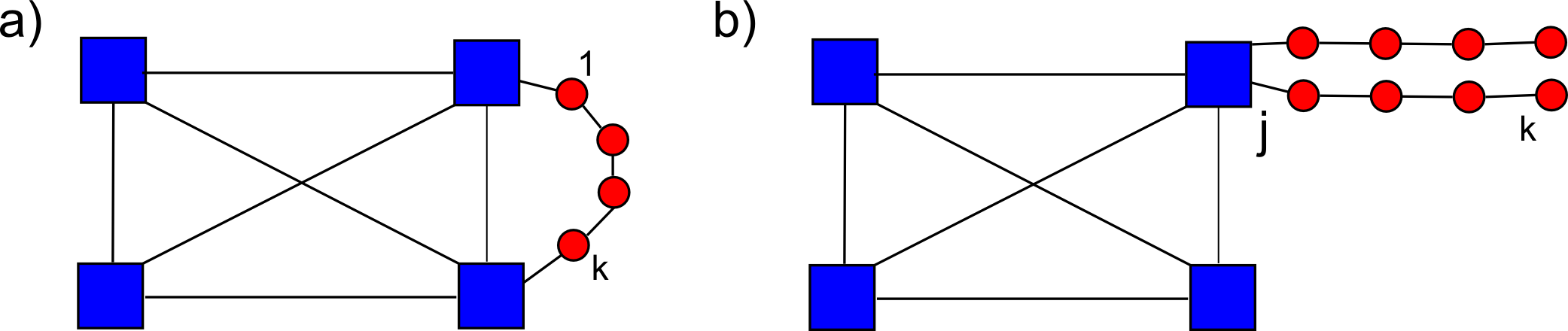}\protect\caption{\label{fig:loop example}Non optimal networks. The type-A clique is
in blue squares, the type-B players are in red circles. a) The network
described in Example \ref{example}. b) \label{fig:A-poor-equilibrium}A
poor equilibrium, as described in \cite{Meirom2013}.}
\end{figure}

\begin{example}
\label{example}Assume for simplicity that $c_{A}=c_{B}=c$. Consider
a line of length $k$ of type B nodes, $(1,2,3...,k)$ such that $\sqrt{8c}>k+1>\sqrt{2c}$
or equivalently $\left(k+1\right)^{2}<8c<4\left(k+1\right)^{2}$.
In addition, the links $(j_{1},1)$ and $(j_{2},k)$ exist, where
$j_{1},j_{2}\in T_{A}$, i.e., the line is connected at both ends
to different nodes of the type-A clique, as depicted in Fig \ref{fig:loop example}.
In \cite{Meirom2013} we show that this is a stabilizable graph.

\begin{IEEEproof}
We show that this structure is stabilizable. For simplicity, assume
$mod(k-1,4)=0$ ($k$ is odd and $\frac{k-1}{2}$ is odd).

\begin{figure}
\centering{}\includegraphics[width=1\columnwidth]{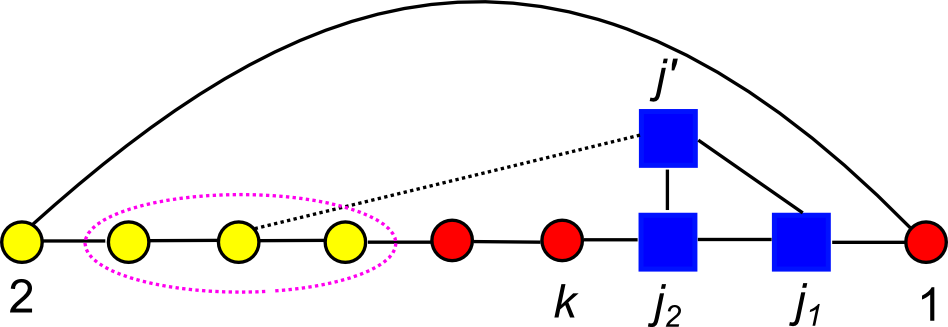}\protect\caption{\label{fig:example}A line of $k=7$ nodes. By removing the link $(1,2)$
only the distances from player 1 to the yellow players are affected.
By establishing the link $(j,4)$ only the distances from player $j'$
to the players encircled by the purple dashed ellipse are affected.}
\end{figure}

Any link removal $(x_{1},x_{2})$ in the circle $(j_{1},1...k,j_{2},j_{1})$
will result in a line with nodes $x_{1}$ and $x_{2}$ at its ends
(Fig. \ref{fig:example}). We now show that if it beneficial for player
$i$ to remove the edge $(i,i-1)$ for $i\leq\left\lceil \frac{k}{2}\right\rceil $,
then it is also true that it is beneficial for node $i-1$ to remove
the edge $(i-1,i-2)$. For player $i$, the removal of the edge $(i,i-1)$
affects only the distance to the other players on the loop. This is
also true for node $i-1$ and the removal of the edge $(i-1,i-2)$.
Denote the distances after the removal of the edge $(i,i-1)$ by $d_{i}\left(\cdot,\cdot\right)$.
A simple geometrical observation shows that 
\[
d\left(i,y\right)-d_{i}\left(i,y\right)\geq d\left(i-1,y\right)-d_{i-1}\left(i-1,y\right)
\]

for every $y=1...k$. In other words, the increase in the distance
cost for node $i$ after the removal of $(i,i-1)$ is greater than
the additional distance cost of player $i-1$ after the removal of
the edge $(i-1,i-2)$. Therefore, the type-B players that have the
most incentive to disconnect a link are either node $1$ or node $k$
( Fig. \ref{fig:example}). Therefore, if players $1$ or $k$ would
not deviate, no type-B player will deviate as well.

W.l.o.g, we discuss node 1. Since $c_{B}<A$, it is not beneficial
for it to disconnect the link $(j_{1},1)$. Assume the link $(1,2)$
is removed. A simple geometric observation shows that the distance
to nodes $\{2,..,\frac{k+3}{2}\}$ is affected, while the distance
to all the other nodes remains intact (Fig. \ref{fig:example}). The
mean increase in distance is $\frac{k+1}{2}$ and the number of affected
nodes is $\frac{k+1}{2}$ . However, 
\[
\Delta C(1,E-12)=c-\frac{(k+1)}{2}^{2}<0
\]
 and player $1$ would prefer the link to remain. The same calculation
shows that it is not beneficial for player $j_{1}$ to disconnect
$(j_{1},1)$.

Clearly, if it not beneficial for $j\in T_{A},\, j\neq j_{1},j_{2}$
to establish an additional link to a type-B player then it is not
beneficial to do so for $j_{1}$ or $j_{2}$ as well. The optimal
additional link connecting $j$ and a type-B player is $\mathcal{E}=(j,\frac{k+1}{2})$,
that is, a link to the middle of the ring (Fig. \ref{fig:example}).
A similar geometric observation shows that by establishing this link,
only the distances to nodes $\left\{ \frac{k-1}{4},...,\frac{3k+1}{4}\right\} $
are affected (Fig. \ref{fig:example}). The reduction in cost is
\[
\Delta C(j,E+\mathcal{E})=c-\frac{(k+1)}{8}^{2}>0
\]

and it is not beneficial to establish the link. 

In order to complete the proof, we need to show that no additional
type-B to type-B links will be formed. By establishing such link,
the distance of at least one of the parties to the type-A clique is
unaffected. The previous calculation shows that by adding such link
the maximal reduction of cost due to shortening the distance to type-B
players is bounded from above by $\frac{(k+1)}{8}^{2}$. Therefore,
as before, no additional type-B to type-B links will be formed.

This completes the proof that this structure is stabilizable.\end{IEEEproof}

\end{example}

Next, we bound from below the number of equilibria. For simplicity,
we discuss the case where $c_{A}=c_{B}=c$. We accomplish that by
considering the number of equilibria where the type-B players are
organized in a forest (multiple trees) and the allowed forest topologies.
The following lemma restricts the possible sets of trees in an equilibrium.
Intuitively, this lemma states that we can not have two ``heavy''
trees, ``heavy'' meaning that there is a deep sub-tree with many
nodes, as it would be beneficial to make a shortcut between the two
sub-trees.

\begin{figure}
\centering{}\includegraphics[width=0.7\columnwidth]{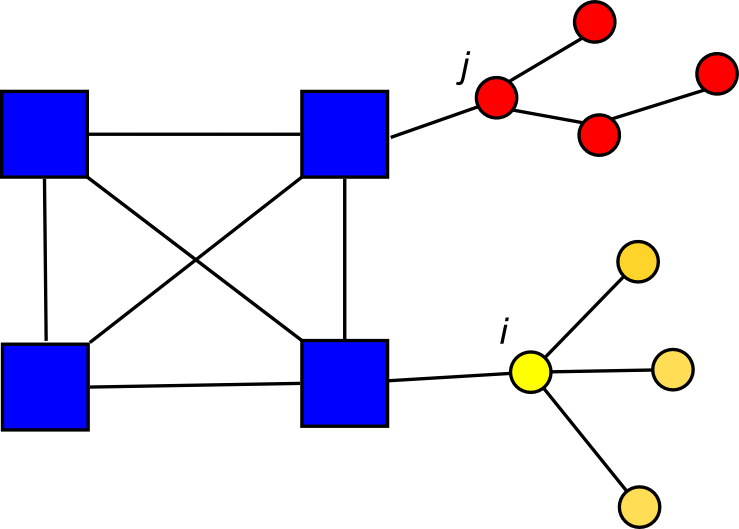}\protect\caption{\label{fig:forest}Node $j$ is on the third level of the tree of
formed by starting a BFS from node $i$, as discussed in lemma \ref{lem:subtree}.
The forest of type-B nodes is composed of two trees, in yellow and
red (lemma \ref{lem:forest sub-graph}). Their roots are $i$ and
$j,$ correspondingly. The maximal depth in this forest is three.}
\end{figure}

\begin{lem}
\label{lem:subtree}Assume $c_{A}=c_{B}=c$. Consider the BFS tree
formed starting from node $i.$ Assume that node $j$ is $k$ levels
deep in this tree. Denote the sub-tree of node $j$ in this tree by
$\mathcal{T}_{i}(j)$ (Fig. \ref{fig:forest}) In a link stable equilibrium,
the number of nodes in sub-trees satisfy either \textup{$|\mathcal{T}_{i}(j)|<c/k$
}\textup{\emph{or }}\textup{$|\mathcal{T}_{j}(i)|<c/k$.}\end{lem}
\begin{IEEEproof}
Assume $|\mathcal{T}_{i}(j)|>c/k$. Consider the change in cost of
player $i$ after the addition of the link $(i,j)$
\begin{eqnarray*}
 &  & \Delta C(i,E+ij)\\
 & = & c+\sum_{x_{\alpha}\in\mathcal{T}i(j)}d'(i,x_{\alpha})(1+\delta_{x_{\alpha},A}(A-1))\\
 &  & +\sum_{x_{\alpha}\notin\mathcal{T}_{i}(j)}d'(i,x_{\alpha})(1+\delta_{x_{\alpha},A}(A-1))\\
 &  & -\sum_{x_{\alpha}\in\mathcal{T}_{i}(j)}d(i,x_{\alpha})(1+\delta_{x_{\alpha},A}(A-1))\\
 &  & -\sum_{x_{\alpha}\notin\mathcal{T}_{i}(j)}d(i,x_{\alpha})(1+\delta_{x_{\alpha},A}(A-1))\\
 & = & c+\sum_{x_{\alpha}\in\mathcal{T}_{i}(j)}\left(d'(i,x_{\alpha})-d(i,x_{\alpha})\right)(1+\delta_{x_{\alpha},A}(A-1))\\
 &  & +\sum_{x_{\alpha}\notin\mathcal{T}_{i}(j)}\left(d'(i,x_{\alpha})-d(i,x_{\alpha})\right)(1+\delta_{x_{\alpha},A}(A-1))\\
 & < & c+\sum_{x_{\alpha}\in\mathcal{T}_{i}(j)}\left(d'(i,x_{\alpha})-d(i,x_{\alpha})\right)\\
 & = & c-k|T_{i}(j,k)|\\
 & < & 0
\end{eqnarray*}

since the distance was shorten by $k$ for every node in the sub-tree
of $j$. 

Therefore, it is beneficial for player $i$ to establish the link.
Likewise, if $|\mathcal{T}_{j}(i)|<c/k$ then it would be beneficial
for player $j$ to establish the link and the link will be established.
Hence, one of the conditions must be violated.
\end{IEEEproof}
The following lemma considers the structure of the type-B players'
sub-graph. It builds on the results of lemma \ref{lem:subtree} to
reinforce the restrictions on trees, showing that trees must be shallow
and small. 
\begin{lem}
\label{lem:forest sub-graph}Assume $c_{A}=c_{B}=c$. If the sub-graph
of type-B nodes is a forest (Fig. \ref{fig:forest}), then there is
at most one tree with depth greater than $\sqrt{c/2}$ and there is
at most one tree with more than $c/2$ nodes. The maximal depth of
a tree in the forest is $\sqrt{2c}-1$. Every type-B forest in which
every tree has a maximal depth of $\min\left\{ \sqrt{c/2},\sqrt{c}-3\right\} $
and at most $\min\left\{ c/2,\sqrt{c}\right\} $ nodes is stabilizable.\end{lem}
\begin{IEEEproof}
Assume there are two trees $S_{1},S_{2}$ that have depth greater
than $\sqrt{c/2}$ . The distance between the nodes at the lowest
level is greater than $2\sqrt{c/2}+1$ as the trees are connected
by at least one node in the type-A clique, $d(i,j)\geq2$ (Fig. \ref{fig:forest}).
This contradicts with Lemma \ref{lem:The-longest-distance}. 

Assume there are two trees $S_{1},S_{2}$ with roots $i,j$ that have
more than $c/2$ nodes. In the BFS tree in which node $i$ is the
root, node $j$ is at least in the second level (as they are connected
by at least one node in the type-A clique). This contradicts with
Lemma \ref{lem:subtree}.

Finally, following the footsteps of Lemma \ref{lem:subtree} proof,
consider two trees $\mathcal{T}_{i}$ and $\mathcal{T}_{j}$, with
corresponding roots $i$ and $j$ (i.e., nodes $i$ and $j$ have
a direct link with the type-A clique). Consider a link $(x,y)$, where
$x\in\mathcal{T}_{i}$ and $y\in\mathcal{T}_{j}$. At least one of
them does not reduce its distance to the type-A clique by establishing
this link. W.l.o.g, we'll assume this is true for player $x$. Therefore,
\begin{eqnarray*}
 &  & \Delta C(x,E+xy)\\
 & = & c+\sum_{z\in\mathcal{T}_{j}}\left(d'(z,x)-d(z,x)\right)\\
 & > & 0
\end{eqnarray*}

as the maximal reduction in distance is $\sqrt{c}$ and the maximal
number of nodes in $\mathcal{T}_{j}$ is also $\sqrt{c}$. Therefore,
it is not beneficial for player $x$ to establish this link, and the
proof is completed.\end{IEEEproof}

Finally, the next proposition provides a lower bound on the number
of link-stable equilibria by a product of $|T_{A}|$ and a polynomial
with a high degree ($\approx2^{\sqrt{c}}$) in $|T_{B}|$.
\begin{prop}
Assume $c_{A}=c_{B}=c$. The number of link-stable equilibria in which
the sub-graph of $T_{B}$ is a forest is at least $\omega(|T_{A}||T_{B}|^{N_{c}})$,
where $N_{c}=o(2^{\frac{\sqrt{c}}{2}}/\sqrt{c})$ is a function of
$c$ only. Therefore, the number of link-stable equilibria is at least
\textup{$o(|T_{A}||T_{B}|^{N_{c}}).$}\end{prop}

\begin{IEEEproof}
We bound the number of link-stable equilibria from below by considering
the number of link-stable graph, in which the minor players are organized
in trees of up to depth $\sqrt{c/2}$ and exactly $\sqrt{c}$ nodes. 

The number of such trees can be bounded from below by the number of
trees with $\sqrt{c}$ elements, depth $\sqrt{c/2}$ and only one
non-leaf node at each level of tree, which is
\[
N_{C}\geq\left(\begin{array}{c}
\left\lfloor \sqrt{c}\right\rfloor \\
\left\lfloor \sqrt{c/2}\right\rfloor 
\end{array}\right)\in\omega(2^{\frac{\sqrt{c}}{2}}/\sqrt{c})
\]

Therefore, the minor players can be organized in at least $\left(\begin{array}{c}
|T_{B}|+N_{c}\\
N_{c}
\end{array}\right)$ different configurations. Using Striling's approximation, we have
$\left(\begin{array}{c}
|T_{B}|+N_{c}\\
N_{c}
\end{array}\right)\in\omega(|T_{B}|^{N_{c}})$. Each tree can be connected to either one of the type-A nodes, and
therefore the number of possible configurations is at least $\omega(|T_{A}||T_{B}|^{N_{c}})$
. \end{IEEEproof}

To sum up, while there are many equilibria, in all of them nodes cannot
be too far apart, i.e., a small-world property. Furthermore, the trees
formed are shallow and are not composed of many nodes.

\subsection{\label{sub:PoA and PoS}Price of Anarchy \& Price of Stability }

As there are many possible link-stable equilibria, a discussion of
the price of anarchy is in place. First, we explicitly find the optimal
configuration. Although we establish a general expression for this
configuration, it is worthy to also consider the limiting case of
a large network, $|T_{B}|\gg1,|T_{A}|\gg1$. Moreover, typically,
the number of major league players is much smaller than the other
players, hence we also consider the limit $|T_{B}|\gg|T_{A}|\gg1$.

\begin{figure}
\centering{}\includegraphics[width=0.7\columnwidth]{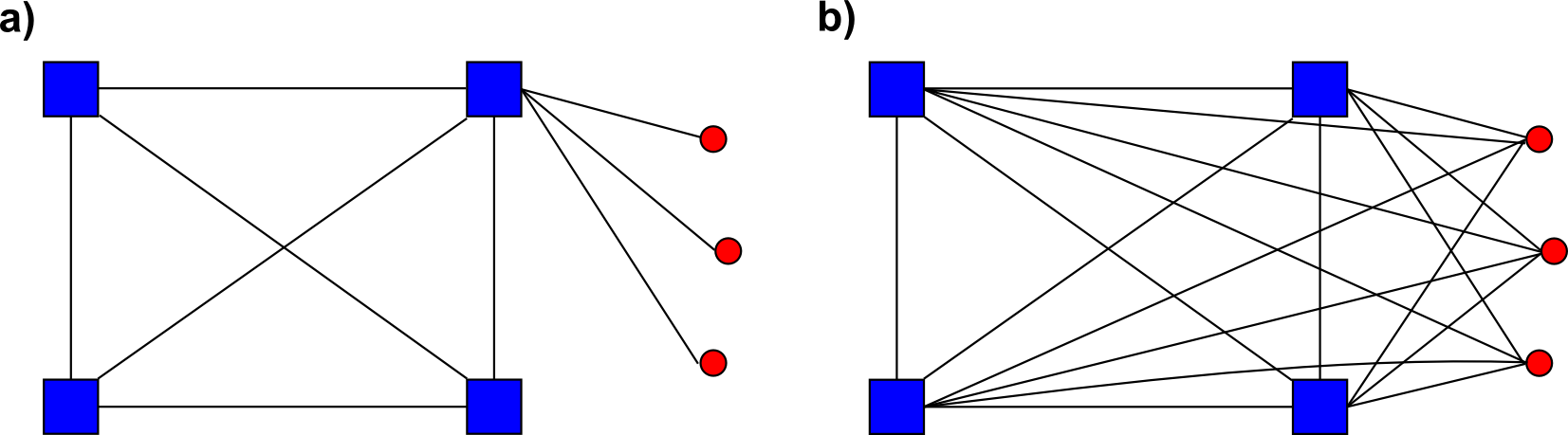}\protect\caption{\label{fig:The-optimal-solution}\label{fig:The-optimal-state - monetary transfers}\label{fig:optimal and stable}The
optimal solution, as described in Lemma \ref{lem:optimal solution}.
If $\left(A+1\right)/2<c$ the optimal solution is described by a),
otherwise by b). When monetary transfers (section \ref{sec:Monetary-transfers})
are allowed, both configurations are stabilizable. Otherwise, only
a) is stabilizable.}
\end{figure}

\begin{prop}
\label{lem:optimal solution}Consider the network where the type $B$
nodes are connected to a specific node $j\in T_{A}$ of the type-A
clique. The social cost in this stabilizable network (Fig. \ref{fig:The-optimal-solution-1}(a))
is 
\begin{eqnarray*}
\mathcal{S} & = & 2|T_{B}|\left(|T_{B}|-1+c+\left(A+1\right)(|T_{A}|-1/2)\right)+|T_{A}|\left(|T_{A}|-1\right)\left(c_{A}+A\right).
\end{eqnarray*}

Furthermore, if $|T_{B}|\gg1,|T_{A}|\gg1$ then, omitting linear terms
in $|T_{B}|,|T_{A}|$, 
\[
\mathcal{S}=2|T_{B}|(|T_{B}|+\left(A+1\right)|T_{A}|)+|T_{A}|^{2}\left(c+A\right).
\]
Moreover, if $\frac{A+1}{2}\le c$ then this network structure is
socially optimal and the price of stability is $1$, otherwise the
price of stability is
\[
PoS=\frac{2|T_{B}|(|T_{B}|+\left(A+1\right)|T_{A}|)+|T_{A}|^{2}\left(c_{A}+A\right)}{2|T_{B}|\left(|T_{B}|+\left(\frac{A+1}{2}+c\right)|T_{A}|\right)+|T_{A}|^{2}\left(c_{A}+A\right)}.
\]

Finally, if \textup{$|T_{B}|\gg|T_{A}|\gg1$, }\textup{\emph{then
the price of stability is asymptotically $1$.}}
\end{prop}
\begin{IEEEproof}
This structure is immune to removal of links as a disconnection of
a $(type-B,type-A)$ link will disconnect the type-B node, and the
type-A clique is stable (lemma \ref{lem:The-longest-distance}). For
every player $j$ and $i\in T_{B}$, any additional link $(i,j)$
will result in $\Delta C(j,E+ij)\geq c_{B}-1>0$ since the link only
reduces the distance $d(i,j)$ from 2 to 1. Hence, player $j$ has
no incentive to accept this link and no additional links will be formed.
This concludes the stability proof. 

We now turn to discuss the optimality of this network structure. First,
consider a set of type-A players. Every link reduce the distance of
at least two nodes by at least one, hence the social cost change by
introducing a link is negative, since $2c_{A}-2A<0$. Therefore, in
any optimal configuration the type-A nodes form a complete graph.
The other terms in the social cost are due to the inter-connectivity
of type-B nodes and the type-A to type-B connections. As $deg(i)=1$
for all $i\in T_{B}$ the cost due to link's prices is minimal. Furthermore,
$d(i,j)=1$ and the distance cost to node $j$ (of type A) is minimal
as well. For all other nodes $j'$, $d(i,j')=2$. 

Assume this configuration is not optimal. Then there is a \emph{topologically
different} configuration in which there exists an additional node
$j'\in T_{A}$ for which $d(i,j')=1$ for some $i\in T_{B}$. Hence,
there's an additional link $(i,j)$. The social cost change is $2c+1+A$
. Therefore, if $\frac{A+1}{2}\le c$ this link reduces the social
cost. On the other hand, if $\frac{A+1}{2}>c$ every link connecting
a type-B player to a type-A player improves the social cost, although
the previous discussion show these link are unstable. In this case,
the optimal configuration is where all type-B nodes are connected
to all the type-A players, but there are no links linking type-B players.
This concludes the optimality proof.

The cost due to inter-connectivity of type A nodes is

\[
c_{A}|T_{A}|\left(|T_{A}|-1\right)+A|T_{A}|\left(|T_{A}|-1\right)=|T_{A}|\left(|T_{A}|-1\right)\left(c_{A}+A\right).
\]

The first expression is due to the cost of $|T_{A}|$ clique's links
and the second is due to distance (=1) between each type-A node. The
distance of each type B nodes to all the other nodes is exactly 2,
except to node $j$, to which its distance is 1. Therefore the social
cost due to type B nodes is 
\begin{align*}
 & 2|T_{B}|(|T_{B}|-1)+2c_{B}|T_{B}|+2\left(A+1\right)|T_{B}|\left(\left(|T_{A}|-1\right)+\left(A+1\right)+2\left(A+1\right)(|T_{A}|-1)\right)\\
= & 2|T_{B}|\left(|T_{B}|-1+c_{B}+\left(A+1\right)(|T_{A}|-1/2)\right).
\end{align*}

The terms on the left hand side are due to (from left to right) the
distance between nodes of type B, the cost of each type-B's single
link, the cost of type-B nodes due to the distance (=2) to all member
of the type-A clique bar $j$ and the cost of type $B$ nodes due
to the distance (=1) to node $j$. The social cost is 
\begin{eqnarray*}
\sum C(i) & = & 2|T_{B}|\left(|T_{B}|-1+c+\left(A+1\right)(|T_{A}|-1/2)\right)+|T_{A}|\left(|T_{A}|-1\right)\left(c_{A}+A\right).
\end{eqnarray*}

To complete the proof, note that if $\frac{A+1}{2}>c$ the latter
term in the social cost of the optimal (and unstable) solution is
\[
2|T_{B}|(|T_{B}|-1)+2c|T_{B}|\left(1+|T_{A}|\right)+\left(A+1\right)|T_{B}||T_{A}|=2|T_{B}|\left(|T_{B}|-1+\left(\frac{A+1}{2}+c\right)|T_{A}|\right).
\]

As the number of links is $|T_{B}|\left(1+|T_{A}|\right)$ and the
distance of type-B to type-A nodes is 1. The optimal social cost is
then
\[
2|T_{B}|\left(|T_{B}|-1+\left(\frac{A+1}{2}+c\right)|T_{A}|\right)+|T_{A}|\left(|T_{A}|-1\right)\left(c_{A}+A\right).
\]

Considering all quantities in the limit $|T_{B}|\gg|T_{A}|\gg1$ completes
the proof.\end{IEEEproof}

Next, we evaluate the price of anarchy. In order to do that, we use
the following two lemmas. The first lemma evaluates a lower bound
by considering the social cost in the stabilizable topology presented
in Fig \ref{fig:A-poor-equilibrium}(b), composed of a type-A clique
and long lines of type-B players. Later on, an upper bound is obtained
by examining the social cost in any topology that satisfies Lemma
\ref{lem:The-longest-distance}. 

For simplicity, in the following lemma we assume that $|T_{B}|=\min\left\{ \left\lfloor \sqrt{4c_{A}}\right\rfloor ,\left\lfloor \sqrt{4c_{B}/5}\right\rfloor \right\} m$
where $m\in\mathcal{\mathbb{N}}$ and $c_{B}\leq20c_{A}$.
\begin{lem}
Consider the network where the type B nodes are composed of $m$ long
lines of length $k=\min\left\{ \left\lfloor \sqrt{4c_{A}}\right\rfloor ,\left\lfloor \sqrt{4c_{B}/5}\right\rfloor \right\} $,
and all the lines are connected at $j\in T_{A}$. The total cost in
this stabilizable network is 
\begin{eqnarray*}
S & = & |T_{A}|\left(|T_{A}|-1\right)\left(c_{A}/2+A\right)+2c_{B}|T_{B}|+\left(A+1\right)|T_{B}|\left(|T_{A}|-1\right)\left(k+3\right)/2\\
 &  & +|T_{B}|\left(\left(A+1\right)\left(k+1\right)/2+2k-4\right)+2|T_{B}|^{2}(k+2)^{2}-2m
\end{eqnarray*}

if $|T_{B}|\gg1,|T_{A}|\gg1$ then 
\begin{eqnarray*}
S & = & \left(A+1\right)|T_{A}|\left|T_{B}\right|o(c)\\
 &  & +|T_{A}|^{2}\left(c+A\right)+\left|T_{B}\right|^{2}o(c)
\end{eqnarray*}
\end{lem}
\begin{IEEEproof}
First, for the same reason as in Prop. \ref{lem:optimal solution},
this network structure is immune to removal of links. Consider $j'\in T_{A},j'\neq j$.
Let us observe the chain $(j,1,2,3,...k)$ where $1,2,3...k\in T_{B}$.
Clearly, the most beneficial link in $j'$ concern is $(j',k)$. By
establishing it, its cost's change is 
\begin{eqnarray*}
\Delta C(j',E+j'k) & = & c_{A}-\frac{\left(k+1\right)\left(k-1\right)+mod(k+1,2)}{4}\\
 & > & c_{A}-\frac{k^{2}}{4}>0
\end{eqnarray*}
by lemma \ref{lem:shortcut benefit} and noting the distance to player
$j$ is unaffected. Therefore there is no incentive for player $j'$
to add the link $(j',k)$. The same calculation indicates that no
additional link $(i,i')$ will be formed between two nodes on the
same line.

Consider two lines of length $k$, $(j,x_{1},x_{2},...x_{k})$ and
$(j,y_{1},y_{2},...y_{k})$. Consider the addition of the link $(x_{k},y_{\left\lfloor k+1/2\right\rfloor })$.
That is, the addition of a link from an end of one line to the middle
player on another line. This link is optimal in $x_{k}'$s concern,
as it minimizes the sum of distances from it to players on the other
line. The change in cost is (see appendix) 
\[
\Delta C(x_{k},E+x_{k}y_{\left\lfloor k+1/2\right\rfloor })=c_{B}-5k^{2}/4>0.
\]

For any other player $i=1..k$ in the line $(j,x_{1},x_{2},...x_{k})$
we have 
\begin{eqnarray*}
\Delta C(x_{k},E+x_{k}y_{\left\lfloor k+1/2\right\rfloor }) & \leq & \Delta C(x_{k},E+x_{k}y_{i})
\end{eqnarray*}
since the player that gains the most from establishing a link to a
node on the another line is the player that is furthest the most from
that line, i.e., the player at the end of the line. This concludes
the stability proof.

The total cost due to type-B nodes is 
\begin{align*}
 & 2c_{B}|T_{B}|+\left(A+1\right)|T_{B}|\left(|T_{A}|-1\right)\left(k+3\right)/2\\
 & +|T_{B}|\left(A+1\right)\left(k+1\right)/2+2m(k-1)^{2}+2(m-1)mk^{2}(k+2)^{2}.
\end{align*}
The terms represent (from left to right) the links' cost, the cost
due to the type-B players' distances to the type-A clique's nodes
(except $j$), the cost due to the type-B players' distances from
node $j$, the cost due to intra-line distances and the cost due inter-lines
distances. By using the relation $|T_{B}|=km$ we have
\begin{align*}
=2 & c_{B}|T_{B}|+\left(A+1\right)|T_{B}|\left(|T_{A}|-1\right)\left(k+3\right)/2\\
 & +|T_{B}|\left(A+1\right)\left(k+1\right)/2+2\left(|T_{B}|-m\right)(k-1)+2|T_{B}|^{2}(k+2)^{2}-2|T_{B}|m(k+2)\\
= & 2c_{B}|T_{B}|+\left(A+1\right)|T_{B}|\left(|T_{A}|-1\right)\left(k+3\right)/2\\
 & +|T_{B}|\left(\left(A+1\right)\left(k+1\right)/2+2k-4\right)+2|T_{B}|^{2}(k+2)^{2}-2m.
\end{align*}

The total cost due to type-A is as before
\[
|T_{A}|\left(|T_{A}|-1\right)\left(c_{A}+A\right).
\]

Therefore, 
\begin{eqnarray*}
\sum C(i) & = & |T_{A}|\left(|T_{A}|-1\right)\left(c_{A}/2+A\right)+2c_{B}|T_{B}|+\left(A+1\right)|T_{B}|\left(|T_{A}|-1\right)\left(k+3\right)/2\\
 &  & +|T_{B}|\left(\left(A+1\right)\left(k+1\right)/2+2k-4\right)+2|T_{B}|^{2}(k+2)^{2}-2m\\
 & \rightarrow & |T_{A}|^{2}\left(c_{A}+A\right)+\left(A+1\right)|T_{A}|\left|T_{B}\right|o(k)+\left|T_{B}\right|^{2}o(k).
\end{eqnarray*}
 
\end{IEEEproof}
The most prevalent situation is when $|T_{B}|\gg|T_{A}|\gg1$. In
this case we can bound the price of anarchy to be at least $o(k^{2})=o(c_{B})$. 

The next lemma bounds the price of anarchy from above by bounding
the maximal total cost in the a link-stable equilibrium.
\begin{lem}
\label{lem:The-worst-social utility}The worst social utility in a
link stable equilibrium is at most 
\begin{eqnarray*}
 &  & |T_{A}|^{2}\left(c_{A}+A\right)+|T_{B}|^{2}\left(c_{B}+\left\lfloor 2\sqrt{c_{B}}\right\rfloor \right)\\
 & + & \left(A+1\right)\left\lfloor 2\sqrt{c}\right\rfloor |T_{A}||T_{B}|
\end{eqnarray*}
\end{lem}
\begin{IEEEproof}
The total cost due to the inter-connectivity of the type-A clique
is identical for all link stable equilibria and is $|T_{A}|\left(|T_{A}|-1\right)\left(c_{A}+A\right)$.
The maximal distance between nodes $i,j$ according to Lemma \ref{lem:The-longest-distance}
is $\left\lfloor 2\sqrt{c_{B}}\right\rfloor $ and therefore the maximal
cost due to the distances between type-B nodes is $\left\lfloor 2\sqrt{c_{B}}\right\rfloor |T_{B}|\left(|T_{B}|-1\right)$.
Likewise, the maximal cost due to the distance between type-B nodes
and the type-A clique is $\left(A+1\right)\left\lfloor 2\sqrt{c}\right\rfloor |T_{A}||T_{B}|$.
Finally, the maximal number of links between type-B nodes is $|T_{B}|\left(|T_{B}|-1\right)/2$
and the total cost due to this part is $|T_{B}|\left(|T_{B}|-1\right)c_{B}$.

Adding all the terms we obtain the required result.
\end{IEEEproof}
To summarize, we state the previous results in a proposition.
\begin{prop}
\label{thm:summary-of-results}If $c_{B}<A$ and $|T_{B}|\gg|T_{A}|\gg1$
the price of anarchy is $\Theta(c_{B})$. 
\end{prop}

\subsection{\label{sec:Monetary-transfers}Monetary transfers }

Monetary transfers allow for a redistribution of costs. It is well
known in the game theoretic literature that, in general, this process
increases the social welfare.Indeed, the next lemma, shows that in this setting, the maximal distance
between players is smaller, compared to Lemma \ref{lem:The-longest-distance}.
\begin{lem}
Allowing monetary transfers, the longest distance between nodes i,$j\in T_{B}$
is $max\{\left\lfloor 2\sqrt{c}\right\rfloor ,1\}$. The longest distance
between nodes $i,j\in T_{A}$ is bounded by 
\[
\max\left\{ \left\lfloor 2\left(\sqrt{(A-1)^{2}+c}-(A-1)\right)\right\rfloor ,1\right\} 
\]
The longest distance between node $i\in T_{A}$ and node $j\in T_{B}$
is bounded by
\[
\max\left\{ \left\lfloor \left(\sqrt{(A-1)^{2}+4c}-(A-1)\right)\right\rfloor ,1\right\} 
\]
\end{lem}
\begin{IEEEproof}
Assume $d(i,j)=k\geq\left\lfloor 2\sqrt{c}\right\rfloor >1$ and $j\in T_{B}$.
Then there exist nodes $(x_{0}=i,x_{1,}x_{2},..x_{k}=j)$ such that
$d(i,x_{\alpha})=\alpha$. By adding a link $(i,j)$ the change in
cost of node $i$ is, according to lemma \ref{lem:The-longest-distance}
is

\begin{align*}
 & \Delta C(i,E+ij)\\
= & c-\frac{k\left(k-2\right)+mod(k,2)}{4}\\
< & c-\frac{k\left(k-2\right)}{4}\\
< & 0
\end{align*}
 Therefore, it is of the interest of player $i$ to add the link. 

Consider the case that $j\in T_{A}$ and 
\[
d(i,j)=k\geq\left\lfloor 2\left(\sqrt{(A-1)^{2}+c}-(A-1)\right)\right\rfloor >1
\]
 The change in cost after the addition of the link $(i,j)$ is 
\begin{align*}
 & c-\sum_{\alpha=1}^{k}d(i,x_{\alpha})(1+\delta_{x_{\alpha},A}(A-1))\\
 & +\sum_{\alpha=1}^{k}d'(i,x_{\alpha})(1+\delta_{x_{\alpha},A}(A-1))\\
= & c-\sum_{\alpha=1}^{k}\left(d(i,x_{\alpha})-d'(i,x_{\alpha})\right)(1+\delta_{x_{\alpha},A}(A-1))\\
< & c-\sum_{\alpha=1}^{k}d(i,x_{\alpha})+\sum_{\alpha=1}^{k}d'(i,x_{\alpha})\\
= & c-\left(1+k\right)k/2-kA+1+\left(1+k/2\right)k/2+A-1\\
= & c-k^{2}/4-k(A-1)
\end{align*}

If $i,j\in T_{B}$ then
\[
\Delta C(i,E+ij)+\Delta C(j,E+ij)<2\left(c-k^{2}/4\right)<0
\]

for $k\geq\left\lfloor 2\sqrt{c}\right\rfloor $ and the link will
be established.

Likewise, if $i,j\in T_{A}$ and 
\[
k\geq\left\lfloor 2\left(\sqrt{(A-1)^{2}+c}-(A-1)\right)\right\rfloor 
\]
 the link $(i,j)$ will be formed.

If $i\in T_{A}$, $j\in T_{B}$ and 
\[
k\geq\left\lfloor \left(\sqrt{(A-1)^{2}+4c}-(A-1)\right)\right\rfloor 
\]
 then 
\begin{eqnarray*}
\Delta C(i,E+ij)+\Delta C(j,E+ij) & < & 2c-k^{2}/2-k(A-1)\\
 & < & 0
\end{eqnarray*}

This concludes our proof.\end{IEEEproof}
 Indeed, the next proposition indicates an improvement on Proposition
\ref{lem:optimal solution}. Specifically, it shows that the optimal
network is always stabilizable, even when $\frac{A+1}{2}>c$. Without
monetary transfers, the additional links in the optimal state (shown
in \cite{Meirom2016}) (Fig. \ref{fig:The-optimal-state - monetary transfers}),
connecting a major league player with a minor league player, are unstable
as the type-A players lack any incentive to form them. By allowing
monetary transfers, the minor players can compensate the major players
for the increase in their costs. It is worthwhile to do so only if
the social optimum of the two-player game implies it. The existence
or removal of an additional link does not inflict on any other player,
as the distance between every two players is at most two.

\begin{figure}
\centering{}\includegraphics[width=0.5\columnwidth]{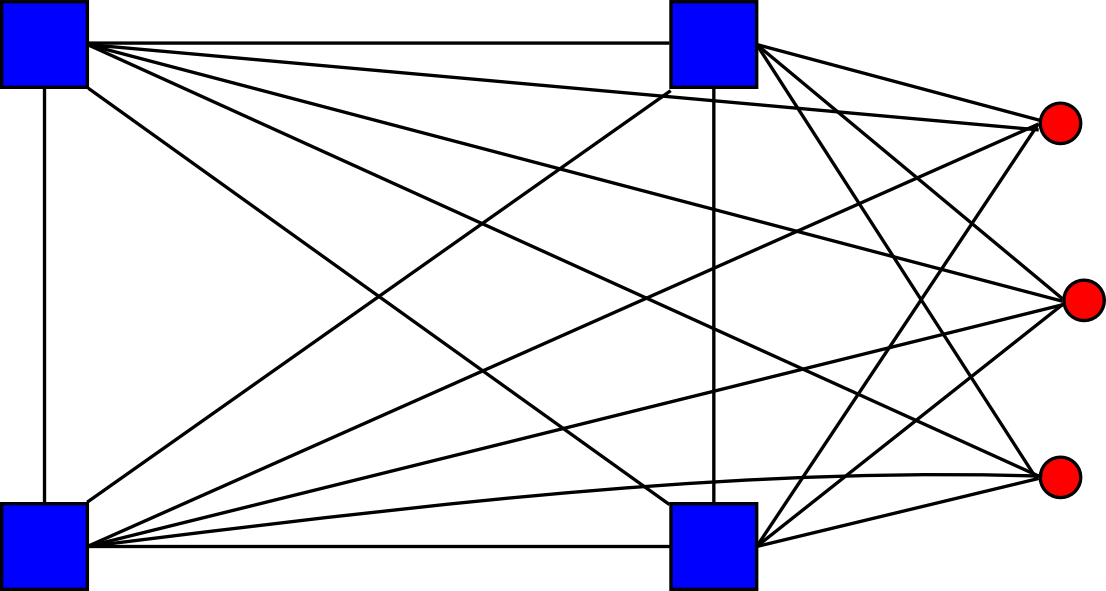}\protect\caption{The optimal state if $\frac{A+1}{2}>c$ . This network is stabilizable
iff utility transfer is allowed, as discussed in Prop. \ref{lem:optimal solution}
and Prop. \ref{prop:optimality under monetary}. }
\end{figure}

\begin{prop}
\label{prop:optimality under monetary}\textup{\emph{The price of
stability is $1$.}} If $\frac{A+1}{2}\leq c\,,$ then Proposition
\ref{lem:optimal solution} holds. Furthermore, if $\frac{A+1}{2}>c$,
then the optimal stable state is such that all the type $B$ nodes
are connected to all nodes of the type-A clique. In the latter case,
the social cost of this stabilizable network is $\mathcal{S}=2|T_{B}|\left(|T_{B}|+\left(\frac{A+1}{2}+c\right)|T_{A}|\right)+|T_{A}|^{2}\left(c+A\right).$
Furthermore, if $|T_{B}|\gg1,|T_{A}|\gg1$ then, omitting linear terms
in $|T_{B}|,|T_{A}|$, $\mathcal{S}=2|T_{B}|(|T_{B}|+\left(A+c\right)|T_{A}|)+|T_{A}|^{2}\left(c+A\right).$
\end{prop}
\begin{IEEEproof}
For the case $\frac{A+1}{2}\leq c$ , it was shown in Prop. \ref{lem:optimal solution}
that the optimal network is a network where all the type $B$ nodes
are connected to a specific node $j\in T_{A}$ of the type-A clique
(Fig. \ref{fig:The-optimal-state - monetary transfers}(a)) and that
this network is stabilizable. Therefore, we only need to address its
stability under monetary transfers. We apply the criteria described
in Corollary \ref{lem:edges with monetary transfers.} and show that
for every two players $i,j'$ such that $(i,j')\notin E$ we have
\[
\Delta C(i,E+ij)+\Delta C(j',E+ij')>0.
\]

If $i\in T_{B}$ then 
\[
\Delta C(i,E+ij')=c_{B}-1>0.
\]

If $i\in T_{A}$ we have that 
\[
\Delta C(i,E+ij')=c_{A}-A>0.
\]

Then, for $i\in T_{B}$ and either $j'\in T_{A}$ or $j'\in T_{B}$
we have $\Delta C(i,E+ij)+\Delta C(j',E+ij')>0$ and the link would
not be established. For every edge $(i,j)\in E$ we have that both
$\Delta C(i,E+ij)<0$ and $\Delta C(j,E+ij)<0$ (Prop. \ref{lem:optimal solution})
and therefore
\[
\Delta C(i,E+ij)+\Delta C(j,E+ij)<0.
\]

Assume $\frac{A+1}{2}>c$. It was shown in Proposition \ref{lem:optimal solution}
that the optimal network is a network where every type $B$ player
is connected to all the members of the type-A clique (Fig. \ref{fig:The-optimal-state - monetary transfers}(b)).
Under monetary transfers, this network is stabilizable, since for
for $i\in T_{B}$ , $j\in T_{A}$
\begin{eqnarray*}
 &  & \Delta C(i,E+ij)+\Delta C(j',E+ij')\\
 & = & 2c-A-1<0
\end{eqnarray*}

and the link $(i,j)$ will be formed. The previous discussion shows
that it is not beneficial to establish links between two type-B players.
Therefore, this network is stabilizable.

In conclusion, in both cases, the price of stability is 1.\end{IEEEproof}

Next, we show that, under mild conditions on the number of type-A
nodes, the price of anarchy is $3/2$, i.e., \emph{a fixed number}
that does not depend on any parameter value. As the number of major
players increases, the motivation to establish a direct connection
to a clique member increases, since such a link reduces the distance
to all clique members. As the incentive increases, players are willing
to pay more for this link, thus increasing, in turn, the utility of
the link in a major player's perspective. With enough major players,
all the minor players will establish direct links. Therefore, any
stable equilibrium will result in a very compact network with a diameter
of at most three. This is the main idea behind the following lemma
and theorem.

\begin{IEEEproof}
Assume $d(i,j)=k>2$, where $i\in T_{B}$ and $j\in T_{A}$ but node
$j$ is not the nearest type-A node to \emph{i}. Therefore, there
exists a series of nodes $(x_{0}=i,x_{1},...,x_{k-1},x_{k}=j)$ such
that $x_{k-1}$ is a member of the type-A clique.

The change in player $j$'s cost by establishing $(i,j)$ is 
\begin{eqnarray*}
\Delta C(j,E+ij) & = & c_{A}-\sum_{\alpha=1}^{k}d(j,x_{\alpha})(1+\delta_{x_{\alpha},A}(A-1))\\
 &  & +\sum_{\alpha=1}^{k}d'(j,x_{\alpha})(1+\delta_{x_{\alpha},A}(A-1))\\
 & < & c_{A}-\sum_{\alpha=1}^{k}\left(d(j,x_{\alpha})-d'(j,x_{\alpha})\right)\\
 & < & c_{A}-k^{2}/4.
\end{eqnarray*}

The corresponding change in player $i$'s cost is
\[
\Delta C(i,E+ij)<c_{B}-k^{2}/4-(2k-4)A-\left(k-2\right)A\cdot\left(|T_{A}|-2\right).
\]

The first term is the link cost, the second and third terms are due
to change of distance from players $x_{k-1},x_{k}$ and the last term
express the change of distance form the rest of the type-A clique.
As
\[
\Delta C(i,E+ij)<c_{B}-k^{2}/4-\left(k-2\right)A\cdot|T_{A}|
\]

the total change in cost is 
\begin{eqnarray*}
 &  & \Delta C(j,E+ij)+\Delta C(i,E+ij)\\
 & < & 2c-k^{2}/2-\left(k-2\right)A\cdot|T_{A}|\\
 & < & 0
\end{eqnarray*}

for 
\[
k<\sqrt{\left(A|T_{A}|\right)^{2}+4cA|T_{A}|}-A|T_{A}|
\]

Note that as the number of member in the type-A clique, $\left|T_{A}\right|$,
increases, the right expression goes to 0, in contradiction to our
initial assumption. Therefore, in the large network limit the maximal
distance of a type-B node from a node in the type-A clique is 2. In
this case, the maximal distance between two type-A nodes is 1 (as
before), between type-A and type-B nodes is 2 and between two type-B
nodes is 3. The maximal social cost in an equilibrium is
\begin{eqnarray*}
\mathcal{S} & < & 3|T_{B}|\left(|T_{B}|-1\right)+|T_{B}|c_{B}\\
 &  & +2|T_{B}||T_{A}|\left(A+1\right)+|T_{A}|\left(|T_{A}|-1\right)\left(c_{A}+A\right)
\end{eqnarray*}

For $|T_{B}|\gg1,|T_{A}|\gg1$ we have 
\[
S=3|T_{B}|^{2}+2|T_{B}||T_{A}|\left(A+1\right)+|T_{A}|^{2}\left(c_{A}+A\right).
\]

comparing this with the optimal cost in this limit 
\[
S_{opt}=2|T_{B}|^{2}+2|T_{B}||T_{A}|\left(A+1\right)+|T_{A}|^{2}\left(c_{A}+A\right)
\]

we obtain the required result.\end{IEEEproof}

This theorem shows that by allowing monetary transfers, the maximal
distance of a type-B player to the type-A clique depends inversely
on the number of nodes in the clique and the number of players in
general. The number of ASs increases in time, and we may assume the
number of type-A players follows. Therefore, we expect a decrease
of the mean ``node-core distance'' in time. Our data analysis (Section
\ref{sec:Data-Analysis}) indicates that this real-world distance
indeed decreases in time.

\subsection{\label{sub:static-survivability}The effect of survivability}

One may have guessed that reliability requirements, which generally
induce the creation of additional, backup edges, would ease the formation
of the clique. The next proposition shows that this naive assumption
is wrong, and in fact, as the frequency of failure increases, it becomes
increasingly difficult to maintain the major player's clique. Consider
a dense set, in which every player may access all the other players
within two hops by a at least two disjoint paths. A direct link between
two players only reduces their mutual distance by one, and does not
affect any other distance. If this link fails often, it may be used
only partially, and it may not be worthy to pay its cost. Hence, in
this setting, counter intuitively, frequent failures end up with a
sparser network.
\begin{prop}
\label{prop:clique-reliability}Assume the frequency of failures is
high, namely $\delta=1.$ Then, the type-A players form a clique if
and only if $c_{A}<A/2$. Allowing monetary transfers does not change
the result. \end{prop}
\begin{IEEEproof}
We consider a major player's clique and ask under which conditions
the removal of a link is a worthy move. Consider an edge $(i,j)$
in this clique. Since only the shortest distance between players $i$
and $j$ is affected, and is increased by one, the type-A players
clique is stable if and only if $A/\left(1+\delta\right)>c_{A}$.
\end{IEEEproof}
 The next proposition describes a scenario in which, surprisingly,
the additional reliability constraints \emph{reduce} the social cost.

\begin{prop}
\label{lem:reliable optimal 1} Assume $1<c_{A}<A/2$ and symmetric
reliability requirements, namely $\tau=1$. Then, the optimal network
is composed of a type-A clique, where all the type $B$ nodes are
connected to all members of the type-A clique, as depicted in Fig. \ref{fig:The-optimal-solution}.
This network is not stabilizable, and $PoS>1$. Nevertheless, for
$|T_{B}|\gg|T_{A}|\gg1$, we have $PoS\rightarrow1$. In addition,
the Price of Reliability is smaller than one. 
\end{prop}
The main idea behind this result is that in the optimal, yet unstable
solution, every minor player established a link with all the major
players (Fig. \ref{fig:The-optimal-solution}). This configuration
is unstable as it is over-saturated with links, and the optimal stable
solution with survivability considerations is obtained by diluting
this network so that every minor player will connect to just two major
players. If the reliability requirements are further removed, then
the additional dilution occurs, increasing the social cost. In other
words, the stable configuration is under-saturated with edges, and
the additional survivability requirements facilitate the formation
of additional links.
\begin{IEEEproof}
� � We shall now prove the first part of this theorem, namely that
the network described in Fig. \ref{fig:The-optimal-solution}, is
optimal in terms of social cost. In this network, every type-B player
is connected to every type-A player. We denote this network by $G.$

First, consider a network in which the type-A players are not in a
clique. Similar to Proposition \ref{prop:clique-reliability}, there
exists two players $i,j\in T_{A}$ such that $d(i,j)\geq2$. By establishing
a link $(i,j)$ the change in social cost is
\[
\Delta\mathcal{S}\leq-2\left(\frac{A}{1+\delta}-c_{A}\right)<0.
\]
Therefore a network which includes this link has a lower social cost.
Hence, in the optimal network, all the type-A players are in clique.
A similar calculation for a missing link between player $i\in T_{A}$
and $j\in T_{B}$ shows that
\[
\Delta\mathcal{S}\leq-\left(\frac{A+1}{1+\delta}-c_{A}-c_{B}\right)<0
\]

and establishing this link reduces the social cost as well. In conclusion,
in an optimal network a type-A player is connected to all the other
players.

Consider node $i\in T_{B}.$ Its distance from every other node $j\in T_{B}$
is $d(i,j)=2$ and $d'(i,j)=2$. The minimal distances between two
minor players $i,j\in T_{B}$ are $d(i,j)=1$ and $d'(i,j)=2$ whereas
in $G$ we have $d(i,j)=d'(i,j)=2$. Assume that a network with a
lower social cost exists, and denote it by $G'$. In $G'$ there must
exists a link between two type-B players, $i,j\in T_{B}$. However,
removing this link reduces the social cost, as

\[
S(E+ij)-S(E)=2\left(\frac{A}{1+\delta}-c\right)>0
\]
Therefore, $G'$ is not optimal, in contradiction to our assumption.
This proves that $G$ is optimal.

Next, we are going to show $G$ is unstable. Consider $i\in T_{B}$
and $x\in T_{A}$. By removing $(i,x)$ we cost change of player $x$
is 
\[
\Delta C(x,E-ix)=-2\left(\frac{1}{1+\delta}-c\right)>0
\]

And therefore it is beneficial for player $x$ to remove this link.
Hence, this network in not stabilizable.

Finally, we are going to show that the $PoR<1$. The outline is as
follows. First, we are going to show that the network depicted in
Fig. $\ref{fig:optimal and stable}$ is stable. The optimal stable
network (without reliability requirements) in this parameter regime
was explicitly derived Proposition \ref{lem:optimal solution}. We
are going to show that the cost in the latter network is higher than
the former, which bound the $PoR$ from above by a value smaller than
one.

Consider a network in which all the minor players are connected to
two major players, $x,y\in T_{A}$. We are going to show that this
network is stable. Clearly, as shown before, the type-A clique is
stable. In addition, for any player $i\in T_{B}$ neither $x,y$ nor
$i$ has the incentive to remove either $(x,i)$ or $(y,i)$ as it
would violate the reliability requirement and would lead to unbounded
cost. Hence, this network is stable. Denote this network by $\tilde{G}.$
In Proposition \ref{lem:optimal solution} it was shown that the optimal
stable network (without reliability requirements) in this configuration
is the network in which all type-A players form a clique, and all
the type-B players are connected to a single type-A player. We denote
the latter network by $G''.$We have
\[
\mathcal{S}(\tilde{G}.)-S(G'')=|T_{B}|\left(\frac{A+1}{1+\delta}-c_{A}-c_{B}\right)<0.
\]

But, 
\[
PoR=\frac{\tilde{S}_{optimal}}{\tilde{S}_{optimal}^{(bare)}}\leq\frac{S(\tilde{G}.)}{\mathcal{S}(G'')}=1+\frac{\mathcal{S}(\tilde{G}.)-S(G'')}{\mathcal{S}(G'')}<1.
\]

This concludes the proof.
\end{IEEEproof}
In conclusion, if the failure frequency is high, the survivability
requirements will induce dilution of the clique of major players.
However, the opposite effect occurs along the graph cut-set between
the minor players set and the major players set, where the additional
constraints lead to an increased number of links connecting major
and minor players.

So far we have assumed that the reliability requirements are symmetric.
As explained in the Introduction, in some cases it is reasonable to
assume that players will require a backup route only to the major
players, i.e., non-symmetric reliability constraints. We shall now
show that in this case, the social cost may deteriorate considerably.
Hence, from a system designer point of view, it is much more important
to incentivize a configuration where the reliability requirements
are symmetric, than to reduce failure frequency globally. This result
stems from the following lemma.
\begin{lem}
\label{lem:finite distances}If $\tau\neq0$ then for every two nodes
i,j, $d(i,j)\leq d'(i,j)\leq2+2c_{B}$. If $\tau=0$ there exists
stable equiliberia such that there are no two disjoint paths between
some nodes $i,j$.\end{lem}
\begin{IEEEproof}
First, we are going to show that $d'(i,j)$ is finite (step 1). Then,
in the second step, we shall bound $d'(i,j)$ from above.

Step 1: If $i$ and $j$ are disconnected, then it is worthy for both
of them to establish the link $(i,j)$. Therefore, there exists a
path from $i$ to $j$ and $d(i,j)$ is finite. Assume $d(i,j)\neq1$.
If there aren't two disjoint paths between nodes $i$ and $j,$ it
is worthy for both nodes to establish a direct link. Hence if $d(i,j)\neq1$
the distance $d'(i,j)$ is finite. We now discuss the case $d(i,j)=1$
and show there must exists an additional disjoint path, so $d'(i,j)$
is finite. We prove by negation.

Take node $x\neq i,j$. As before, both $d(i,x)$ and $d(x,j)$ are
finite. Consider the following two cases:

A) $d(i,x)=1$, $d(x,j)=1$. Then, the trajectory $(i,x,j)$ is disjoint
from the path. Hence $d'(i,j)$ is finite.

B) $d(i,x)=1$, $d(x,j)\neq1$ or $d(i,x)\neq1$, $d(x,j)=1$. Without
loss of generality, we assume the first case. According to the previous
discussion, there are at least two disjoint paths from $x$ to $j.$
Since they are disjoint, only one of them may contain the edge $(i,j)$.
Therefore, there exist a path $(i,x,...j)$ that is disjoint from
the edge $(i,j)$ and therefore $d'(i,j)$ is finite.

C) $d(i,x)\neq1$, $d(x,j)\neq1$ . Applying the same reasoning as
in B), there exists a path from $i$ to $x$ that is disjoint from
the edge $(i,j)$. Likewise, there exists a path from $j$ to $x$
that is disjoint from the edge $(i,j).$ Therefore there exist a path
from $i$ to $j$ that is disjoint from $(i,j).$

Step 2: Step 1 has shown that every two nodes are connected by a cycle.
Let us assume the longest shortest cycle connecting two nodes is of
length $l>2+2c_{B}>3$, namely there exists a cycle $(x_{0},x_{1},x_{2},...,x_{\left\lfloor l/2\right\rfloor },...x_{l},x_{0})$.

By establishing the link $(x_{0},x_{\left\lfloor l/2\right\rfloor })$
the cost of player $x_{0}$ due to distances from other players is
reduced by at least $\left(l^{2}-1+mod(l+1,2)\right)/4$, hence the
link will be established as
\[
\Delta C\left(x_{0},E+(x_{0},x_{\left\lfloor l/2\right\rfloor }\right)\leq\left(l^{2}-1+mod(l+1,2)\right)-c_{B}\leq0
\]

and similarly $\Delta C\left(x_{0},E+(x_{0},x_{\left\lfloor l/2\right\rfloor }\right)\leq0$
. Since $d(i.j)\leq d'(i.j)\leq l$ the proof is complete.
\end{IEEEproof}
If there exist players with no two disjoint paths in an equilibrium,
then the social cost becomes unbounded. This immediately results in
an \emph{unbounded} Price of Anarchy, as indicated by the following
theorem.
\begin{thm}
\label{thm:price of anarchy}Consider a network such that $T_{B}\gg T_{A}\gg1$.
The Price of Anarchy is as follows:

A) In a setting with asymmetric reliability requirements $(\tau=0)$:
unbounded.

B) In a setting with symmetric reliability requirements $(\tau\neq0)$:
bounded by $o(c)$. \end{thm}
\begin{IEEEproof}
A) We shall prove this by showing that there exists a stable equilibrium
with unbounded social cost. Consider a network in which the major
players (type-A players) form a clique, while every minor player is
connected only to a single type-A player $j\in T_{A}$. We shall show
that this network is stabilizable. As discusses before, the type-A
clique is stable. Consider $i\in T_{B}.$ By forming the link $(i,x)$
the change of cost of player $x\in T_{A}$, $x\neq j$ is
\[
\Delta C(x,ix)=c_{A}-1>0.
\]

Hence, this link will be formed. No additional links between minor
players will be established, since such a link only reduces the distance
between the participating parties by one, $c_{B}-1>0$ and does not
provide an additional disjoint path to the type-A clique. Therefore,
the social cost is $\mathcal{S=\omega}(\mathcal{Q})\rightarrow\infty$. 

B) The total cost due to the inter-connectivity of the type-A clique
is identical for all link stable equilibria and is $|T_{A}|\left(|T_{A}|-1\right)\left(c+\left(1+\delta/2\right)A\right)$.
This cost is composed of $|T_{A}-1|$ links per node, and the distance
cost to every other major nodes, 
\[
d(i,j)+d'(i,j)=1+\delta/2.
\]
Next, we evaluate the cost due to the type-B nodes inter-distances.
According to Lemma \ref{lem:finite distances}, both $d(i,j)\leq4c_{B}$
and $d'(i,j)\leq4c_{B}$, so the cost due to the inter-distances between
a type-B player and every other player is bounded from above by $4c_{B}\left(|T_{A}|+|T_{B}|\right)$.
When summed up over all minor players, this contributes a term $4c_{B}|T_{B}|\left(|T_{A}|+|T_{B}|\right)$
to the social cost. Likewise, the cost of links that at least one
of their ends is a type B player is at most $c_{B}|T_{B}|\left(|T_{A}|+|T_{B}|\right)$. 

Therefore, the maximal cost in all link stable equilibria is bounded
from above by 
\[
|T_{A}|\left(|T_{A}|-1\right)\left(c+\left(1+\delta/2\right)A\right)+5c_{B}|T_{B}|\left(|T_{A}|+|T_{B}|\right).
\]

The optimal network configuration is described in Proposition \ref{lem:reliable optimal 1}.
It is straightforward to evaluate the social cost in this configuration.
The distance cost due to inter-distances of the type-B players is
\[
2|T_{B}|\left(|T_{B}-1\right),
\]

while the cost of all links which connect type-B players to type-A
players is
\[
|T_{B}||T_{A}|\left(c_{B}+c_{A}\right).
\]

Finally, the social cost due to the type-A clique remains the same,
so finally we obtain that the minimal social cost is

\begin{eqnarray*}
\mathcal{S}_{optimal} & = & |T_{A}|\left(|T_{A}|-1\right)\left(c+\left(1+\delta/2\right)A\right)\\
 &  & +2|T_{B}|\left(|T_{B}-1\right)+2|T_{B}||T_{A}|\left(c_{B}+c_{A}\right)
\end{eqnarray*}

By taking the limits $|T_{A}|\rightarrow\infty,|T_{B}|/|T_{A}|\rightarrow\infty$
we obtain that $PoA\leq5c_{B}/2$.
\end{IEEEproof}
A stable equilibrium with infinite social cost can be easily achieved
by considering a network where all minor players are connected to
a single, designated, major player. There exists a single path of
at most two hops between every minor player to every major player.
However, as the stability requirements are asymmetric, the major players
have no incentive to establish additional routes to any minor player,
and the reliability requirements of the minor players remain unsatisfied.

If monetary transfers are feasible, players may compensate other players
for the cost of additional links such that all the additional constraints
are satisfied. Therefore, in this setting, in contrast to the previous
setting, there always exists a fallback route between every two players,
regardless of the symmetric or asymmetric nature of the additional
survivability constraints. Hence, symmetry is less important than
in the previous scenario. Furthermore, the following result suggests
that every player is connected to every other player by a cycle, and
that the maximal cycle length decays with the number of major players.
As the number of ASs increases in time, this predicts that the maximal
cycle length should decrease in time. We shall verify this prediction
in Section \ref{sec:Data-Analysis}.
\begin{prop}
\label{prop:max distance}Assume $1<c<A/2$ . Then, every two players
are connected by a cycle, and the maximal cycle length is bounded
by 
\[
\max\left\{ 2\left(\left\lfloor \sqrt{\left(A|T_{A}|\right)^{2}+5c}-A|T_{A}|\right\rfloor +1\right),4\right\} .
\]

\begin{IEEEproof}
\textit{\emph{Lemma \ref{lem:finite distances} showed that the maximal
distance between players is bounded. We are going to tighten this
result in the regime where monetary transfers are feasible. Denote
the maximal distance between a type-A player and a type B player by
$k_{A}$.}}

\textit{\emph{First, we are going to show that the maximal cycle length
connecting a major player and a minor player is $2k_{A}+1.$ This
follows from a simple geometric argument. Consider two players, $i\in T_{A}$
and $j\in T_{B}$. Assume $k_{A}>2$. If the cycle length is $2k_{A}+2$
or greater, there exists a type-B node that its distance $k_{A}+1,$
in contradiction to the assumption that the maximal distance between
a major player and and a minor player is $k_{A}$. Denote the maximal
distance between two minor players by $k_{B}.$ A similar argument
shows that maximal cycle length between two minor players is $2k_{B}+1$.}}

\textit{\emph{Next, are going to show that the maximal distance connecting
a major player and a minor player is $k_{A}\leq l\leq\max\left\{ 2\left\lfloor \sqrt{\left(A|T_{A}|\right)^{2}+5c}-A|T_{A}|\right\rfloor ,2\right\} $
. We prove by negation. Assume that the distance between player $j\in T_{A}$
and $i\in T_{B}$ is $l$. Denote the nodes on the path as $(x_{0}=i,x_{1},x_{2}....,x_{l}=j)$.
Then, by establishing a link between them, the distance between $j$
and $\{x_{0},x_{1}....x_{\left\lfloor l/2\right\rfloor }\}$ (similarly,
and distance between player $i$ and players $\{x_{\left\lceil l/2\right\rceil }...x_{l-1}\}$)
is reduced. In addition, player $i$ reduces its distance to every
node of the type-A clique by $l$. Lemma 25 of \cite{Meirom2014}
shows that the total reduction in distance is $\left(l^{2}-1+mod(l+1,2)\right)/4$.
Then, by establishing the link $(i,j)$ we have
\[
\Delta C(i,E+ij)+\Delta C(j,E+ij)\leq c_{A}+c_{B}-2\left(l^{2}-1\right)-lA|T_{A}|
\]
}}

\textit{\emph{and as $l\geq2\left\lfloor \sqrt{\left(A|T_{A}|\right)^{2}+5c}-A|T_{A}|\right\rfloor $
this expression is negative. Therefore, the link will be established,
and the maximal shortest distance between a major player and a minor
player is smaller than $l$. This concludes the proof.}}\end{IEEEproof}

\end{prop}

\section{\label{sec:Dynamic-Analysis}Dynamic Analysis}

The Internet is a rapidly evolving network. In fact, it may very well
be that it would never reach an equilibrium as ASs emerge, merge,
and draft new contracts among them. Therefore, a dynamic analysis
is a necessity. We first define the dynamic rules. Then, we analyze
the basin of attractions of different states, indicating which final
configurations are possible and what their likelihood is. We shall
establish that reasonable dynamics converge to \emph{just a few} equilibria.
Furthermore, we investigate the speed of convergence, and show that
convergence time is \emph{linear} in the number of players. Lastly,
we identify prevalent network motifs \cite{Milo2002a}, i.e., small
sub-graphs that emerge during the natural evolution of the network,
that arise due to survivability constraints. In Section \ref{sec:Data-Analysis}
we show that these motifs are indeed ubiquitous in the real AS topology,
and that the frequency of their occurrences is few folds\emph{ }more
than expected in a random network.

\subsection{Setup \& Definitions}

We split the game into \emph{turns}, where at each turn only a single
player is allowed to remove or initiate the formation of links. At
each point in time, or turn, the players that already joined the game
form a subset $N'\subset T_{A}\cup T_{B}$. We shall implicitly assume
that the cost function is calculated with respect to the set $N'$
of players that are already present in the network. Each turn is divided
into \emph{moves, }at each of which a player either forms or removes
a single link. A player's turn is over when it has no incentive to
perform additional moves. Note that disconnections of several links
can be done unilaterally and hence iteratively. 

During player's $i$ turn, all the other players will act in a greedy,
rather than strategic, manner. For example, although it may be that
player $j$ prefers that a link $(i,j')$ would be established for
some $j'\neq j$, it will accept the establishment of the less favorable
link $(i,j)$, as long as its formation is beneficial to it. In other
words, the active player has the advantage of initiation and the other
players react to its offers. There are numerous scenarios in which
players cannot fully forecast other players' moves and offers, e.g.,
when information is asymmetric or when only partial information is
available \cite{5173479}. In these settings, it is likely that a
greedy strategy will become the modus operandi of many players. This
is a prevalent strategy also when the system evolves rapidly and it
is difficult to assess the current network state and dynamics. Formally,
we express this dynamic rule as follows.
\begin{defn}
Dynamic Rule \#1: Assume it is player $i$'s turn and let the set
of links at its $m$th move be denoted as $E_{m}$. In player $i$'s
$m$th move, it may remove a link $(i,j)\in E_{m}$ or, if player
$j$ agrees, it may establish the link $(i,j)\notin E_{m}$. Player
$j$ would agree to establish $(i,j)$ iff $C(j;E_{m}+(i,j))-C(j;E_{m})<0$.
\end{defn}

In a dynamic network formation game, a key question is: Can a player
temporarily disconnects itself from the graph, only to reconnect after
getting to a better bargaining position? Or must a player stay connected?
If the timescale in which the costs are evaluated is comparable to
the timescale in which the dynamics occur, then, clearly, a player
will not disconnect from the network voluntarily. However, if the
latter is much shorter, it may, for a very brief time, disconnect
itself from the graph in order to perform some strategic move. If
player $i$ cannot temporarily increase its cost, then it will only
act such that on \emph{each move} (rather than on \emph{each turn)
}its cost will reduce. The following rules address the two alternative
limits. 
\begin{defn}
Dynamic Rule \#2a: A link $(i,j)$ will be added if $i$ asks to form
this link and $C(j;E_{m}+ij)<C(j;E_{m})$. In addition, any link $(i,j)$
can be removed in move $m.$

Dynamic Rule \#2b: In addition to Dynamic Rule \#2a, at the end of
its move is cost is reduced. Namely, player $i$ may remove a link
$(i,j)$ only if $C(i;E_{m}-ij)<C(i;E_{m})$ and may establish a link
$(i,j)$ if both $C(j;E_{m}+ij)<C(j;E_{m})$ and $C(i;E_{m}+ij)<C(i;E_{m})$
\end{defn}
If the game follows the stricter Dynamic Rule \#2b, a player's cost
must be reduced \emph{at each move}, hence such multi-move plan is
not possible.

Finally, the following lemma will be useful in the next section.
\begin{lem}
\label{lem:decay time}Assume $N$ players act consecutively in a
(uniformly) random order\textup{\emph{ at integer times, which we'll
denote by $t$.}} the probability $P(t)$ that a specific player did
not act $k\mathcal{\in N}$ times by $t\gg N$ decays exponentially.\end{lem}

\subsection{\label{sub:dynamical Results}Basic Model - Results}

After mapping the possible dynamics, we are at a position to consider
the different equilibria's basins of attraction. Specifically, we
shall establish that, in most settings of the basic model, the system
converges to the optimal network, and if not, then the network's social
cost is asymptotically equal to the optimal social cost. The main
reason behind this result is the observation that a disconnected player
has an immense bargaining power, and may force its optimal choice.
As the highest connected node is usually the optimal communication
partner for other nodes, new arrivals may force links between them
and this node, forming a star-like structure. There may be few star
centers in the graphs, but as one emerges from the other, the distance
between them is small, yielding an optimal (or almost optimal) cost.

We outline the main ideas of the proof. The first few type-B players,
in the absence of a type-A player, will form a star. The star center
can be considered as a new type of player, with an intermediate importance, as presented in Fig. \ref{fig:credible threat corr.}.
We monitor the network state at any turn and show that the minor players
are organized in two stars, one centered about a minor player and
one centered about a major player (Fig. \ref{fig:credible threat corr.}(a)).
Some cross links may be present(Fig.\,\ref{fig:cross-tiers}). By
increasing its client base, the incentive of a major player to establish
a direct link with the star center is increased. This, in turn, increases
the attractiveness of the star's center in the eyes of minor players,
creating a positive feedback loop. Additional links connecting it
to all the major league players will be established, ending up with
the star's center transformation into a member of the type-A clique.
On the other hand, if the star center is not attractive enough, then
minor players may disconnect from it and establish direct links with
the type-A clique, thus reducing its importance and establishing a
negative feedback loop. The star will become empty, and the star's
center $x$ will be become a stub of a major player, like every other
type-B player. The latter is the optimal configuration, according to proposition
\ref{lem:optimal solution}. We analyze the optimal choice of the
active player, and establish that the optimal action of a minor player
depends on the number of players in each structure and on the number
of links between the major players and the minor players' star center
$x$. The latter figure depends, in turn, on the number of players
in the star. We map this to a two dimensional dynamical system and
inspect its stable points and basins of attraction of the aforementioned
configurations. This dynamic process shows how an effectively new major player emerges
out of former type-B members in a natural way. Interestingly, Theorem
\ref{cor:credible threat part 3} also shows that there exists a transient
state with a better social cost than the final state. In fact, in
a certain scenario, the transient state is better than the optimal
stable state.
\begin{thm}
\label{cor:credible threat part 3}If the game obeys Dynamic Rules
\#1 and \#2a, then, in any playing order:

a) The system converges to a solution in which the total cost is at
most 

\begin{eqnarray*}
\mathcal{S} & = & |T_{A}|\left(|T_{A}|-1\right)\left(c+A\right)+2c_{B}|T_{B}|+\left(A+1\right)\left(3|T_{A}||T_{B}|-|T_{A}|+|T_{B}|\right)+2\left(|T_{B}|-1\right)^{2};
\end{eqnarray*}

furthermore, by taking the limit $|T_{B}|\gg|T_{A}|\gg1$, we have
$\mathcal{S}/\mathcal{S}_{optimal}\rightarrow1$ .

b) Convergence to the optimal stable solution occurs if either:

1) \textup{$A\cdot k_{A}>k+1$, }\textup{\emph{where $k\geq0$ is
the number of type-B nodes that first join the network, followed later
by $k_{A}$ consecutive type-A nodes (``initial condition'').}}

2) $A\cdot|T_{A}|>|T_{B}|$ (``final condition'').

c) In all of the above, if every player plays at least once in O(N)
turns, convergence occurs after O(N) steps. Otherwise, if players
play in a uniformly random order, the probability the system has not
converged by turn $t$ decays exponentially with $t$.\end{thm}
\begin{IEEEproof}
We first consider the case $c_{A}\geq2$. Denote the first type-A
player that establish a link with a type-B player as $k$. First,
we show that the network structure is composed of a type-A (possibly
empty) clique, a set of type-B players $S$ linked to player $x\in T_{B}$,
and an additional (possibly empty) set of type-B players $L$ connected
to the type-A player $k$. See Fig. \ref{fig:credible threat corr.}(a)
for an illustration. In addition, there is a set $D$ of type-A nodes
that are connected to node $x$, the star center. After we establish
this, we show that the system can be mapped to a two dimensional dynamical
system. Then, we evaluate the social cost at each equilibria, and
calculate the convergence rate. We first assume $(k,x)\in E$ and
then discuss the case $(k,x)\notin E$ in. We prove by induction.
At turn $t\leq2$, this is certainly true. Denote the active player
at time $t$ as $r.$ 

\begin{figure}
\centering{}\includegraphics[width=0.75\columnwidth]{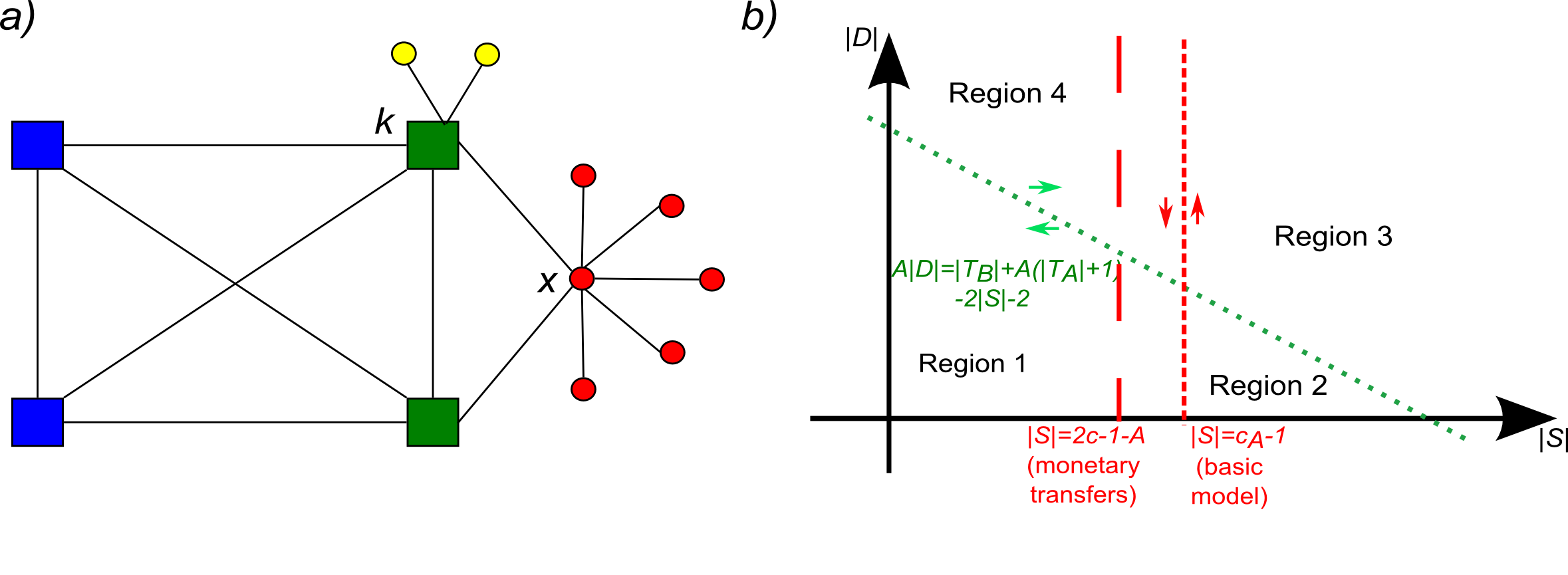}\protect\caption{\label{fig:credible threat corr.}a) The network structures described
in Theorem \ref{cor:credible threat part 3}. The type-A clique contains
$|T_{A}|=4$ nodes (squares), and there are $|S|=5$ nodes in the
star (red circles). There are $|L|=2$ nodes that are connected directly
to node $k$ (yellow circles). The number of type-A nodes that are
connected to node 1, the star center, is $|D|=2$ (green squares).
b) The phase state of Theorem \ref{cor:credible threat part 3}. The
dotted green line is the $|S|$ increase / decrease nullcline. The
dotted (dashed) red line is the nullcline for the increase / decrease
in $|D|$ when monetary transfers are forbidden (allowed). (Proposition
\ref{prop:monetary-dyanmics-1}). \label{fig:The-phase-state}}
\end{figure}

\textbf{If $(k,x)\in E$} : Consider the following cases:

Case 1, $r\in T_{A}$: Since $1<c_{A}<A,$ all links to the other
type-A nodes will be established (lemma \ref{lem:optimal solution})
or maintained, if $r$ is already connected to the network. Clearly,
the optimal link in $r$'s concern is the link with the star center
$x$. If $c_{B}<A$ every minor player will accept a link with a major
player even if it reduces its distance only by one. Therefore, the
link $(r,x)$ is formed if the change of cost of the major player
$r$,
\begin{equation}
\Delta C(r,E+rx)=c_{A}-|S|-1\label{eq: term1}
\end{equation}
is negative. In this case, the number of type A players connected
to the star's center, $|D|$, will increase by one. If this expression
is positive and player $r$ is connected to at least another major
player (as otherwise the graph is disconnected), the link will be
dissolved and $|D|$ will be reduced by one. It is not beneficial
for $r$ to form an additional link to any type-B player, as they
only reduce the distance from a single node by one (see the discussion
in lemma \ref{lem:optimal solution}). 

If $c_{B}\geq A$ then the star center will not form a link to $r$,
as such link reduce its distance to a single major player by one.
A link to any other minor player will not be beneficial to player
$r$, as it reduces its distance to a single minor player by at most
two, and $c_{A}\geq2$. The graph will remain the same.

Case 2, $r\in T_{B},\, r\neq x$ : First, assume that $r$ is a newly
arrived player, and hence it is disconnected. Obviously, in its concern,
a link to the star's center, player $x$, is preferred over a link
to any other type-B player. Similarly, a link to a type-A player that
is linked with the star's center is preferred over a link with a player
that maintains no such link.

We claim that either $(r,k)$ or $(r,x)$ exists. Denote the number
of type-A player at turn $t$ as $m_{A}.$ The link $(r,x)$ is preferred
in $r$'s concern if the expression 

\begin{equation}
C(r,E+rk)-C(r,E+rx)=-A(1+m_{A}-|D|)+1+|S|-|L|\label{eq: term 2}
\end{equation}

is positive, and will be established as otherwise the network is disconnected.
If the latter expression is negative, $(r,k)$ will be formed. The
same reasoning as in case 1 shows that no additional links to a type-B
player will be formed. Otherwise, if $r$ is already connected to
the graph, than according to Dynamic Rule \#2a, $r$ may disconnect
itself, and apply its optimal policy, increasing or decreasing $|L|$
and $|S|$.

Case 3, $r=x$, the star's center: $r$ may not remove any edge connected
to a type-B player and render the graph disconnected. On the other
hand, it has no interest in removing links to major players. On the
contrary, it will try to establish links with the major players, and
these will be formed if eq. \ref{eq: term1} is negative. An additional
link to a minor player connected to $k$ will only reduce the distance
to it by one and since $c_{B}>2$ player $x$ would not consider this
move worthy.

The dynamical parameters that govern the system dynamics are the number
of players in the different sets, $|S|$, $|L|$, and $|D|$. Consider
the state of the system after all the players have player once. Using
the relations $|S|+|L|+1=|T_{B}|,\: m_{A}=|T_{A}|$ we note the change
in $|S|$ depends on $|S|$ and $|L|$ while the change in $|D|$
depends only on $|S|.$ We can map this to a 2D dynamical, discrete
system with the aforementioned mapping. In Fig.\,\ref{fig:The-phase-state}
the state is mapped to a point in phase space $(|S|,|L|)$. The possible
states lie on a grid, and during the game the state move by an single
unit either along the $x$ or $y$ axis. There are only two stable
points, corresponding to $|S|=0,|D|=1$, which is the optimal solution
(Fig. \ref{fig:The-optimal-solution}(a)), and the state $|S|=|T_{B}|-1$
and $|D|=|T_{A}|$. 

If at a certain time expression \ref{eq: term1} is positive and expression
\ref{eq: term 2} is negative (region 3 in Fig.\,\ref{fig:The-phase-state}(b)),
the type-B players will prefer to connect to player $x$. This, in
turn, increases the utility a major player gains by establishing a
link with player $x.$ The greater the set of type-A that have a direct
connection with $x$, having $|D|$ members, the more utility a direct
link with $x$ carries to a minor player. Hence, a positive feedback
loop is established. The end result is that all the players will form
a link with $x$. In particular, the type-A clique is extended to
include the type-B player $x$. Likewise, if the reverse conditions
appliy, a feedback loop will disconnect all links between node $x$
to the clique (except node $k$) and all type-B players will prefer
to establish a direct link with the clique. The end result in this
case is the optimal stable state. The region that is relevant to the
latter domain is region 1. 

However, there is an intermediate range of states, described by region
2 and region 4, in which the player order may dictate to which one
of the previous states the system will converge. For example, starting
from a point in region 4, if the type-A players move first, changing
the $|D|$ value, than the dynamics will lead to region 1, which converge
to the optimal solution. However, if the type-B players move first,
then the system will converge to the other equilibrium point.

\textbf{If $(k,x)\notin E$: }We now discuss explicitly the case where,
at some point, the link $(k,x)$ is removed and assume that $c_{A}>2$.
In this case, the nullcline described by eq. \ref{eq: term 2} is
replaced by 
\begin{equation}
C(r,E+rk)-C(r,E+rx)=-A(1+m_{A}-|D|)+2\left(1+|S|-|L|\right).\label{eq: term 2 -kx removed}
\end{equation}

This changes the regions according to Fig. \ref{fig:The-dynamical-regions}.
Region 1, which is the basin of attraction for the optimal configuration,
increases its area, on the expense of region 4. The dynamical discussion
as described for the case $(k,x)\in E$ is still applicable, and if
the player play in a specific order, than the state vector $(|S|,|D|)$
will be in either region 1 or region 3 after $\Theta(1)$ turns. If
the players play in random order, then the system might not converge
only if player $k$ will play in every $\Theta(1)$ turns. This probability
decays exponentially, according to lemma \ref{lem:decay time}.

We now extended our discussion to the case $c_{A}\leq2.$ Here, we
claim that the structure s more complex - there could be more than
one major player with minor player clients, and some cross links between
different tiers may be present, as in Fig. \ref{fig:cross-tiers}.
We denote the set of player $k\in T_{A}$ clients as $L_{k}$. We
define player $k^{*}$ as a player such that 
\begin{eqnarray*}
k* & = & \arg\max_{k\in T_{A}}\left|L_{k}\right|
\end{eqnarray*}
namely, $k^{*}$ is a player with most direct clients.

Case 1: the active player is $r\in T_{A}-D$. As before, the most
beneficial link is a link to the star's center. As the l.h.s of Eq.
\ref{eq: term1} is negative, it will form a link with the star's
center if the latter would agree. This will take place iff $c_{B}\leq A$.
Otherwise, if 
\[
c_{B}\leq A\left(\left|T_{A}\right|-\left|D\right|-1\right)+2\left|L_{k}\right|
\]
it may form a link with some $j\in S$, as in Fig. \ref{fig:cross-tiers}(c),
as such link is beneficial for player $j\in S$ as well. No other
links will be form, as they only reduce the distance to a singly type
B player by one, and if already $r\in D$ and $|S|\neq\emptyset$
no other links will be removed (similar to the case $c_{A}>2$).

Case 2: the active player is $r\in T_{B},\, r\neq x$. As in the case
for $c_{A}>2$, after establishing its preferred connection according
to Eq. \ref{eq: term 2} (with $L\leftarrow L_{k*})$ , the active
player may establish its preferred link - either as a client of the
star center $x$, namely $r\in S$, or a client of $k^{*}$, $r\in L_{k*}$.
Additionally, If $r\in S$, it may establish a link with $k*$ if

\[
c_{B}\leq A\left(\left|T_{A}\right|-\left|D\right|-1\right)+2\left|L_{k^{*}}\right|
\]

as under this condition and $c_{A}<2$ such link is beneficial for
both $r$ and $k^{*}$. Note that such link is preferred over any
link to one of $k$'s clients. In this case, $r$ is a member of both
$S$ and $L_{k*}$, and we address this by the transformation $|S|\leftarrow|S|$,
$|L_{k*}|\leftarrow|L_{k*}|+1$. Similarly, if $r\in L$ then it will
establish links with the star center $x$ if and only if $c_{B}\leq2$.
The analogous transformation is $|S|\leftarrow|S|+1$, $|L_{k*}|\leftarrow|L_{k*}|$
. Additionally, consider a client $i\in L_{K}$ of a major player
$k\in D$ clients. If $c_{B}\leq2$, than a link between between $i$
and star's leaf, $j\in S$ is feasible and will be temporarily formed
when $i$ is the active player (see Fig. \ref{fig:cross-tiers}).
This link will be replaced by a link to player $k$ during player
$j$'s turn, according to the previous discussion. Similarly, a link
connecting client $i\in L_{K}$ of a major player $k\in T_{A}-D$
with the star center $x$ may be temporarily established, and will
be replaced by the link $(x,k)$ when $x$ will play.  

Case 3: $r=x$, is the same as in the scenario where $c_{A}>2$.

\begin{figure}
\centering{}\includegraphics[width=0.6\columnwidth]{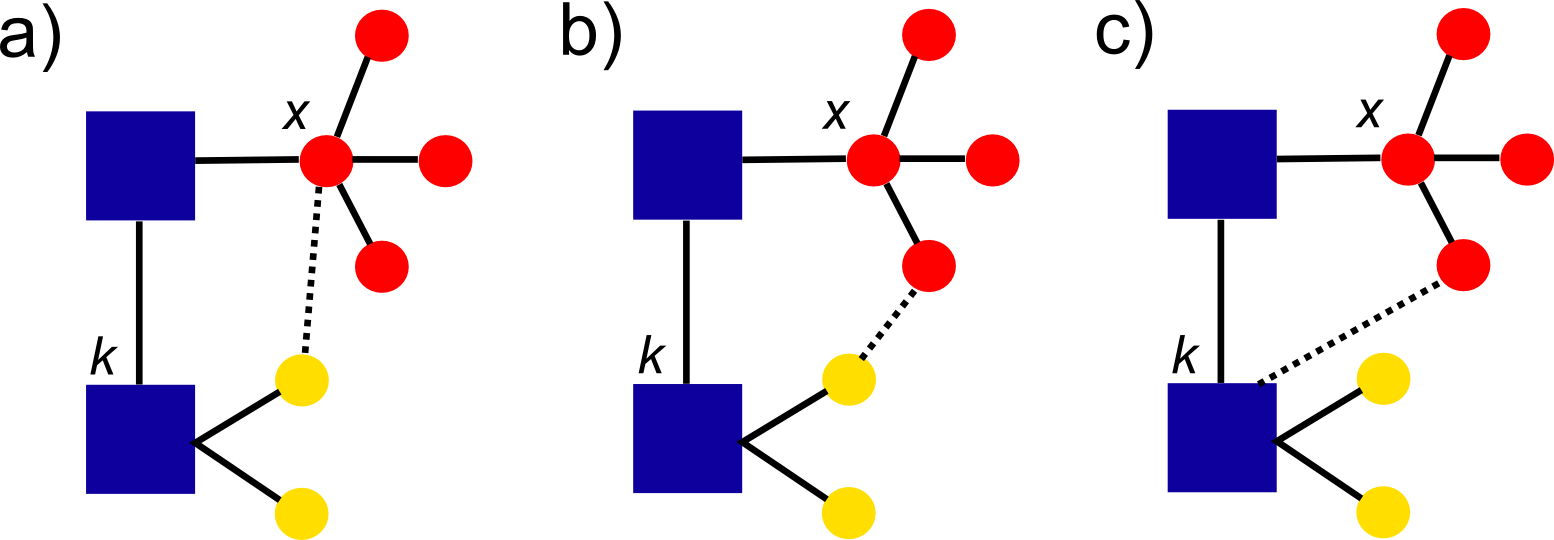}\protect\caption{\label{fig:cross-tiers}Additional feasible cross-tiers links. The
star players $S$ are in red, the set $L$ is in yellow. a) a link
between the star center and $i\in L$. b) a cross-tier link $(i,j)$
where $i\in S,j\in L$. c) a minor player - major player link, $(i,j)$
where $i\in T_{A}$ and $j\in S.$ }
\end{figure}

\begin{figure}
\centering{}\includegraphics[width=0.6\paperwidth]{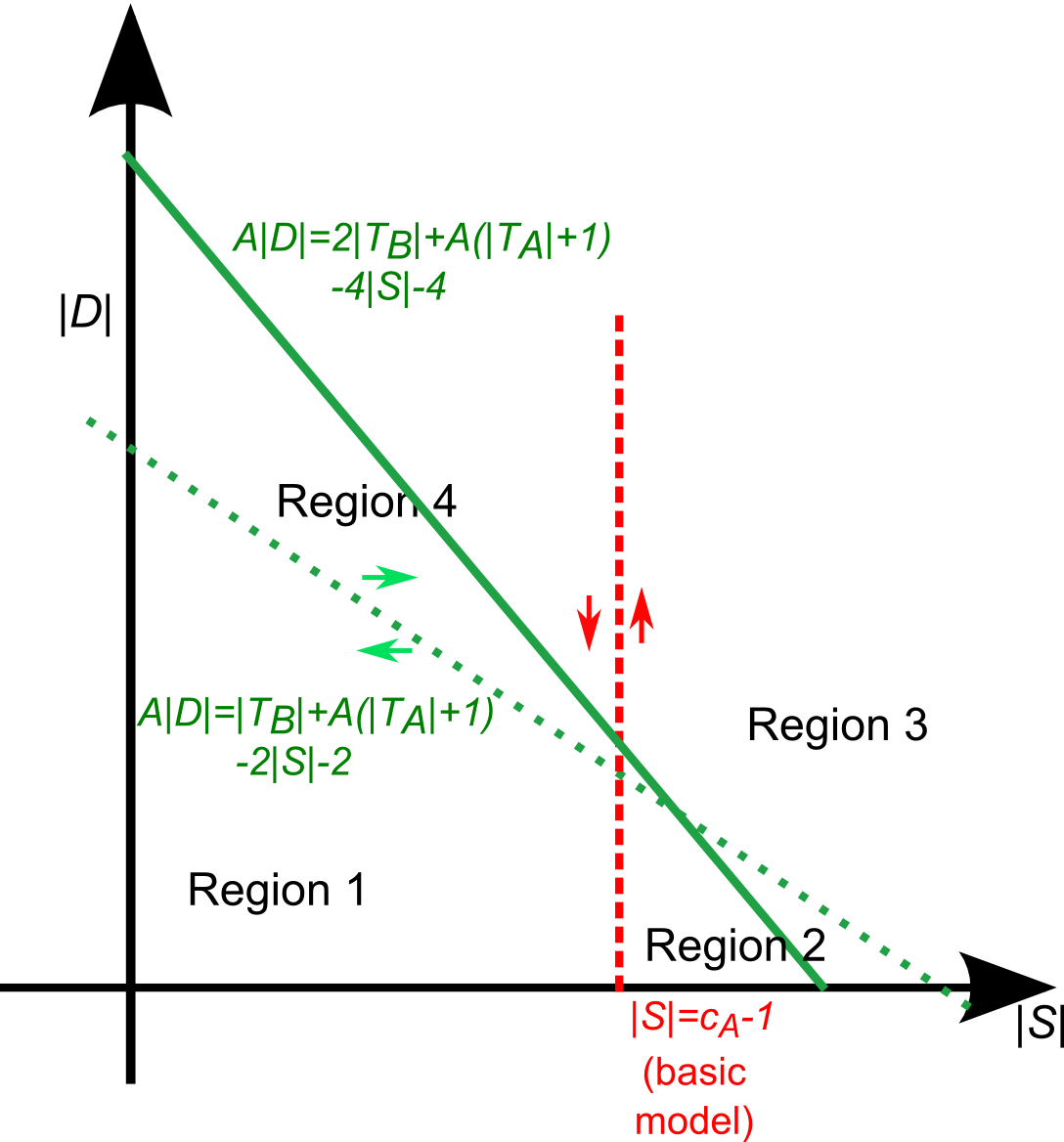}\protect\caption{\label{fig:The-dynamical-regions}The dynamical regions with or without
the link $(k,x)$. The dashed green represents the new type-B player
preference nullcline when $(k,x)\in E$, according to eq. \ref{eq: term 2}.
According to eq. \ref{eq: term 2 -kx removed}, the nullcline when
$(k,x)\protect\notin E$ is the solid green line.}
\end{figure}

We now turn to calculate the social cost at the different equilibria.
If $|D|=|T_{A}|$ and $|S|=|T'_{B}|-1$, The network topology is composed
of a $|T_{A}|$ members clique, all connected to the center $x$,
that, in turn, has $|T_{B}|-1$ stubs. The total cost in this configuration
is 
\begin{eqnarray}
S & = & |T_{A}|\left(|T_{A}|-1\right)\left(c_{A}+A\right)+2c_{B}|T_{B}|+\left(A+1\right)|T_{A}|+2\left(|T_{B}|-1\right)\nonumber \\
 &  & +2\left(|T_{B}|-1\right)\left(A+1\right)+2\left(|T_{B}|-2\right)\left(|T_{B}|-1\right)+\left(c_{B}+c_{A}\right)|T_{A}|/2\label{eq:cost at star}
\end{eqnarray}

where the costs are, from the left to right: the cost of the type-A
clique, the cost of the type-B star's links, the distance cost $(=1)$
between the clique and node $x$, the distance $(=1)$ cost between
the star's members and node $x$, the distance $(=2)$ cost between
the clique and the star's member, the distance $(=2)$ cost between
the star's members, and the cost due to major player link's to the
start center $x$. Adding all up, we have for the total cost
\begin{eqnarray}
\mathcal{S} & \leq & |T_{A}|\left(|T_{A}|-1\right)\left(c+A\right)+2c_{B}|T_{B}|+\left(A+1\right)\left(3|T_{A}||T_{B}|+|T_{B}|\right)+2\left(|T_{B}|-1\right)^{2}.\label{eq:social cost}
\end{eqnarray}

Convergence is fast, and as soon as all players have acted three times
the system will reach equilibrium. If every player plays at least
once in $o(N$) turns convergence occurs after $o(N)$ turns, otherwise
the probability the system did not reach equilibrium by time $t$
decays exponentially with $t$ according to lemma \ref{lem:decay time}
(\cite{Meirom2013}).

Taking the limit $T_{B}\rightarrow\infty$ and $T_{B}\in\omega\left(T_{A}\right)$
in eq. \ref{eq:social cost}, we get $\mathcal{S}/\mathcal{S}_{optimal}\rightarrow1$.
This concludes the proof.

\end{IEEEproof}

If the star's center has a principal role in the network, then links
connecting it to all the major league players will be established,
ending up with the star's center transformation into a member of the
type-A clique. This dynamic process shows how an effectively new major
player emerges out of former type-B members in a natural way. Interestingly,
Theorem \ref{cor:credible threat part 3} also shows that there exists
a transient state with a better social cost than the final state.
In fact, in a certain scenario, the transient state is better than
the optimal stable state.

So far we have discussed the possibility that a player may perform
a strategic plan, implemented by Dynamic Rule \#2a. However, if we
follow Dynamic Rule \#2b instead, then a player may not disconnect
itself from the graph. The previous results indicate that it is not
worthy to add additional links to the set of type-B nodes. Therefore,
no links will be added except for the initial ones, or, in other words,
renegotiation will always fail. The dynamics will halt as soon as
each player has acted once. Formally:
\begin{prop}
\label{prop:credible theat part 4}If the game obeys Dynamic Rules
\#1 and \#2b, then the system will converge to a solution in which
the total cost is at most 
\begin{eqnarray*}
\mathcal{S} & = & |T_{A}|\left(|T_{A}|-1\right)\left(c_{A}+A\right)+3|T_{B}|^{2}+2c_{B}|T_{B}|+2|T_{A}||T_{B}|\left(A+1\right)\,.
\end{eqnarray*}
Furthermore, for $|T_{B}|\gg|T_{A}|\gg1$, we have $\mbox{\ensuremath{\mathcal{S}/\mathcal{S}_{optimal}\leq3/2.}}$
Moreover, if every player plays at least once in O(N) turns, convergence
occurs after O(N) steps. Otherwise, if players play in a uniformly
random order, the probability the system has not converged by turn
$t$ decays exponentially with $t$.
\end{prop}
\begin{IEEEproof}
We discuss the case $c_{A}\geq2.$ The proof follows along the same
lines of the previous theorem (Theorem \ref{cor:credible threat part 3}).
We claim that at any given turn, the network structure is composed
of the same structures as before (See Fig. \ref{fig:credible threat corr.}(a)).
We prove by induction. Clearly, at turn one the induction assumption
is true. 

The dynamics of a is a newly arrived players are the same under Dynamic
Rules \#2a or \#2b. Therefore, if the active player is a newly arrived
player, the analysis in Theorem \ref{cor:credible threat part 3}
applies. Hence, we only need to discuss the change in policies of
existing players. 

The only difference from the dynamics described in Theorem \ref{cor:credible threat part 3}
is that the a type-B players may not disconnect itself. Therefore,
the only changes are in case 2 of Theorem \ref{cor:credible threat part 3},
in which the active player is $r\in T_{B},\, r\neq x$. Here, player
$r$ may not disconnect itself, and therefore, if it already connected,
it may not establish an additional link to star center (assuming $r\in L$)
or to any type-A player (if $r\in S$) as the utility to the other
party is at most two, and $c_{B}\geq c_{A}\geq2$. In other words,
depending on the sign of Eq. \ref{eq: term 2} at a type-B player
arrival, it will become either a leave of node $k$, namely, member
of $|L|$, or members of the star $|S|$, and then becomes stagnant.
A similar analysis holds for the case $(k,x)\notin E$, where the
decision boundary is replaced Eq. \ref{eq: term 2 -kx removed} rather
than Eq. \ref{eq: term 2}.

We extended the discussion to a scenario where $c_{A}<2$. As before,
we need to address only case 2, in which the active player is $r\in T_{B},\, r\neq x$.
If $r\in S,$ as $c_{A}\leq2$, player $r$ may establish links with
a player $j\in T_{B}-D$. This may allow it to disconnect from $S$.
Therefore, player $r\in S$ may choose to be either a member of $S$
or a member $L_{k*}$, where now 
\[
k^{*}=\arg\max_{k\in T_{A}-A}\left|L_{k}\right|
\]

according to the sign of Eq. \ref{eq: term 2 -kx removed}. The rest
of the analysis is the same as in Theorem \ref{cor:credible threat part 3}

The maximal distance between a type-A player and a type B player is
$2$. The maximal value of the type B - type B term is the social
cost function is when $|L|=|S|=|T_{B}|/2$. In this case, this term
contributes $3|T_{B}|^{2}$ to the social cost. Therefore, the social
cost is bounded by 
\begin{equation}
\mathcal{S}=|T_{A}|\left(|T_{A}|-1\right)\left(c_{A}+A\right)+3|T_{B}|^{2}+2c_{B}|T_{B}|+2|T_{A}||T_{B}|\left(A+1\right)\label{eq:bound for 2b}
\end{equation}

where we included the type-A clique's contribution to the social cost
and used $c_{B}\geq c_{A}.$ 

Taking the limit $N\rightarrow\infty$ in eq. \ref{eq:bound for 2b}
and using $T_{A}\in\omega(1)$, $T_{B}\in\omega(T_{A})$, we obtain
$\mbox{\ensuremath{\mathcal{S}/\mathcal{S}_{optimal}\leq3/2}}$. \end{IEEEproof}

Theorem \ref{cor:credible threat part 3} and Proposition \ref{prop:credible theat part 4}
shows that the intermediate network structures of the type-B players
are not necessarily trees, and additional links among the tier two
players may exist, as found in reality. Furthermore, our model predicts
that some cross-tier links, although less likely, may be formed as
well. If Dynamic Rule \#2a is in effect, These structures are only
transient, otherwise they might remain permanent.

\subsection{Monetary Transfers Dynamics\label{sub: monetary Dynamics}}

We now consider the dynamic process of network formation under the
presence of monetary transfers. For every node $i$ there may be several
nodes, indexed by \emph{j, }such that $\Delta C(j,ij)+\Delta C(i,ij)<0,$
and player \emph{i }needs to decide on the order of players with which
it will ask to form links. Each player's decision is myopic, and is
based solely on the current state of the network. Hence, the order
of establishing links is important, and may depends on the pricing
mechanism. There are several alternatives and, correspondingly, several
possible ways to specify player \emph{i's }preferences\emph{, }each
leading to a different dynamic rule. 

Perhaps the most naive assumption is that if for player $j,$ $\Delta C(j,ij)>0$,
then the price it will ask player $i$ to pay is $P_{ij}=\max\{\Delta C(j,ij),0\}.$
In other words, if it is beneficial for player $j$ to establish a
link, it will not ask for a payment in order to do so. Otherwise,
it will demand the minimal price that compensates for the increase
in its costs. This dynamic rule represents an efficient market. This
suggests the following preference order rule.

\begin{defn}
Preference Order \#1: Player $i$ will establish a link with a player
$j$ such that 
\[
\Delta C(i,ij)+\min\{\Delta C(j,ij),0\}
\]

is minimal. The price player $i$ will pay is 
\[
P_{ij}=\max\{\Delta C(j,ij),0\}
\]
\end{defn}

As established by the next theorem, Preference Order \#1 leads to
the optimal equilibrium fast. In essence, if the clique is large enough,
then it is worthy for type-B players to establish a direct link to
the clique, compensating a type-A player, and follow this move by
disconnecting from the star. Therefore, monetary transfers increase
the fluidity of the system, enabling players to escape from an unfortunate
position. Hence, we obtain an improved version of Theorem \ref{cor:credible threat part 3}.
\begin{thm}
Assume the players follow Preference Order \#1 and Dynamic Rule \#1,
and either Dynamic Rule \#2a or \#2b. If $\frac{A+1}{2}>c$, then
the system converges to the optimal solution. If every player plays
at least once in O(N) turns, convergence occurs after o(N) steps.
Otherwise, e.g., if players play in a random order, convergence occurs
exponentially fast. 
\end{thm}
\begin{IEEEproof}
Assume it is player $i$'s turn. For every player $j$ such that $(i,j)\notin E$,
we have that $d(i,j)\geq2$. By establishing the link $(i,j)$ the
distance is reduced to one and 
\begin{eqnarray*}
 &  & \Delta C(j,E+ij)+\Delta C(i,E+ij)\\
 & \leq & 2c+\left(1-d(i,j)\right)\left(2+\delta_{i,A}(A-1)+\delta_{j,A}(A-1)\right).
\end{eqnarray*}

This expression is negative if either $i\in T_{A}$ or $j\in T_{A}$,
as 
\[
2c-A-1<0
\]
Therefore, if player $i\in T_{A}$ it will form links all other players,
whereas if $i\in T_{B}$ it will form links with all the type-A players.
Finally, after every player has played twice, every type-B player
has established links to all members of the type-B clique. Therefore,
the distance between every two type-B players is at most two. Consider
two type-B players, $i,j\in T_{B}$ for which the link $(i,j)$ exists.
If the link is removed, the distance will grow from one to two, per
the previous discussion. But,

\begin{eqnarray*}
 &  & \Delta C(j,E+ij)+\Delta C(i,E+ij)\\
 & = & 2c+2\left(1-2\right)\\
 & > & 0
\end{eqnarray*}
 Hence, this link will be dissolved. This process will be completed
as soon as every type-B player has played at least three times. If
every player plays at least once in $o(N$) turns convergence occurs
after $o(N)$ turns, otherwise the probability the system did not
reach equilibrium by time $t$ decays exponentially with $t$ according
to lemma \ref{lem:decay time}. 

The resulting network structure is composed of a type-A clique, and
every type-B player is connected to all members of the type-A clique
(Fig.\ref{fig:The-optimal-solution}(b)). As discussed in Prop. \ref{prop:optimality under monetary},
this structure is optimal and stable.\end{IEEEproof}

Yet, the common wisdom that monetary transfers, or utility transfers
in general, should increase the social welfare, is contradicted in
our setting by the following proposition. Specifically, there are
certain instances, where allowing monetary transfers yields a decrease
in the social utility. In other words, if monetary transfers are allowed,
then the system may converge to a sub-optimal state.
\begin{prop}
Assume $\frac{A+1}{2}\leq c$. Consider the case where monetary transfers
are allowed and the game obeys Dynamic Rules \#1,\#2a and Preference
Order \#1. Then:

a) The system will either converge to the optimal solution or to a
solution in which the social cost is 
\begin{eqnarray*}
\mathcal{S} & = & |T_{A}|\left(|T_{A}|-1\right)\left(c_{A}+A\right)+2\left(|T_{B}|-1\right)^{2}+\left(A+1\right)\left(3|T_{A}||T_{B}|-|T_{A}|+|T_{B}|\right)+2c|T_{B}|.
\end{eqnarray*}
For $|T_{B}|\rightarrow\infty$, $|T_{B}|\in\omega\left(|T_{A}|\right)$
we have $S/S_{optimal}\rightarrow1\,$. In addition, if one of the
first $\left\lfloor c-1\right\rfloor $ nodes to attach to the network
is of type-A then the system converges to the optimal solution. 

b) For some parameters and playing orders, the system converges to
the optimal state if monetary transfers are forbidden, but when transfers
are allowed it fails to do so. This is the case, for example, when
the first $k$ players are of type-B, and $2c-A-1<k<c-1$.
\end{prop}
\begin{IEEEproof}
a) We claim that, at any given turn $t$, the network is composed
of the same structures as in Theorem \ref{cor:credible threat part 3}.
We use the notation described there. See Fig. \ref{fig:credible threat corr.}
for an illustration. First, assume that the link $(k,x)$ exists.

We prove by induction. At turn $t=1$ the induction hypothesis is
true. We discuss the different configurations at time $t$. 

1. $r\in T_{A}$: As in Theorem \ref{cor:credible threat part 3},
all links to the other type-A nodes will be established or maintained,
if $r$ is already connected to the network. The link $(r,x)$ will
be formed if the change of cost of player $r$,
\begin{equation}
\Delta C(j,E+ij)+\Delta C(i,E+ij)=2c-A-|S|-1\label{eq: term1-1-1}
\end{equation}
is negative. In this case $|S|$ will increase by one. If this expression
is positive and $(r,x)\in E,$ the link will be dissolved and $|D|$
will be reduced. It is not beneficial for $r$ to form an additional
link to any type-B player, as they only reduce the distance from a
single node by 1 and $\frac{A+1}{2}\leq c$. 

2. $r\in T_{B}$ :The discussion in Theorem \ref{cor:credible threat part 3}
shows that a newly arrived may choose to establish its optimal link,
which would be either $(r,k)$ or $(r,x)$ according to the sign of
expression \ref{eq: term 2} in Theorem \ref{cor:credible threat part 3}.
As otherwise the graph is disconnected, such link will cost nothing.
Similarly, if $r$ is already connected, it may disconnect itself
as an intermediate state and use its improved bargaining point to
impose its optimal choice. Hence, the formation of either $(r,k)$
or $(r,x)$ is not affected by the inclusion of monetary transfers
to the basic model. Assume the optimal move for $r$ is to be a member
of the star, $r\in S$. If 
\begin{eqnarray}
\Delta C(k,E+kr)+\Delta C(r,E+kr) & = & 2c-A|m_{A}|-1-|L|\label{eq:monetary, S}
\end{eqnarray}
 is negative, than this link will be formed. In this case, $r$ is
a member of both $S$ and $L$, and we address this by the transformation
$|S|\leftarrow|S|$, $|L|\leftarrow|L|+1$ and $|T_{B}|\leftarrow|T_{B}|+1.$
Similarly, if $r\in L$ than it will establish links with the star
center $x$ if and only if 
\begin{eqnarray}
2c & < & |S|+1.\label{eq:monetary, L}
\end{eqnarray}
The analogous transformation is, $|S|\leftarrow|S|+1$, $|L|\leftarrow|L|$
and $|T_{B}|\leftarrow|T_{B}|+1.$ The is also true if $r=x$ and
the latter condition is satisfied. We have shown in Theorem \ref{cor:credible threat part 3}
that such links only reduce the social cost, do not incite link removals,
and do not effect the considerations of new type-B player. 

Consider the case that at some point the link $(k,x)$ was removed.
The new player preference nullcline is described by eq. \ref{eq: term 2 -kx removed}.
Now, if $r\in S$, it has an increased incentive to establish a link
with $k$, as the nodes in $L$ are farther away from it. In this
case, the condition to establish the link $(k,r)$ is 
\begin{eqnarray*}
\Delta C(k,E+kr)+\Delta C(r,E+kr) & = & 2c-A|m_{A}|-1-2|L|<0
\end{eqnarray*}
(compared to eq. \ref{eq:monetary, S}) . Similarly, if $r\in L$
the criteria for establishing the link $(r,x)$ is $2c<2|S|+1$ (compared
to eq. \ref{eq:monetary, L}). The transformation described above
should be applied in either case.

As before, as soon as all the players have played two time, the system
will be in either region 1 or region 3, and from there convergence
occurs after every player has played once more.

b) If dynamic rule \#2a is in effect, the nullcline represented by
eq. \ref{eq: term1-1-1} is shifted to the left compared to the nullcline
of eq. \ref{eq: term1}, increasing region 3 and region 2 on the expanse
of region 1 and region 4. Therefore, there are cases where the system
would have converge to the optimal state, but allowing monetary transfers
it would converge to the other stable state. Intuitively, the star
center may pay type-A players to establish links with her, reducing
the motivation for one of her leafs to defect and in turn, increasing
the incentive of the other players to directly connect to it. Hence,
monetary transfers reduce the threshold for the positive feedback
loop discussing in Theorem \ref{cor:credible threat part 3}.\end{IEEEproof}

The latter proposition shows that the emergence of an effectively
new major league player, namely the star center, occurs more frequently
with monetary transfers, although the social cost is hindered. 

A more elaborate choice of a price mechanism is that of ``strategic''
pricing. Specifically, consider a player $j^{*}$ that knows that
the link $(i,j^{*})$ carries the least utility for player $i$. It
is reasonable to assume that player $j$ will ask the minimal price
for it, as long as it is greater than its implied costs. We will denote
this price as $P_{ij^{*}}$. Every other player $x$ will use this
value and demand an additional payment from player $i$, as the link
$(i,x)$ is more beneficial for player $i$. Formally,

\begin{defn}
Pricing mechanism \#1: Set $j^{*}$ as the node which maximize $\Delta C(i,E+ij*)$.
Set $P_{ij^{*}}=max\{-\Delta C(j*,E,ij*),0\}$. Finally, set
\[
\alpha_{ij}=\Delta C(i,E+ij)-\left(\Delta C(i,E+ij^{*})+P_{ij^{*}}\right)
\]

The price that player $j$ will require in order to establish \emph{$(i,j)$
}is\emph{ }
\[
P_{ij}=max\{0,\alpha_{ij},-\Delta C(j,E+ij)\}
\]
\end{defn}

As far as player $i$ is concerned, all the links $(i,j)$ such that
$P_{ij}=\alpha_{ij}$ carry the same utility, and this utility is
greater than the utility of links for which the former condition is
invalidated. Some of these links have a better connection value, but
they come at a higher price. Since all the links carry the same utility,
we need to decide on some preference mechanism for player $i$. The
simplest one is the ``cheap'' choice, by which we mean that, if
there are a few equivalent links, then the player will choose the
cheapest one. This can be reasoned by the assumption that a new player
cannot spend too much resources, and therefore it will choose the
``cheapest'' option that belongs to the set of links with maximal
utility.

\begin{defn}
Preference Order \#2: Player $i$ will establish links with player
$j$ if player $j$ minimizes 
\[
\Delta\tilde{C}(i,ij)=\Delta C(i,ij)+P_{ij}
\]

and 
\[
\Delta\tilde{C}(i,ij)<0
\]

If there are several players that minimize $\Delta\tilde{C}(i,ij)$,
then player $i$ will establish a link with a node (player) that minimizes
$P_{ij}$. If there are several nodes that satisfy the previous condition,
one will be chosen randomly.\end{defn}
Note that low-cost links have a poor ``connection value'' and therefore
the previous statement can also be formulated as a preference for
links with low connection value. 

\begin{defn}
Preference Order \#1: Player $i$ will establish a link with a player
$j$ such that 
\[
\Delta C(i,ij)+\min\{\Delta C(j,ij),0\}
\]

is minimal. The price player $i$ will pay is 
\[
P_{ij}=\max\{\Delta C(j,ij),0\}
\]
\end{defn}

\begin{defn}
``Strategic'' Pricing mechanism: Set $j^{*}$ as the node that maximizes
$\Delta C(i,E+ij*)$. Set $P^{*}=\max\{-\Delta C(j*,E+ij*),0\}$.
Denote the excess utility of the link $(i,j)$ as $\alpha_{ij}=\Delta C(i,E+ij)-\left(\Delta C(i,E+ij^{*})+P^{*}\right).$
The price player $j$ requires in order to establish \emph{$(i,j)$
}is\emph{ }$P_{ij}=\max\{0,\alpha_{ij},-\Delta C(j,E+ij)\}$.
\end{defn}
We proceed to consider the dynamic aspects of the system under such
conditions.

An immediate result of this definition is the following.
\begin{lem}
If node $j^{*}$ satisfies
\[
\Delta C(j*,ij*)<0
\]
then the link $(i,j^{*})$ will be formed. If there are few nodes
that satisfy this criterion, a link connecting $i$ and one of this
node will be picked at random.\end{lem}
\begin{IEEEproof}
As 
\[
P_{ij^{*}}=max\{-\Delta C(j*,E,ij*),0\}=0
\]
 $\Delta\tilde{C}(i,ij)$ is maximal when $\Delta C(i,ij)$ is, which
is for node $j^{*}$.
\end{IEEEproof}
The resulting equilibria following this preference order are very
diverse and depend heavily on the order of acting players. The only
general statement that can are of the form of Lemma \ref{lem:The-longest-distance-1}.
Before we elaborate, let us provide another useful lemma.
\begin{lem}
\label{lem:shortcut}Assume that according to the preference order
player $i$ will establish the link $(i,j^{*})$. If there is a node
$x$ such that
\begin{eqnarray*}
\Delta C(i,(E+ij*)+ix) & < & 0\\
\Delta C(x,(E+ij*)+ix) & < & 0\\
\Delta C(i,(E+ix)+ij^{*})+\Delta C(j^{*},(E+ix)+ij^{*}) & > & 0
\end{eqnarray*}

Then effectively the link $(i,x)$ will be formed instead of $(i,j^{*})$.\end{lem}
\begin{IEEEproof}
The first two inequalities state that after establishing the link
$(i,j^{*})$ the link $(i,x)$ will be formed as well. However, the
last inequality indicates that after the former step, it is worthy
for player $i$ to disconnect the link $(i,j^{*})$. 
\end{IEEEproof}
We proceed to consider the dynamic aspects of the system under such
conditions.
\begin{prop}
\label{prop:monetary-dyanmics-1}Assume that:

A) Players follow Preference Order \#2 and Dynamic Rule \#1, and either
Dynamic Rule \#2a or \#2b. 

B) There are enough players such that $2c<T_{A}\cdot A+T_{B}^{2}/4$.

C) At least one out of the first $m$ players is of type-A, where
$m$ satisfies $m\geq\sqrt{A^{2}+4c-1}-A$.

Then, if players play in a non-random order, the system converges
to a state where all the type-B nodes are connected directly to the
type-A clique, except perhaps lines of nodes with summed maximal length
of $m$. In the large network limit, $\mathcal{S}/\mathcal{S}_{optimal}<3/2+c$
.

D) If $2c>(A-1)+|T_{B}|/|T_{A}|$ then the bound in (C) can be tightened
to $\mathcal{S}/\mathcal{S}_{optimal}<3/2$.
\end{prop}
\begin{IEEEproof}
We prove by induction. Assume that the first type-B player is player
$k_{A}$ and the first type-A player is $k_{B}$. We first prove that
the structure up-to the first move of $\max\{k_{A},k_{B}\}$ is a
type-A clique, and an additional line of maximal length $\left|k_{B}-k_{A}\right|$
of type-B players connected to a single type-A player.

If $k_{A}>k_{B}=1$, there is a set of type-B players that play before
the first type-A joins the game, we claim they form a line. For the
first player it is true. We denote the sequence of players $1...x$
and assume the assumption holds up to player $x-1$ . Consider the
newly arrived player $x$. The least utility it may obtain is by establishing
a link to a node at the end of the line, and therefore it will connect
first to one of the ends, as this would be the cheapest link. W.l.o.g,
assume that it connects to $x-1$. The most beneficial additional
link it may establish is $(1,x)$ but according to Corollary \ref{lem:shortcut benefit}
\begin{eqnarray*}
 &  & \Delta C(x,E+1x)+\Delta C(1,E+1x)\\
 & = & 2\Delta C(1,E+1x)\\
 & = & 2c--\frac{x\left(x-2\right)+mod(x,2)}{2}\\
 & > & 2c-\frac{m\left(m-2\right)+1}{2}\\
 & > & 0
\end{eqnarray*}

and therefore no additional link will be formed. Player $k_{A}$ will
establish a link with a node at one of the line's ends, say to player
$k_{A}-1$, and as 
\begin{eqnarray*}
 &  & \Delta C(k_{A},E+1k_{A})+\Delta C(1,E+1k_{A})\\
 & > & 2c-\frac{m\left(m-2\right)+1}{2}-(m-2)(A-1)
\end{eqnarray*}

there would be no additional links between player $k_{A}$ and a member
of the line.

If $k_{A}<k_{B}=1$ then the first $k_{B}-1$ type-A player will form
a clique (same reasoning as in lemma \ref{prop:optimality under monetary}),
and player $k_{B}$ (of type-B) will form a link to one of the type-A
players randomly. This completes the induction proof.

For every new player, the link with the least utility is the link
connecting the new arrival and the end of the type-B strand. For a
type-A player, using lemma \ref{lem:shortcut}, immediately after
establishing this link it will be dissolved and the new player will
join the clique. Type B players will attach to the end of the line
and the line's length will grow. Assume that when player $x$ turn
to play there are $m_{x}$ type-A players in the clique. By establishing
the link $(i,x)$ we have (Fig. \ref{fig:suggested shortcut}), 
\begin{eqnarray*}
\Delta C(x,E+ix) & < & c_{B}-m_{X}\cdot A\\
\Delta C(i,E+ix) & = & c_{A}-\frac{\left(x+1\right)\left(x-1\right)+mod(x+1,2)}{2}.
\end{eqnarray*}

For $x\geq m\geq\sqrt{A^{2}+4c-1}-A$
\[
\Delta C(x,ix)+\Delta C(i,ix)<2c-m_{X}\cdot A-x^{2}/4<0
\]

and therefore according to lemma \ref{lem:shortcut} player $x$ will
connect directly to one of the nodes in the clique instead of line\textquoteright s
end. After all the players have played once, the structure formed
is a type-A clique, a line of maximum length $m$, and type-B nodes
that are connected directly to at least one of the members of the
clique. In a large network, the line's maximal length is $o(\sqrt{T_{B}})$
and is negligible in comparison to terms that are $o(T_{B})$. 

\begin{figure}
\centering{}\includegraphics[width=0.8\columnwidth]{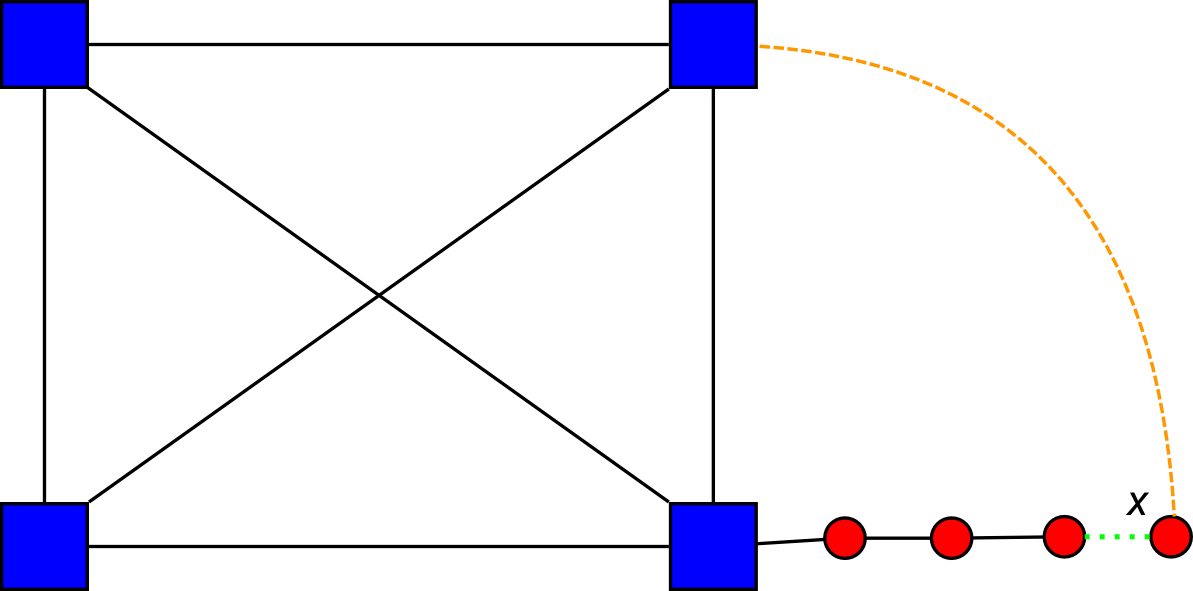}\protect\caption{\label{fig:suggested shortcut}The suggested dynamics in Prop. \ref{prop:monetary-dyanmics-1}.
Here, $x=4$ and $m_{X}=4$. The link to the clique is dashed in orange,
the canceled link is in dotted green.}
\end{figure}

The only possible deviation in this sub-graph is establishing additional
links between a type-B player and clique members (other than the one
it is currently linked to). This will only reduce the overall distance.
Hence we can asymptotically bound the social cost by 
\begin{eqnarray*}
\mathcal{S} & < & 3|T_{B}|^{2}+2|T_{B}||T_{A}|\left(A+1\right)\\
 &  & +2|T_{B}||T_{A}|c+|T_{A}|^{2}\left(c_{A}+A\right)
\end{eqnarray*}

where the first term expresses the (maximal) distance between type-B
nodes, the second the distance cost between the type-A clique and
the type-B nodes, the third is the cost of links between type-A players
and type-B players and the third is the cost of the type-A clique.
Comparing this to the optimal solution (Proposition \ref{prop:maximal distance from clique with money-1})
under the assumption that $T_{B}>T_{A}$, we have
\[
\mathcal{S}/\mathcal{S}_{optimal}<\frac{3}{2}+c.
\]

This concludes the proof.

D) We can improve on the former bound by noting that according to
Preference Order \#2, when disconnecting from the long line a player
will reconnect to the type-A player that has the least utility in
his concern (and hence requires the lowest price). In other words,
it will connect to one of the nodes that carry the least amount of
type-B nodes at that moment. Therefore, to each type-A node roughly
$|T_{B}|/|T_{A}|$ type-B nodes will be connected. This allows us
to provide the following corollary. 

According to the aforementioned discussion by establishing a link
between node $j\in T_{B}$ and $i\in T_{A}$ (where $j$ in not connected
directly to $i$) the change of cost is 
\begin{eqnarray*}
\Delta C(i,ij) & = & c_{A}-A-\frac{|T_{B}|}{|T_{A}|}\\
\Delta C(j,ij) & = & c_{B}-1
\end{eqnarray*}

but
\[
\Delta C(i,ij)+\Delta C(j,ij)=2c-(A-1)-\frac{|T_{B}|}{|T_{A}|}>0
\]

and the link would not be establish. Hence, we can neglect the term
$2|T_{B}||T_{A}|c$ and 
\[
\sum C(i)<3|T_{B}|^{2}+2|T_{B}||T_{A}|\left(A+1\right)+|T_{A}|^{2}\left(c_{A}+A\right)
\]

By comparing with the optimal solution we obtained the required result.\end{IEEEproof}

\begin{figure}
\centering{}\includegraphics[width=0.8\columnwidth]{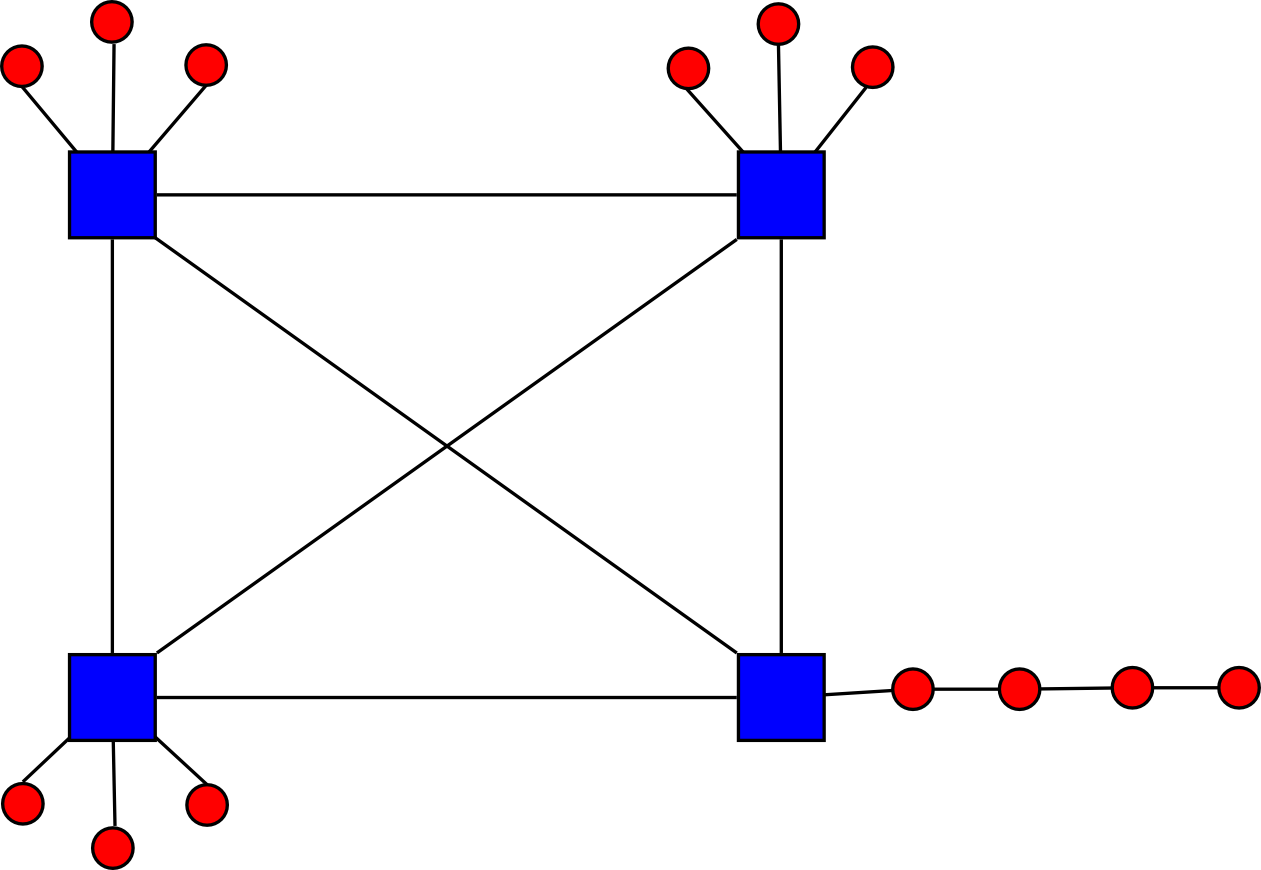}\protect\caption{The network structure described in Prop. \ref{prop:monetary-dyanmics-1}.
The type-B players connect almost uniformly to the different members
of the type-A clique. In addition, there is one line of type-B players. }
\end{figure}
In order to obtain the result in Proposition \ref{prop:maximal distance from clique with money},
we had to assume a large limit for the number of type-A players. Here,
on the other hand, we were able to obtain a similar result yet without
that assumption, i.e., solely by dynamic considerations. 

It is important to note that, although our model allows for monetary
transfers, in \emph{every} resulting agreement between major players
no monetary transaction is performed. In other words, our model predicts
that the major players clique will form a \emph{settlement-free} interconnection
subgraph, while in major player - minor player contracts transactions
will occur, and they will be of a transit contract type. Indeed, this
observation is well founded in reality.

\subsection{\label{sub:dynamical Results - reliability}Network Motifs}

\begin{figure}
\centering{}\includegraphics[width=1\columnwidth]{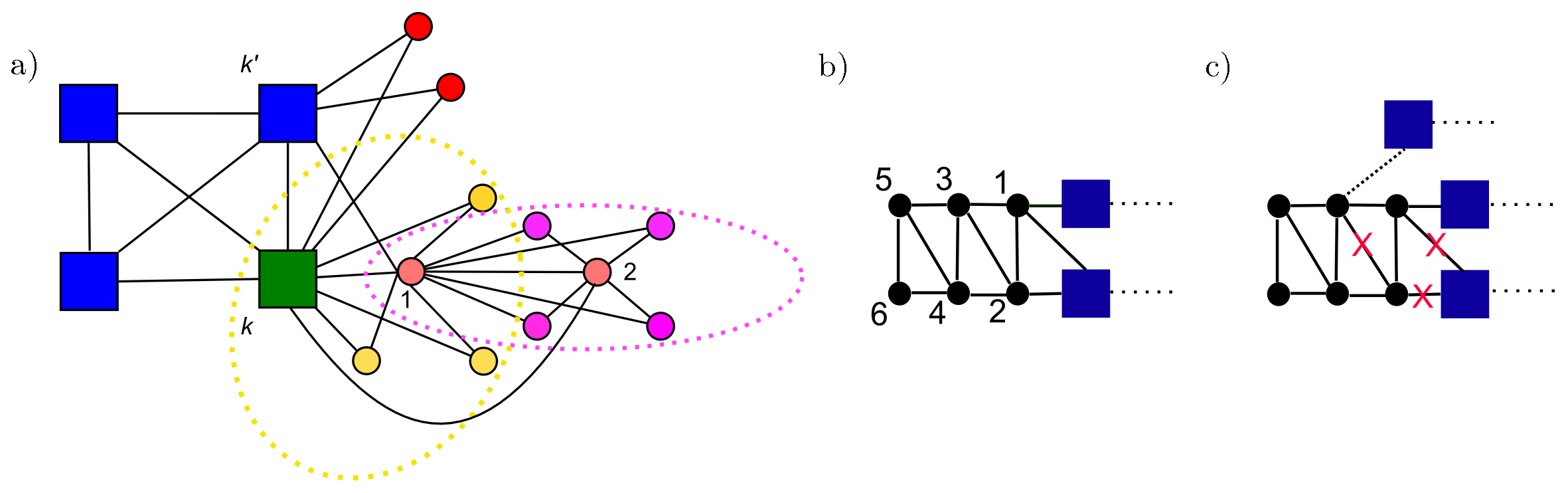}\protect\caption{\label{fig:generalized star 2}Network Motifs. a) A network configuration
which includes a ``double-star'' structure of minor players. Every
node in the primary star (encircled in yellow) is linked to a major
player, node $k$ (in green). A direct link connects the two star
centers, denoted by 1 and 2 (in pink). The members in the secondary
star (in purple) are connected to both star centers. In addition,
there secondary star center is also connected to the major player
$k$. There might be additional minor players outside the stars (in
red). \label{fig:The-entangled-loops} b) The ``entangled cycles''
motif. Six minor player are connected in an ``entangled cycles''
subgraph. The first two nodes have direct connection to some major
player, and access the rest of the network by the major player's additional
links, represented by the dotted line. c) If at some point, a link
between a player in this subgraph and some external player is formed
(in this example, a major player), some links may be removed without
violating the reliability requirement and without increasing the distance
cost appreciably. The removed links marked by a red X. and a generalized star about node $k,k'$ of (the red type-B nodes).
In this example, $m_{A}=4$, $m_{SA}=2$,$m_{S'A}=1$,$m_{B}=2,m_{S}=3,m_{S'}=4$
. }
\end{figure}

We now show that, under survivability constraints, during the natural
evolution of the network a ``double star'' sub-graph, or network
motif, often emerges. In the ``double star'' motif, as depicted
in Fig. \ref{fig:generalized star 2}, there exists a primary and
a secondary star. All the minor players are connected to the primary
star's center. Part of the players are also connected to the other
star's center, forming the secondary star. Consider a region where
it is immensely difficult to establish a link to a major player, either
due to geographical distance, link prices or perhaps additional physical
links are simply not accessible. Nevertheless, in order to maintain
a reliable connection, there must be at least two links that connect
this region to the Internet backbone via some major players. In order
to provide a stable, fault tolerant service, every player in this
region will form links with the players hosting the endpoints of these
links, forming the double star sub-graph. Assume now that link prices
reduce over time, or that the importance of a fast connection to the
Internet core increases in time. In this case, players may decide
to establish direct links with the major players, and remove either
one of both links connecting them to the star centers. Note though,
that players will be reluctant to disconnect from the star center
if the number of nodes in the star is large.

Moreover, the next theorem also shows that, eventually, and fairly
quickly, the system will converge to either the optimal stable state,
or to a state in which the social cost is a low multiple of the optimal
social cost.
\begin{thm}
\label{thm:symmetric dynamic}Assume symmetric reliability requirements,
i.e., $\tau=1$. If the players follow Dynamic Rules \#1 and \#2a,
then, in any playing order:

A) The system converges to either the optimal stable state,  depicted in Fig. \ref{fig:optimal and stable},
or to the network depicted in Fig. \ref{fig:generalized star 2}.

B) In the large network limit, namely, when $|T_{B}|\gg|T_{A}|\gg1$,
the social costs ratio satisfy $\mathcal{S}/\mathcal{S}_{optimal}<3/2+\epsilon$,
with $\epsilon\rightarrow0$.

C) if players play in a uniformly random order, the probability that
the system has not converged by turn $t$ decays exponentially with
$t$. Otherwise, if every player plays at least once in O(N) turns,
convergence occurs after O(N) steps.\end{thm}
\begin{IEEEproof}
We claim that the network, at any time, has the following structure
(Fig. \ref{fig:generalized star 2}): A type-A players' clique, and
\emph{at most} two type-B star centers $1,2\in T_{B}$. The larger
ball's center is labelled as $1$, and the smaller is $2$. We denote
the set of type-B players which are members of the star centered about
node 1 (2) as $S_{1}$ and (correspondingly, $S_{2}$). We have $\left|S_{1}\right|\geq\left|S_{2}\right|.$
In addition, some type-B players might be linked to two nodes $k,k'\in T_{A}$,
the set of these type-B players is denoted by $L$. The set of type-A
players that have a direct link with the star center $1$ (star center
$2$) is denoted by $D_{1}$ (correspondingly $D_{2}$). Players $\{1,2,k\}$
form a clique. Assuming the second star exists, it is connected to
either players $1$ or $k'$ (or both). Finally, there may be additional
links between players is $L$ and $D_{2}$. An example of this network
structure is presented in Fig. \ref{fig:generalized star 2}. 

\begin{table}
\centering{}%
\begin{tabular}{|c|c|c|c|}
\hline 
$d(i,j)$ & $i\in n_{1}$ & $i\in n_{2}$ & $i\in n_{3}$\tabularnewline
\hline 
\hline 
$j=1$ & 1 & 1 & 2\tabularnewline
\hline 
$j\in S_{1}$ & 2 & 2 & 2\tabularnewline
\hline 
$j=2$ & 2 & 1 & 2\tabularnewline
\hline 
$j\in S_{2}$ & 2 & 2 & 3\tabularnewline
\hline 
$j\in L$ & 2 & 2 & 2\tabularnewline
\hline 
$j=k$  & 1 & 2 & 1\tabularnewline
\hline 
$j=k$' & 2 & 2 & 1\tabularnewline
\hline 
$j\in D_{2}$, $j\neq k,k'$ & 2 & 2 & 2\tabularnewline
\hline 
$j\in T_{A}$, $j\notin D_{2}$ & 2 & 3 & 2\tabularnewline
\hline 
\end{tabular}\protect\caption{\label{tab:d(i,k)}The shortest distance $d(i,j)$ between two nodes,
as discussed in Prop. \ref{thm:symmetric dynamic}. Note that if $k'\in D_{2}$,
than column 9 applies instead of column 8 for $i\in S_{2}$.}
\end{table}

\begin{table}
\begin{centering}
\begin{tabular}{|c|c|c|c|}
\hline 
$d'(i,j)$ & $i\in n_{1}$ & $i\in n_{2}$ & $i\in n_{3}$\tabularnewline
\hline 
\hline 
$j=1$ & 2 & 2 & 2\tabularnewline
\hline 
$j\in S_{1}$ & 2 & 3 & 2\tabularnewline
\hline 
$j=2$ & 2 & 2 & 3\tabularnewline
\hline 
$j\in S_{2}$ & 3 & 2 & 3\tabularnewline
\hline 
$j\in L$ & 3 & 3 & 2\tabularnewline
\hline 
$j=k$  & 2 & 2 & 2\tabularnewline
\hline 
$j=k$' & 2 & 3 & 2\tabularnewline
\hline 
$j\in D_{2}$, $j\neq k,k'$ & 2 & 2 & 2\tabularnewline
\hline 
$j\in D_{1}$, $j\notin D_{2}$ & 2 & 3 & 2\tabularnewline
\hline 
$j\notin D_{1},$$j\in T_{A}$ & 3 & 3 & 2\tabularnewline
\hline 
\end{tabular}
\par\end{centering}

\centering{}\protect\caption{\label{tab:d'(i,k)}The second shortest distance $d'(i,j)$ between
two nodes, as discussed in Prop. \ref{thm:symmetric dynamic}. Note
that if $k'\in D_{2}$, than column 9 applies instead of column 8
for $i\in S_{2}$.}
\end{table}

We prove by induction. After the first three players played the induction
base case is true. We first assume that the link $(1,2)$ and consider
the case $(1,2)\notin E$ later. Denote the active player by $r$.
Consider the following cases:

1. $r\in T_{A}$: As $c<A/2$, $r$ will form (or maintain) links
with every type-A player. If additional links to minor players are
to be formed, it is clearly better for player $r$ to establish links
first with either player $1$ or player $2$ than to any player in
$i\in S_{1}$ or $i\in S_{2}$. We first consider the case where $r\notin D_{1}\cup D_{2}$
and split into two cases:

1A) If no two disjoint paths exists between player $r$ to players
$1$ or $2$, namely, if $|T_{A}|\geq2$ and either $|D_{1}|<\min\{|T_{A}|,2\}$
or $|D_{2}|\leq\min\left\{ |T_{A}|,2\right\} $, then it is beneficial
for both $r$ and the star centers to link such that reliability requirements
will be satisfied. 

1B) If $D_{2}\neq\emptyset$ then by establishing the edge $(r,1)$
we have
\begin{eqnarray*}
\Delta C(r,E+1r) & = & c-\frac{1+|S_{1}|}{1+\delta}
\end{eqnarray*}

while by establishing $(r,2)$ the change of cost is:

\begin{eqnarray}
\Delta C(r,E+2r) & = & c-\frac{1+|S_{2}|}{1+\delta}.\label{eq:2r edge}
\end{eqnarray}

As $|S_{2}|\leq|S_{1}|$, player $r$ will prefer to establish first
a link with player 1 rather than with player 2. The link will be formed
if $\Delta C(1,E+1r)\leq0$, which is true if 
\begin{eqnarray}
|D_{1}| & \leq & 1\nonumber \\
 & \mathcal{\text{or}}\label{eq:a-b link condition}\\
c_{B} & < & A/\left(1+\delta\right).\nonumber 
\end{eqnarray}
If condition \ref{eq:2r edge} holds, then player $r$ will attempt
to form a link with the star center $2$, and succeed if condition
\ref{eq:a-b link condition} holds as well. Establishing a link to
any other type-B node is clearly an inferior option, as if a player
had decided to establish links with either node $1$ or node $2,$
an additional link to one of their leafs will only reduce the distance
to it by 1 and hence is not a worthy course of action. Therefore,
$D_{2}\subseteq D_{1}$.

2. $r\in T_{B}$ and $r\neq1,2$. A player may disconnect itself and
then choose its two optimal links. Clearly, among the type-A clique,
player $r's$ best candidates are players $k,k'$, while among the
type-B players, the cost reduction is maximal by linking to nodes
$1,2$. In addition, a link to node $k$ is at least as preferred
a link to $k'$, while a link to node 1 is always preferred over a
link to node $2$ as long as $n_{1}\geq n_{2}$. Therefore, we only
need to consider three possible moves by player $r$ - linking to
$k,k'$ (the red players in Fig. \ref{fig:generalized star 2}), linking
to $k,1$ or linking to $1,2$. 

Using tables \ref{tab:d(i,k)} and \ref{tab:d'(i,k)} we can compare
the cost of connecting to nodes $1,2$ versus the costs of connecting
to nodes $1$ and $k$. A simple calculation shows that if the expression
\[
A-1+(|D_{1}|-|D_{2}|)A+\delta\frac{|S_{1}|-|S_{2}|+A+A(\left|T_{A}\right|-|D_{2}|)}{1+\delta}
\]

is positive, than linking to players $1$ and $k$ is preferred over
linking to player $1$ and $2$. This expression is positive, as $|D_{1}|\geq|D_{2}$
and $|S_{2}|\leq|S_{1}|$.

In a similar fashion, we can compare between the choices of linking
to nodes $1$ and $k$ or forming links with nodes $k$ and $k'.$
If the expression 
\[
1+|S_{2}|-A+\delta\frac{1-|L|-A(\left|T_{A}\right|-|D_{1}|)}{1+\delta}
\]

is positive, then establishing links to $1$ and $k$ is preferred,
otherwise the alternative will be chosen. Two special cases are of
interest: If $|S_{2}|<A-1$ and $\delta=0$, then every player will
prefer to link to $k$ and $k'$. That is, we'll get the optimal configuration.
If $\delta=1$, $T_{A}=D_{1}$, i.e, node 1 is a member of the clique,
$|L|=0$ and $|S_{2}|\gg1$, then a minor player will connect to players
$1$ and $k$. Note that if there are no type-A players present when
$r$ plays, then it must connect to the star centers $1$ and $2$. 

In addition, after $r$ formed two links, no player $x\in T_{A}\cup\{1,2\}\cup D_{1}\cup L$
will agree to form the link $(x,r)$, as the induced change of cost
is
\[
\Delta C(x,E+xr)\geq c_{A}-1>0.
\]
If $c_{B}\leq2$ and $\delta\rightarrow0$ then player $r\in L$ and
player $x\in D_{2}$ will establish the link $(x,r)$. By symmetry,
this also happens when $r\in D_{2}$ and $x\in L$.

3. If $r=1,$ then it must maintain links to all nodes in $S_{2}\cup S_{1}\cup\{2\}$
(if there are any) in order to satisfy the reliability criteria. Additional
links to type-A players will be formed according to the discussion
in case 1, while no other links to type-B player will be formed according
to the discussion in case 2. 

4. If $r=2$, then it must maintain links to all nodes in $S_{2}$
(if there are any) in order to satisfy the reliability criteria. A
similar calculation to the one in case $2$ shows that player $2$'s
best course of action is to remain connected to player 

In order to complete the induction proof all that is left is to address
the case $(1,2)\notin E$. In this case, player $2$ must be connected
to at least one additional player $i\notin S_{2}$ in order to have
two disjoint path to every player in the network. Clearly, its optimal
choice is player $k'$. In order to maintain reliable path to every
$i\in S_{2}\cup\{2\}$ player $k'$ must agree to form this link.
Player $2$ will disconnect $(1,2)$ if either $\Delta C(2,E-12)\leq0$
or $\Delta C(2,E-12+2k')\leq0$ . However, if either of this conditions
hold, Tables \ref{tab:d(i,k)} and \ref{tab:d'(i,k)} show that every
$i\in T_{B}/\{2\}$ would prefer to link to player $k'$ rather than
player $2$. In particular, this is also true for every node $i\in S_{2}$.
Therefore, as soon as every member in $S_{2}$ played once, the star
will be empty and no additional star will raise again. Therefore,
there will be at most one star left in the network.

This completes the proof on the momentarily structure of the network.
Note that if $|S_{1}|=|S_{2}|=\emptyset$ we obtain the optimal stable
solution. 

B) Since every minor player has either two or three links, the contribution
to the social cost due to minor players' links is $o(|T_{B}|),$ while
the contribution to the social costs due to the inter-distances between
minor players is $o\left(T_{B}^{2}\right)$. In the limit $|T_{A}|\rightarrow\infty,|T_{B}|/|T_{A}|\rightarrow\infty$,
the dominant term in the cost function is the term proportional to
$|T_{B}|^{2}$. Both $d(i,j)$ and $d'(i,j)$ for $i,j\in T_{B}$
are bounded by $3$, and comparing this results with the optimal social
cost (Proposition \ref{lem:reliable optimal 1}) we have
\[
\frac{\mathcal{S}}{\mathcal{S}_{optimal}}=\frac{3T_{B}^{2}+o(T_{B}^{2})}{2T_{B}^{2}+o(T_{B}^{2})}\leq\frac{3}{2}+\epsilon
\]

with $\epsilon\rightarrow0$ in the limit $|T_{A}|\rightarrow\infty,|T_{B}|/|T_{A}|\rightarrow\infty$. 

C) The discussion of case 2 shows that in a large network, it is suboptimal
for a minor player to be a member of $S_{2}$. Given the the opportunity,
it will prefer to become a member in either $S_{1}$ or $L$. Therefore,
after every player has played at least twice, $S_{2}=\emptyset$,
and after every player has played four times the system will reach
equilibrium. Lemma 13 in \cite{Meirom2014} then shows that the probability
that the system has not converged by turn $t$ decays exponentially
with $t$. 
\end{IEEEproof}
In section \ref{sec:Static-analysis} we emphasized the importance
of symmetry in the reliability requirements for reducing the Price
of Anarchy. The next theorem affirms this assertion, and shows that
if the constraints are asymmetric, the system converges to a state
with an unbounded social cost on a large set of possible dynamics
and initial conditions.
\begin{thm}
Assume asymmetric reliability requirements, namely $\tau=0$. If the
players follow Dynamic Rules \#1 and either Dynamic Rule \#2a or \#2b,
then the system converges to a state with an unbounded social cost.
\end{thm}
For technical reason, we define the cost of a node with violated survivability
constrains as $Q$, and take $Q\rightarrow\infty$, such that $Q$
is greater than any property, namely 
\[
Q\in\omega\left(\exp\left(T_{A}+T_{B}+A+c_{A}+c_{B}\right)\right).
\]

\begin{IEEEproof}
The idea behind the proof is to show that the reliability requirements
of at least one minor player remain unsatisfied, the therefore the
social costs, as a sum over individual costs, is unbounded.

First assume that $c_{A}>2$. We shall now show that at any given
turn, the network is composed of a type-A (possibly empty) clique,
a set of type-B players $S$ linked to player $x$, acting as a star
center, and an additional (possibly empty) set of type-B players $L$
connected to the type-A player $k$. Player $x$ is also connected
to player $k$. Some major player may establish links with the star
center $x.$ The set of these major players is denoted by $D$. Note
that the cost of every player in either $S$ or $L$ is $Q\rightarrow\infty$,
since every path from each of these players to the type-A clique crosses
player $k$. 

We prove by induction. At turn $t\leq2$, this is certainly true.
Denote the active player at time $t$ as $r.$ Consider the following
cases:

1. $r\in T_{A}$: Since $1<c_{A}<A,$ all links to the other type-A
nodes will be established or maintained, if $r$ is already connected
to the network. Clearly, the optimal link in $r$'s concern is the
link with the star center $x$. Therefore, player $r$ will attempt
to establish the link $(r,x)$ if
\begin{equation}
\Delta C(r,E+rx)=c_{A}-|S|-1\label{eq: term1-1}
\end{equation}
is negative. If $c_{B}<A/\left(1+\delta\right)$ or that $|D|\leq1$
then $x$ will accept this link. Regardless, if player $r$ has formed
the link it will not establish a link with $i\in S$, as it only reduces
its distance to player $i$ by a single hop, and 
\begin{eqnarray}
\Delta C(r,E+ir) & = & c_{A}-1\leq0.\label{eq:no single hop}
\end{eqnarray}

If Eq. \ref{eq: term1-1} is positive, then, player $r$ has no incentive
to establish a link with any $i\in S$, as $|S|\geq1$ and the by
establishing $(i,r)$ the sole change is the reduction of $d(i,r)$
from $3$ to $1$, 
\[
\Delta C(r,E+ir)=c_{A}-2\geq c_{A}-|S|-1=\Delta C(r,E+rx)\geq0.
\]

Eq. \ref{eq:no single hop} also shows that if $r\neq k$, it will
not form a link with $i\in L$, while if $r=k$ it may not remove
the link $(k,i)$ as otherwise $i$ is disconnected.

2. $r\in T_{B},\, r\neq x$ : First, assume that $r$ is a newly arrived
player, hence it is disconnected. Obviously, in its concern, a link
to the star's center, player $x$, is preferred over a link to any
other type-B player. Similarly, a link to a player $k$ is preferred
over a link to any other type-A player (or if $L=\emptyset$ and $D=T_{A}$,
equivalent to a link to any other type-A player). Therefore, $r$
first link choice would be either $(r,k)$ or $(r,x)$. In other words,
$r\in L$ or $r\in S$. If $r\in L$, than no $i\in T_{A}$ will agree
to establish a link with $r$, as it only reduces its distance from
$r$ by one hop, and doesn't alter $d(i,x)$ for any other $x$. Similarly,
if $r\in S$ than 
\begin{eqnarray*}
\Delta C(i,E+ri) & = & c_{A}-2>0
\end{eqnarray*}

and the link would not be formed.  Likewise, no link between $i\in L$
and $j\in S$ may be formed, as 
\[
\Delta C(i,E+ij)=c_{B}-2>c_{A}-2>0
\]

3. $r=x$, the star's center: $r$ may not remove any edge connected
to a type-B player and render the graph disconnected. On the other
hand, the previous discussion shows it will not establish additional
links to nodes in $L$.

Note that the reliability requirements of players in $L\cup S$ are
invalidated. Therefore, the cost of every player $j\in L\cup S$ is
 $\mathcal{Q}$, and we have $Q\rightarrow\infty$. Hence, at any
given turn, as soon as either $|L|>1$ and $|S|\geq1$, the social
cost is unbounded.

If $c_{A}<2$ a link between player $i\in T_{A}$ and player $r\in T_{B}$
will be formed. However, as soon as player $i$ becomes the active
player, it is beneficial for it to remove the link $(i,r)$ and establish
a link to the star center $(i,x)$ instead. Therefore, after player
$i$'s turn, the cost of player $r$ is again $Q$. In a similar fashion,
if $c_{B}<2$ than player $i\in S$ and player $j\in L$ may establish
the link $(i,j)$. This does not affect any other player and while
it may reduce their costs, it does not provide them with two disjoint
paths to the type-A clique, as they both must traverse player $k$
in order to access it. Therefore, their costs is still $Q$.
\end{IEEEproof}
An additional network motif, the ``entangled cycles'' motif, arises
when monetary transfers are taken into account. This network motif
is composed of a line (i.e., interconnected sequence) of minor players'
nodes, with some cross-links between the nodes along this line, breaking
the hierarchy (Fig. \ref{fig:The-entangled-loops}). The ``entangled
cycle'' of length three is the ``feedback loop'' motif, which was
previously found to exist in a higher frequency than expected in the
Internet graph \cite{Milo2002a}. 

When a new minor player arrives, it will choose the two cheapest links
and will connect to the corresponding players. Clearly, its costs
due to the distance from the rest of the network will be the highest.
As such, when the next player arrives, it will offer the lowest link
price. The new arrival will link to it and to one of its providers.
The process will repeat, until, at some point, an existing player
will decide that this growing branch is too far from it, and will
connect to one of the nodes along this ``entangled cycles''. At
this point, this subgraph will be over-saturated with links as players
may utilize this link to access the Internet core. Hence, some links
will be removed (see, for example Fig. \ref{fig:The-entangled-loops}(b)).
The set of the links that will be removed depends heavily on the playing
order and the temporary network structure. Nevertheless, some cross-links
may remain in order to satisfy reliability constrains. This explains
the following result.
\begin{thm}
\label{thm:entangled cycles motif}Assume the number of major players
is at least $|T_{A}|\geq\sqrt{4c}$. Denote the distance cost of player
$i\in T_{\beta},$ $\beta\in\{A,B\}$, as $D(i)$, namely
\[
D(i)=C_{\beta}(i)-c_{\beta}deg(i)\cdot x_{i}
\]
Assume that a subset $W$ of minor player first join the game and
play consecutively and that the two players with the maximal distance
cost are adjacent. Then:

A) These players will form an ``entangled cycles'' structure of
length $l$, as depicted in Fig. \ref{fig:The-entangled-loops}, and
\[
l\leq2\sqrt{\left(A|T_{A}|\right)^{2}+5A}-2A|T_{A}|.
\]

B) The \textquotedblleft entangled cycles\textquotedblright{} structure
is semi-stable, in the following sense: If, at some later turn, there
exist players $j\notin W$, $i\in W$ such that the link $(i,j)$
is formed (Fig. \ref{fig:The-entangled-loops}(b)), then some links
in the ``entangled cycles'' structure may be removed in subsequent
turns.\end{thm}
\begin{IEEEproof}
Assume that at time $t$ a subset of $W=\{x_{1},x_{2}...x_{m}\}$
players first join the game. Denote the set of players that are connected
to the network at time $t$ as $N'$. We denote by $n_{A}$ ($n_{B})$
the number of type A players (correspondingly, type-B player) at that
moment. Let us denote the players with the highest distance cost term
as $x_{0}$ and $x_{-1}$, namely,
\begin{eqnarray*}
x_{0} & = & \arg\max_{i\in N'}C(i)-c_{B}deg(i)\cdot x_{i}\\
x_{-1} & = & \arg\max_{i\in N'\backslash\left\{ x_{0}\right\} }C(i)-c_{B}deg(i)\cdot x_{i}
\end{eqnarray*}

where for simplicity we assumed that $x_{0},x_{-1}$ are minor player
(type-B players). According to the ``strategic'' pricing mechanism
a new player will establish links with these players, as they will
offer the cheapest links. Note that by connecting to these players,
its survivability requirement are satisfied, as each of these players
maintain two disjoint path to every major player (if the reliability
requirements are asymmetric) or to all other players (in case of symmetric
reliability requirements).

We are now going to prove by induction that at step $j$, player $x_{j}$
will connect to players $x_{j-1}$ and $x_{j-2}$. We shall prove
by this by showing that the distance cost of $x_{j-1}$ and $x_{j-2}$
is maximal.

Assume the induction holds, and every player $\{x_{1},x_{2}...x_{j-1}\}$
is connected only to its two predecessors, which, at the time it played,
had the maximal distance cost. First, note that for every player $x_{j'}=1...j-1$,
we have $D(x_{j'})\geq D(x_{j'-1})$ and $D(x_{j'})>D(x_{j'-2})$,
since the path from $j$' to every player in $N'$ pass through $x_{j'-1}$
or $x_{j'-2}$, and $N'\gg l>|W|$. Therefore, in order to show that
players $x_{j-1}$ and $x_{j-2}$ have the highest distance cost,
it is sufficient to show that $D(x_{j-1})>D(y)$ for every $y\in N'$.
For every player $i\in N$' we have 
\begin{eqnarray*}
d(x_{j-1},i) & \geq & d(x_{0},i)+\left\lfloor j/2\right\rfloor \\
d'(x_{j-1},i) & \geq & d'(x_{0},i)+\left\lfloor j/2\right\rfloor 
\end{eqnarray*}
 since the path that connects player $x_{j-1}$ to the any player
$i\in N'$ crosses either player $x_{0}$ or $x_{-1}$, which are
adjacent. Therefore, the distance cost of player $x_{j-1}$ due to
its distance from every player $i\in N'$, denoted by $\tilde{D}(x_{j-1})$
is at least grater than the corresponding distance cost of player
$x_{0}$, $\tilde{D}(x_{0})$ by at least $\left(An_{A}+n_{B}\right)\left\lfloor j/2\right\rfloor $.
Note that $\tilde{D}(x_{0})\geq\tilde{D}(y)$ for every $y\in N'$
according to the definition of $x_{0}$. However, player $x_{j'-1}$
may be closer to players $\{x_{1},x_{2}...x_{j-2}\}$ than player
$y$. Denote the maximal distance between any two players by $r.$
We have,
\[
D(x_{j-1})-D(y)\geq\left(An_{A}+n_{B}\right)\left\lfloor j/2\right\rfloor -r\cdot j
\]

where $r$ is the maximal distance between two players. In proposition
\ref{prop:max distance} it was shown that 
\begin{eqnarray*}
r & \leq & 2\left\lfloor \sqrt{\left(An_{A}\right)^{2}+5c}-An_{A}\right\rfloor |
\end{eqnarray*}

Plugging in the relation $|T_{A}|\geq\sqrt{4c}$ we obtain 
\[
r\leq\left\lfloor \frac{An_{A}}{2}\right\rfloor 
\]

And therefore $D(x_{j-1})-D(y)\geq0$. Therefore, links will be established
as stated. This shows that at the time player $x_{j}$ joins the game,
the distance cost of player $x_{j-1}$ is maximal. A similar calculation
shows that at this turn, the distance cost $D(x_{j-2})$ is maximal
in the set $\left\{ D(i)|i\in N'\cup\{x_{1},..x_{j-2}\}\right\} $. 

We have thus far shown that the first two links of a new player $x_{j}$
are to players $x_{j-1}$ and $x_{j-2}$. Nevertheless, as the ``entangled
cycles'' motif grows, the incentive of other players to connect to
players in it increases. At some point, player $j$ may be able to
form links with players not in $W$, or an active player $r\notin W$
may decide to connect to some player in $W$ (or vice versa for $r\in W$).
In either of these cases, players in the ``entangled cycles'' motif
may have three disjoint paths to other players, and may therefore
remove a link, should they deemed to do so. In this case, the ``entangled
cycles'' motif will be diluted.
\end{IEEEproof}
This theorem shows that reliability is a major factor in breaking
up tree hierarchy in the Internet. In addition, it also hints that
the hierarchical structure does not break frequently in the top levels
of the Internet, but rather mostly in the intermediate and lower tiers.
Note that according to Theorem \ref{thm:entangled cycles motif}(B),
in large networks the length of the ``entangled motifs'' is short.
Therefore, we do not expect to see excessively long structures, but
rather small ones, having just a few ASs.

\section{\label{sec:Data-Analysis}Data Analysis}

\begin{figure}
\centering{}\includegraphics[width=1\textwidth]{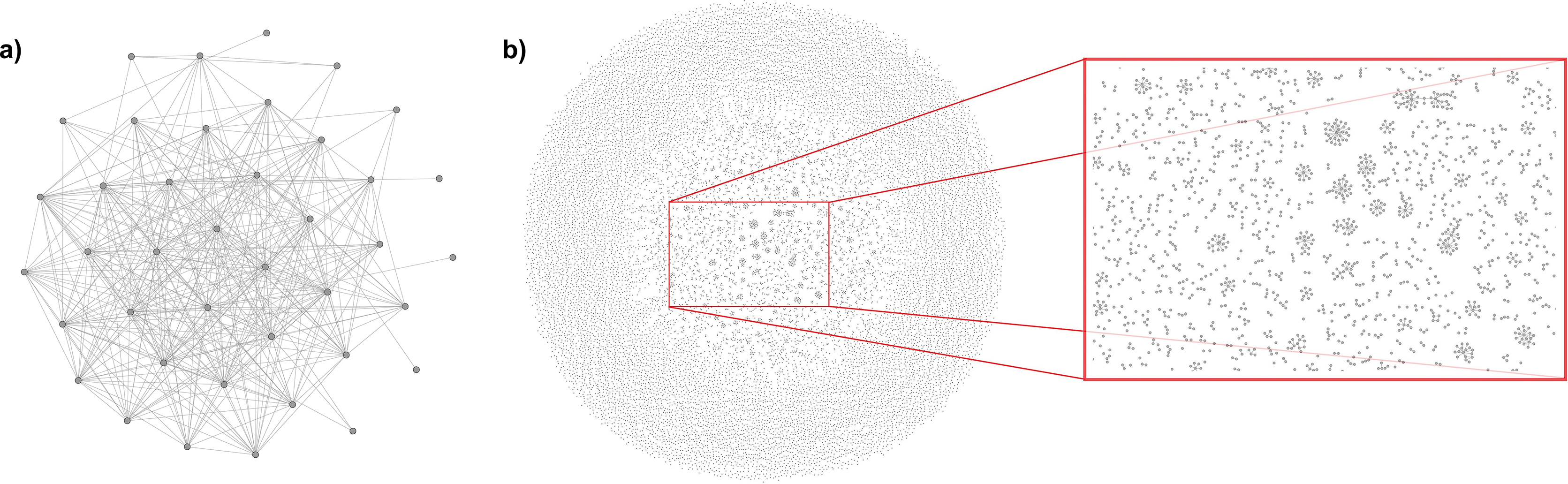}\protect\caption{Structure of the AS topology. a) The sub-graph of the top 40 ASs,
according to CAIDA ranking, in January, 2006. b) The minor nodes sub-graph
was created by omitting nodes in higher k-core $(k\geq3$) and removing
any links from the shell to the core. The subgraph contains 16,442
nodes, which are $\sim75\%$ of the ASs in the networks. Left: The
full network map. Singleton are displayed in the exterior and complex
objects in the center. Right: A zoom-in on a sample (red box) of the
subgraph. The complex structure are mainly short lines and stars (or
star-like objects).}
\label{fig:clique subgraph}
\end{figure}

In this section we compare our theoretical findings with actual monthly
snapshots of the inter-AS connectivity, reconstructed from BGP update
messages \cite{Gregori2011}.

Our model predicts that, for $A>c_{A}>1$ , the type-A (``major league'')
players will form a highly connected subset, specifically a clique
(Section \ref{sub:The-type-A-clique}). The type-B players, in turn,
form structures that are connected to the clique. Figure \ref{fig:clique subgraph}
presents the graph of a subset of the top 100 ASs per January 2006,
according to CAIDA ranking \cite{CAIDA}. It is visually clear that
the inter-connectivity of this subset is high. Indeed, the top 100
ASs graph density, which is the ratio between the number of links
present and the number of possible links, is 0.23, compared to a mean
$0.024\pm0.004$ for a random connected set of 100. It is important
to note that we were able to obtain similar results by ranking the
top ASs using topological measures, such as betweeness, closeness
and k-core analysis.

\begin{figure}
\centering{}\includegraphics[width=1\columnwidth]{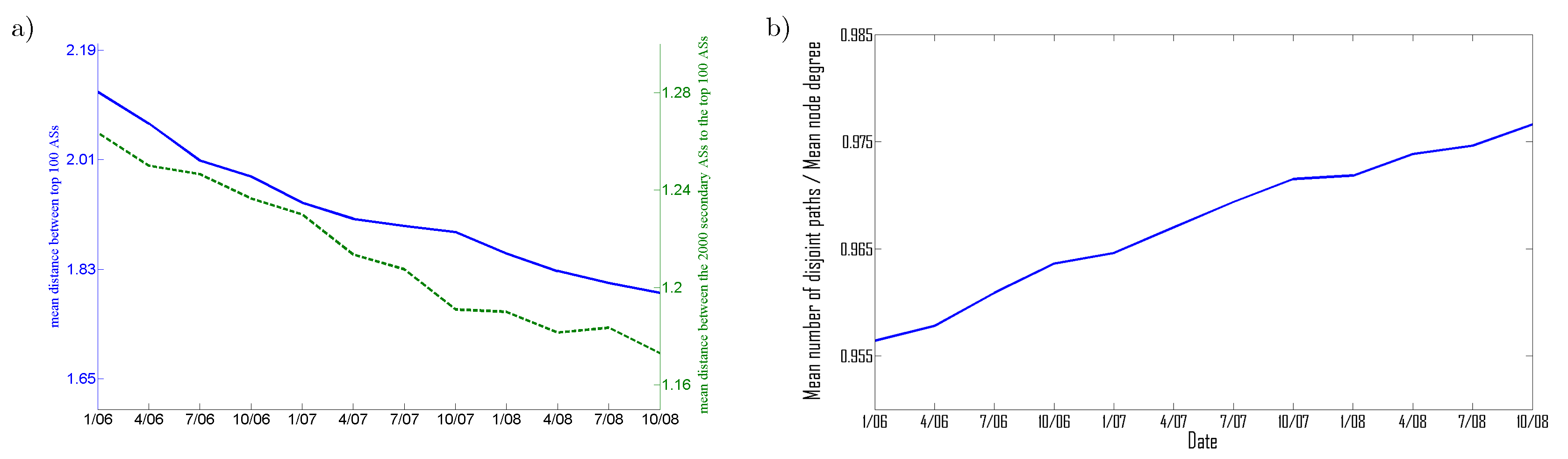}\protect\caption{\label{fig:shell-core distance graph} a) In solid blue: the mean
distance of an AS in the top 100 ASs CAIDA ranking to all the other
top 100 ASs, from January, 2006 to October, 2008. In dashed green:
The mean shortest distance of a secondary AS (ranked 101-2100) from
any top AS (ranked 1-100). b) \label{fig:disjoint path ratio}The
ratio of the mean number of disjoint paths connecting a minor player
to the core to the mean degree of a minor player. This ratio is above
$0.95$, and increases in time, showing that additional links are
likely to part of disjoint paths to the core. This ratio is presented
as a function of time from January 2006 to October 2008.}
\end{figure}

In the dynamic aspect, we expect the type-A players sub-graph to converge
to a complete graph. We evaluate the mean node-to-node distance in
this subset as a function of time by using quarterly snapshots of
the AS graph from January, 2006 to October, 2008. Indeed, the mean
distance decreases approximately linearly. The result is presented
in Fig. \ref{fig:shell-core distance graph}(a). Also, the distance
value tends to $1$, indicating the almost-completeness of this sub-graph.

Although, in principle, there are many structures the type-B players
(``minor players'') may form, the dynamics we considered indicates
the presence of stars and lines mainly (Sections \ref{sub:dynamical Results}
and \ref{sub: monetary Dynamics}). While the partition of ASs to
just two types is a simplification, we still expect our model to predict
fairly accurately the structures at the limits of high-importance
ASs and marginal ASs. A \emph{$k$-core} of a graph is the maximal
connected subgraph in which all nodes have degree of at least $k$.
The \emph{$k$-shell} is obtained after the removal of all the $k$-core
nodes. In Fig. \ref{fig:clique subgraph}, a snapshot of the sub-graph
of the marginal ASs is presented, using a $k$-core separation ($k=3$),
where all the nodes in the higher cores are removed. The abundance
of lines and stars is visually clear. In addition, the spanning tree
of this subset, which consists of 75\% of the ASs in the Internet,
is formed by removing just 0.02\% of the links in this sub-graph,
a strong indication for a forest-like structure.

\begin{figure}
\centering{}\includegraphics[width=1\columnwidth]{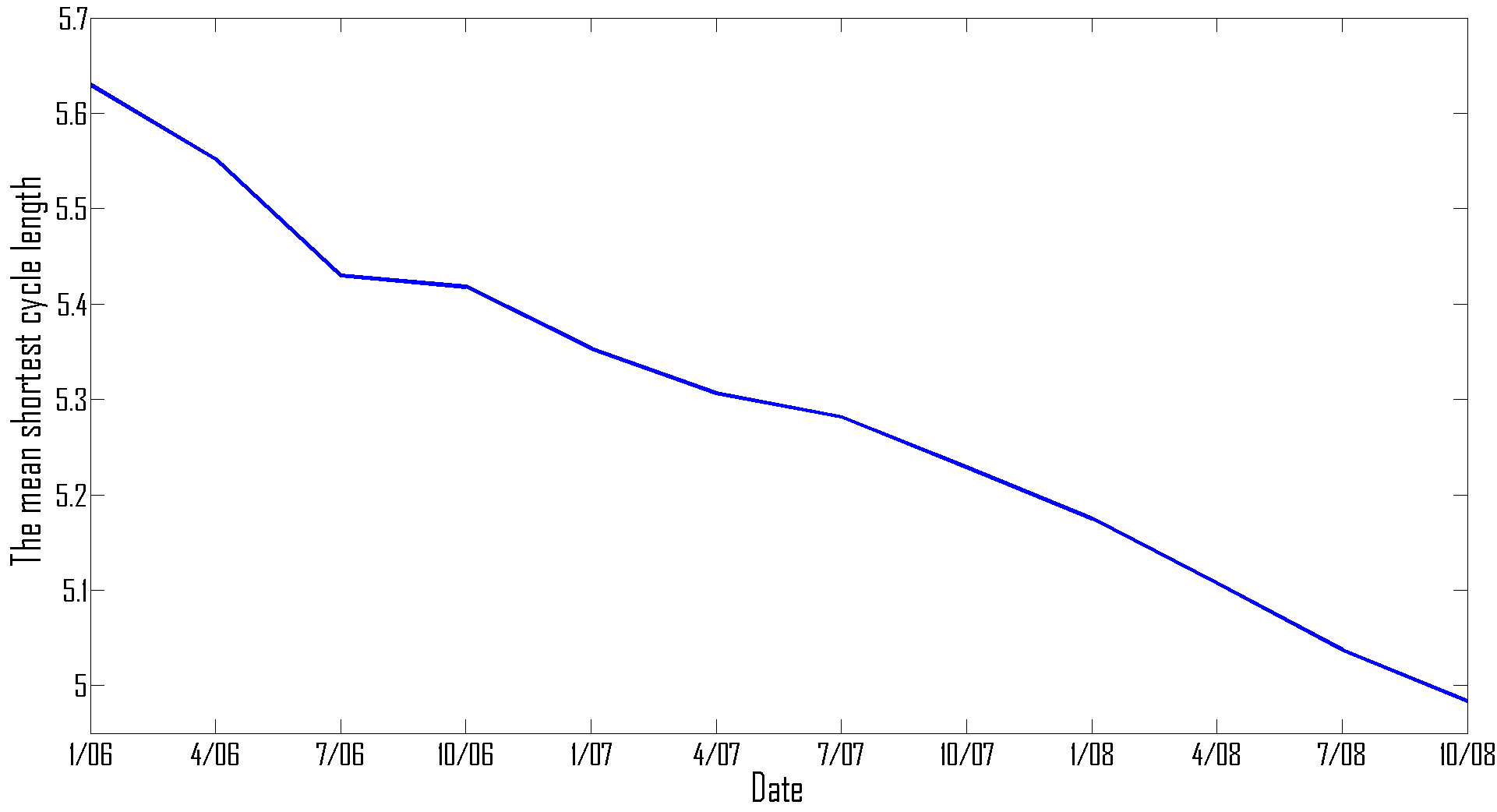}\protect\caption{\label{fig:The-mean cycle-length}The mean length of the shortest
cycle connecting a major player and a minor player as function of
time, from January 2006 to October 2008. The length decrease as time
passes and the network grows, in agreement with our model. }
\end{figure}

For a choice of core $\mathcal{C}$, the \emph{node-core distance}
of a node $i\notin\mathcal{C}$ is defined as the shortest path from
node $i$ to any node in the core. In Section \ref{sub:The-type-A-clique},
we showed that, by allowing monetary transfers, the maximal distance
of a type-B player to the type-A clique (the maximal ``node-core
distance'' in our model) depends inversely on the number of nodes
in the clique and the number of players in general. Likewise, we expect
the mean ``node-core'' distance to depend inversely on the number
of nodes in the clique. The number of ASs increases in time, and we
may assume the number of type-A players follows. Therefore, we expect
a decrease of the aforementioned mean ``node-core distance'' in
time. Fig \ref{fig:shell-core distance graph} shows the mean distance
of the secondary leading 2000 ASs, ranked 101-2100 in CAIDA ranking,
from the set of the top 100 nodes. The distance decreases in time,
in agreement with our model. Similarly, In Section \ref{sub:static-survivability},
we showed that, if the AS are constrained by survivability requirements,
mostly applicable in the intermediate and top tiers of the Internet,
the cycle length connecting a major player to a minor player depends
inversely on the number of major players. The steady decline of the
cycle's length in time is predicted by our model (see section in \cite{Meirom2016})
(Fig. \ref{fig:The-mean cycle-length}).

It is widely assumed that the evolvement of the Internet follows a
``preferential attachment'' process \cite{Barabasi1999}. According
to this process, the probability that a new node will attach to an
existing node is proportional (in some models, up to some power) to
the existing node's degree. An immediate corollary is that the probability
that a new node will connect to any node in a set of nodes is proportional
to the set's sum of degrees. The sum of degrees of the secondary ASs
set is \textasciitilde{}1.9 greater than the sum of degrees in the
core, according to the examined data\cite{Gregori2011}. Therefore,
a ``preferential attachment'' class model predicts that a new node
is likely to attach to the shell rather than to the core. As all the
nodes in the shell have a distance of at least one from the core,
the new node's distance from the core will be at least two. Since
the initial mean ``shell-core'' distance is \textasciitilde{}1.26,
a model belonging to the ``preferential attachment'' class predicts
that the mean distance will be pushed to two, and in general increase
over time. However, this is contradicted by the data that shows (Fig
\ref{fig:shell-core distance graph}) a decrease of the aforementioned
distance. The slope of the latter has the 95\% confidence bound of
$(-3.1\cdot10^{-3},-2.3\cdot10^{-3}$) hops/month, a strong indication
of a negative trend, in disagreement with the ``preferential attachment''
model class. In contrast, this trend is predicted by our model, per
the discussion in Section \ref{sec:Monetary-transfers}. In fact,
if the Internet is described by a random, power law (``scale free'')
network, then the mean distance should grow as $\Theta(\log N)$ or
$\Theta(\log\log N)$ (\cite{Cohen2003}). However, experimental observations
shows that the mean distance grows slower than that (\cite{Pastor-Satorras2001}
), and it fact it may even be reduced with the network size, as predicted
by our model. 

Our analysis showed that in most of the generated topologies, the
minor players are organized in small subgraphs that have direct connection
to the Internet core, namely the major players clique, or the tier-1
subgraph. In order to maintain a reliable connection, in each subgraph
there must be at least two links that connect minor players to the
core. Indeed, we have found out that the ratio between the mean number
of disjoint paths from a minor player to the core and the mean degree
of minor players is more than $0.95$, and it increases in time (Fig.
\ref{fig:disjoint path ratio}). That is, almost every outgoing link
of a minor player is used to provide it with an additional, disjoint
path to the core. In other words, a player is more likely to establish
an additional link, hence increase its degree, if it supplies it with
a new path to the core that does not intersect its current paths.

In section \ref{sec:Dynamic-Analysis} we predicted the ubiquity of
two network motifs, the ``double star'' motif (Fig. \ref{fig:generalized star 2})
and the ``entangled cycles'' motif (Fig. \ref{fig:The-entangled-loops}).
We define the occurrence of a ``double star'' motif as the existence
of a connected pair of nodes, each with degree greater than $m$,
designated as the centers, such that at least $m$ neighbors of one
center are also neighbors of the other center. We generated random
networks according to the Configuration Model (CM), in which each
node is given a number of stubs according to its degree, and stubs
are connected uniformly. Then, we evaluated the mean number of occurrences
of this motif in a random CM network with the same number of nodes
and the same degree distribution as the real inter-AS topology. For
$m=2$, we have found $28.8K$ occurrences in the real-world inter-AS
topology, whereas the mean number of occurrences in the random CM
network was only \textbf{$5.8K\pm1.3K$ }instances (the $\pm$ indicates
standard deviation). Namely, there are more than four times displays
of this subgraph in the Internet than in a random network with the
same degree distribution. Chebyshev's inequality provides a bound
on the \emph{p-value, $p<0.003$.} This low value indicates that it
is highly unlikely that a random CM network explains the frequent
appearance of this network motif. We have tested the prevalence of
this motif with other values of $m$ and the number of occurrences
is consistently a few times more than expected in a random CM network.
Our analysis suggests that reliability considerations is one of the
factors leading to the increased number of incidents.

The unexpected prevalence of the ``feedback loop'', which is a special
case of the ``entangled cycles'' motif, was first reported in \cite{Milo2002a}.
The ``feedback loop'' motif coincides with the \textquotedbl{}entangled
cycles\textquotedbl{} motif of length three. In order to further assess
our results we tested for the occurrence frequency of the ``entangled
cycles'' motif of length four. We compared the number of occurrences
of this motif in the real-world Internet graph to the expected number
of occurrences in a random Configuration Model network. While the
number of instances of this motif in the Internet graph was $27.7M$,
the expected number of occurrences in the random network was only
$1.3M\pm0.8M$. The abundance of this network motif, an order of magnitude
greater than expected (\emph{$p<0.001$}, a relaxed bound based on
Chebyshev's inequality) provides a positive indication to the implications
of survivability requirements.

In summary, we have provided both static and dynamic empirical evidence
that conform with our predictions, suggesting the importance of reliability
considerations on the structure and dynamics of the inter-AS topology.

\section{Conclusions }

Does the Internet resembles a clique or a tree? Is it contracting
or expanding? Can one statement be true on one segment of the network
while the opposite is correct on a different segment? How do reliability
requirements affect the topological structure of the Internet? The
game theoretic model presented in this work, while abstracting many
details away, focuses on the essence of the strategical decision-making
that ASs perform. It provides answers to such questions by addressing
the different roles ASs play.

Our inherently heterogeneous network formation game is flexible, and
may be used in a wide variety of settings. It allows for many variations
and schemes, for example situations in which failures are frequent
or rare, or if monetary transfers are feasible or not. 

The static analysis has indicated that in all equilibria, the major
players form a clique. Our model predicts that the major players clique
will form a \emph{settlement-free} interconnection subgraph, while
in major player - minor player contracts transactions will occur,
and they will be of a transit contract type. This observation is supported
by the empirical evidence,showing the tight tier-1 subgraph, and the
fact these ASs provide transit service to the other ASs. We established
the \emph{Price of Reliability, }which measures the excess social
cost that is required in order to maintain network survivability in
an optimal stable equilibrium. Surprisingly, we showed that it can
be smaller than one, that is, the additional survivability constraints
\emph{add }to the social utility. We have also showed that reliability
requirements have disparate effects on different parts of the network.
While it may support dilution in dense areas, it facilitates edges
formation in sparse areas, and in particular it supports the formation
of edges connecting minor players and major players. 

We discussed multiple dynamics, which represent different scenarios
and playing orders. The dynamic analysis showed that, when the individual
players act selfishly and rationally, the system will converge to
either the optimal configuration or to a state in which the social
cost differs by a negligible amount from the optimal social cost.
This is important as a prospective mechanism design. Furthermore,
although a multitude of equilibria exist, the dynamics restrict the
convergence to a limited set.. We also learned that, as the number
of major players increase, the distance of the minor players to the
core should decrease. This was also confirmed empirically. In our
dynamic analysis we have found the repetitive appearance of small
sub-graphs, or network motifs, namely the ``entangled cycles'' motif
and the ``double star'' motif. Indeed, the number of appearance
of these motifs in the real Inter-AS topology surpassed the expected
number by a few times, indicating that additional factors support
their formation, and as our analysis shows, survivability is one of
them. 

Finally, while our analysis focuses on the inter-AS topology, it may
be applied to other networks as well, that are composed of heterogeneous,
rational agents that are required to maintain some reliability aspects.
Primary examples are trade networks and MVNO operators in the cellular
market.


\begin{thebibliography}{10}
\providecommand{\url}[1]{#1}
\csname url@samestyle\endcsname
\providecommand{\newblock}{\relax}
\providecommand{\bibinfo}[2]{#2}
\providecommand{\BIBentrySTDinterwordspacing}{\spaceskip=0pt\relax}
\providecommand{\BIBentryALTinterwordstretchfactor}{4}
\providecommand{\BIBentryALTinterwordspacing}{\spaceskip=\fontdimen2\font plus
\BIBentryALTinterwordstretchfactor\fontdimen3\font minus
  \fontdimen4\font\relax}
\providecommand{\BIBforeignlanguage}[2]{{%
\expandafter\ifx\csname l@#1\endcsname\relax
\typeout{** WARNING: IEEEtran.bst: No hyphenation pattern has been}%
\typeout{** loaded for the language `#1'. Using the pattern for}%
\typeout{** the default language instead.}%
\else
\language=\csname l@#1\endcsname
\fi
#2}}
\providecommand{\BIBdecl}{\relax}
\BIBdecl

\bibitem{Gregori2013}
E.~Gregori, L.~Lenzini, and C.~Orsini, ``{k-Dense communities in the Internet
  AS-level topology graph},'' \emph{Computer Networks}, vol.~57, no.~1, Jan.
  2013.

\bibitem{Vazquez2002}
A.~V\'{a}zquez, R.~Pastor-Satorras, and A.~Vespignani, ``{Large-scale
  topological and dynamical properties of the Internet},'' \emph{Physical
  Review E}, vol.~65, no.~6, p. 066130, Jun. 2002.

\bibitem{Siganos2003}
G.~Siganos, M.~Faloutsos, P.~Faloutsos, and C.~Faloutsos, ``{Power laws and the
  AS-level internet topology},'' \emph{IEEE/ACM Trans. Netw.}, vol.~11, no.~4,
  Aug. 2003.

\bibitem{Barabasi1999}
A.~Barab\'{a}si, ``{Emergence of Scaling in Random Networks},'' \emph{Science},
  vol. 286, no. 5439, Oct. 1999.

\bibitem{1019306}
Q.~Chen, H.~Chang, R.~Govindan, and S.~Jamin, ``{The origin of power laws in
  Internet topologies revisited},'' in \emph{INFOCOM 2002. Proceedings. IEEE},
  vol.~2, 2002.

\bibitem{Borkar2007}
V.~S. Borkar and D.~Manjunath, ``{Distributed topology control of wireless
  networks},'' \emph{Wireless Networks}, vol.~14, no.~5, Jan. 2007.

\bibitem{Altman1994}
E.~Altman, ``{Flow control using the theory of zero sum Markov games},''
  \emph{IEEE Trans. Automat. Controll}, vol.~39, no.~4, Apr. 1994.

\bibitem{Roy2010}
S.~Roy, C.~Ellis, S.~Shiva, D.~Dasgupta, V.~Shandilya, and Q.~Wu, ``{A Survey
  of Game Theory as Applied to Network Security},'' \emph{HICSS 2010}, 2010.

\bibitem{Orda1993}
A.~Orda, R.~Rom, and N.~Shimkin, ``{Competitive routing in multiuser
  communication networks},'' \emph{IEEE/ACM Trans. Netw.}, vol.~1, no.~5, 1993.

\bibitem{Charilas2010}
D.~E. Charilas and A.~D. Panagopoulos, ``{A survey on game theory applications
  in wireless networks},'' \emph{Computer Networks}, vol.~54, no.~18, Dec.
  2010.

\bibitem{Johari2006}
R.~Johari, S.~Mannor, and J.~N. Tsitsiklis, ``{A contract-based model for
  directed network formation},'' \emph{Games and Economic Behavior}, vol.~56,
  no.~2, Aug. 2006.

\bibitem{Jackson1996}
M.~O. Jackson and A.~Wolinsky, ``{A Strategic Model of Social and Economic
  Networks},'' \emph{Journal of Economic Theory}, vol.~71, no.~1, 1996.

\bibitem{5062080}
A.~Nahir, A.~Orda, and A.~Freund, ``{Topology Design and Control: A
  Game-Theoretic Perspective},'' in \emph{INFOCOM 2009, IEEE}, 2009.

\bibitem{Anshelevich2011}
E.~Anshelevich, F.~Shepherd, and G.~Wilfong, ``{Strategic network formation
  through peering and service agreements},'' \emph{Games and Economic
  Behavior}, vol.~73, no.~1, Sep. 2011.

\bibitem{Alvarez2012}
\BIBentryALTinterwordspacing
C.~\`{A}lvarez and A.~Fern\`{a}ndez, ``{Network Formation: Heterogeneous
  Traffic, Bilateral Contracting and Myopic Dynamics},'' Mar. 2012. [Online].
  Available: \url{http://arxiv.org/abs/1203.5715}
\BIBentrySTDinterwordspacing

\bibitem{Lodhi2012a}
A.~Lodhi, A.~Dhamdhere, and C.~Dovrolis, ``{GENESIS: An agent-based model of
  interdomain network formation, traffic flow and economics},'' in \emph{2012
  Proc. IEEE INFOCOM}.\hskip 1em plus 0.5em minus 0.4em\relax IEEE, Mar. 2012,
  pp. 1197--1205.

\bibitem{Vandenbossche2012}
J.~Vandenbossche and T.~Demuynck, ``{Network Formation with Heterogeneous
  Agents and Absolute Friction},'' \emph{Computational Economics}, Jan. 2012.

\bibitem{Arcaute2013}
E.~Arcaute, K.~Dyagilev, R.~Johari, and S.~Mannor, ``{Dynamics in tree
  formation games},'' \emph{Games and Economic Behavior}, vol.~79, May 2013.

\bibitem{Bala}
V.~Bala and S.~Goyal, ``{A strategic analysis of network reliability},''
  \emph{Rev. Econ. Des.}, 2000.

\bibitem{Haller2005}
H.~Haller and S.~Sarangi, ``{Nash networks with heterogeneous links},''
  \emph{Mathematical Social Sciences}, vol.~50, no.~2, Sep. 2005.

\bibitem{Fabrikant2003}
A.~Fabrikant, A.~Luthra, E.~Maneva, C.~H. Papadimitriou, and S.~Shenker, ``{On
  a network creation game},'' in \emph{PODC '03}.\hskip 1em plus 0.5em minus
  0.4em\relax ACM Press, Jul. 2003, pp. 347--351.

\bibitem{Corbo2005}
J.~Corbo and D.~Parkes, ``{The price of selfish behavior in bilateral network
  formation},'' in \emph{Proc. PODC '05}.\hskip 1em plus 0.5em minus
  0.4em\relax New York, New York, USA: ACM Press, Jul. 2005, p.~99.

\bibitem{Meirom2014}
E.~A. Meirom, S.~Mannor, and A.~Orda, ``{Network formation games with
  heterogeneous players and the internet structure},'' in \emph{Proc. fifteenth
  ACM Conf. Econ. Comput. - EC '14}.\hskip 1em plus 0.5em minus 0.4em\relax New
  York, New York, USA: ACM Press, Jun. 2014, pp. 735--752.

\bibitem{Meirom2015a}
\BIBentryALTinterwordspacing
------, ``{Formation games of reliable networks},'' in \emph{2015 IEEE Conf.
  Comput. Commun.}\hskip 1em plus 0.5em minus 0.4em\relax IEEE, apr 2015, pp.
  1760--1768. [Online]. Available:
  \url{http://ieeexplore.ieee.org/lpdocs/epic03/wrapper.htm?arnumber=7218557}
\BIBentrySTDinterwordspacing

\bibitem{Suurballe1974}
J.~W. Suurballe, ``{Disjoint paths in a network},'' \emph{Networks}, vol.~4,
  no.~2, pp. 125--145, 1974.

\bibitem{Bhandari1999}
R.~Bhandari, \emph{{Survivable Networks: Algorithms for Diverse Routing}},
  1999.

\bibitem{5173479}
E.~Arcaute, R.~Johari, and S.~Mannor, ``{Network Formation: Bilateral
  Contracting and Myopic Dynamics},'' \emph{IEEE Trans. Automat. Control},
  vol.~54, no.~8, 2009.

\bibitem{NisanN.RoughgardenT.TardosE.2007}
T.~E. V.~V. Nisan~N., Roughgarden~T., Ed., \emph{{Algorithmic Game
  Theory}}.\hskip 1em plus 0.5em minus 0.4em\relax Cambridge University Press,
  2007.

\bibitem{Meirom2013}
E.~A. Meirom, S.~Mannor, and A.~Orda, ``{Formation Games and the Internet's
  Structure},'' \emph{http://arxiv.org/abs/1307.4102}, Jul. 2013.

\bibitem{Milo2002a}
R.~Milo, S.~Shen-Orr, S.~Itzkovitz, N.~Kashtan, D.~Chklovskii, and U.~Alon,
  ``{Network motifs: simple building blocks of complex networks.}''
  \emph{Science}, no. 5594, pp. 824--7, Oct.

\bibitem{Gregori2011}
E.~Gregori, A.~Improta, L.~Lenzini, L.~Rossi, and L.~Sani, ``Bgp and inter-as
  economic relationships,'' in \emph{Proc. of the 10th international IFIP TC 6
  conference on Networking}, 2011.

\bibitem{CAIDA}
\BIBentryALTinterwordspacing
``{CAIDA AS Ranking}.'' [Online]. Available: \url{http://as-rank.caida.org/}
\BIBentrySTDinterwordspacing

\bibitem{Cohen2003}
R.~Cohen and S.~Havlin, ``{Scale-Free Networks Are Ultrasmall},'' \emph{Phys.
  Rev. Lett.}, vol.~90, no.~5, p. 058701, Feb. 2003.

\bibitem{Pastor-Satorras2001}
R.~Pastor-Satorras, A.~V\'{a}zquez, and A.~Vespignani, ``{Dynamical and
  Correlation Properties of the Internet},'' \emph{Phys. Rev. Lett.}, vol.~87,
  no.~25, p. 258701, Nov. 2001.

\end{thebibliography}

\pagebreak{}

\section{Appendix : Preliminaries}

\begin{figure}
\centering{}\includegraphics[width=0.7\columnwidth]{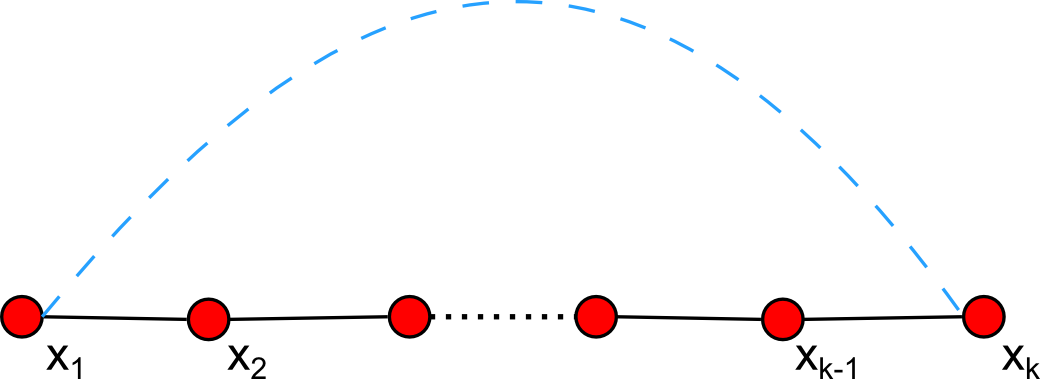}\protect\caption{\label{fig:long line}The scenario described in Lemma \ref{lem:shortcut benefit}.
The additional link is dashed in blue.}
\end{figure}

The next lemma will be useful in many instances. It measures the benefit
of connecting the two ends of a long line of players, as presented
in \ref{fig:long line}. If the line is too long, it is better for
both parties at its end to form a link between them.

\begin{lem}
\label{lem:shortcut benefit}Assume of lone line having $k$ nodes,
$(x_{1},x_{2},...x_{k}).$ By establishing the link $(x_{1},x_{k})$
the sum of distances $\sum_{i}d(x_{1},x_{i})$ $\left(\sum d(x_{k},x_{i})\right)$
is reduced by
\[
\frac{k\left(k-2\right)+mod(k,2)}{4}
\]
\end{lem}
\begin{IEEEproof}
Without the link $(x_{1},x_{k})$ the sum of distances is given by
the algebraic series
\[
\sum_{i}d(x_{1},x_{i})=\sum_{i=1}^{k-1}i=\frac{k\left(k-1\right)}{2}
\]

If $k$ is odd, than the the addition of the link $(x_{1},x_{k})$
we have 
\begin{eqnarray*}
\sum_{i}d(x_{1},x_{i}) & = & 2\sum_{i=1}^{\left\lfloor k/2\right\rfloor }i=\left(\left\lfloor k/2\right\rfloor +1\right)\left\lfloor k/2\right\rfloor \\
 & = & \left(\left(k-1\right)/2+1\right)\left(\left(k-1\right)/2\right)\\
 & = & \frac{k^{2}-1}{4}
\end{eqnarray*}

If $k$ is even, the corresponding sum is 
\begin{eqnarray*}
\sum_{i}d(x_{1},x_{i}) & = & \sum_{i=1}^{k/2}i+\sum_{i=1}^{k/2-1}i\\
 & = & \frac{\left(k/2+1\right)k}{4}+\frac{\left(k/2-1\right)k}{4}\\
 & = & \frac{k^{2}}{4}
\end{eqnarray*}

We conclude that the difference for $k$ even is 
\[
\frac{k^{2}}{4}-\frac{k}{2}=\frac{k\left(k-2\right)}{4}
\]

and for odd $k$ is
\[
\frac{k^{2}}{4}-\frac{k}{2}+\frac{1}{4}=\frac{k\left(k-2\right)+1}{4}
\]
\end{IEEEproof}

\end{document}